\documentclass[english]{IEEEtran}
\usepackage[T1]{fontenc}
\usepackage[latin9]{inputenc}
\usepackage{color}
\usepackage{float}
\usepackage{booktabs}
\usepackage{amsmath}
\usepackage{amsthm}
\usepackage{amssymb}
\usepackage{graphicx}

\makeatletter

\providecommand{\tabularnewline}{\\}
\floatstyle{ruled}
\newfloat{algorithm}{tbp}{loa}
\providecommand{\algorithmname}{Algorithm}
\floatname{algorithm}{\protect\algorithmname}

\theoremstyle{plain}
\newtheorem{thm}{\protect\theoremname}
\theoremstyle{plain}
\newtheorem{lem}[thm]{\protect\lemmaname}

\usepackage{algorithm}
\usepackage{algorithmic}

\newtheorem{remark}{Remark}\newtheorem{proposition}{Proposition}\usepackage{empheq}
\usepackage[compress]{cite}
\usepackage{eucal}
\usepackage{array}
\usepackage{cases}
\usepackage{mathrsfs}
\usepackage{mathrsfs}\usepackage[noblocks]{authblk}
\usepackage{amsfonts}
\allowdisplaybreaks[4]

\usepackage{babel}
\providecommand{\lemmaname}{Lemma}
\providecommand{\theoremname}{Theorem}
\usepackage[noblocks]{authblk}

\makeatother

\usepackage{babel}
\providecommand{\lemmaname}{Lemma}
\providecommand{\theoremname}{Theorem}

\begin{document}
\title{Channel Estimation for RIS-Aided Multi-User mmWave Systems with Uniform
Planar Arrays }
\author{Zhendong Peng, Gui Zhou, Cunhua Pan, Hong Ren, A. Lee Swindlehurst,\textit{
Fellow}, \textit{IEEE}, Petar Popovski, \textit{Fellow}, \textit{IEEE},
and Gang Wu\thanks{\textit{(Corresponding author: Cunhua Pan.)}

Z. Peng was with the University of Electronic Science and Technology
of China. (e-mail: zhendongpeng@ece.ubc.ca). G. Zhou is with the Institute
for Digital Communications, Friedrich-Alexander-University Erlangen-N{\"u}rnberg
(FAU), 91054 Erlangen, Germany (e-mail: gui.zhou@fau.de). C. Pan and
H. Ren are with the National Mobile Communications Research Laboratory,
Southeast University, Nanjing 210096, China. (e-mail: cpan@seu.edu.cn;
hren@seu.edu.cn). A. L. Swindlehurst is with the Center for Pervasive
Communications and Computing, University of California, Irvine, CA
92697, USA (e-mail: swindle@uci.edu). P. Popovski is with the Department
of Electronic Systems, Aalborg University, 9220 Aalborg, Denmark (e-mail:
petarp@es.aau.dk). G. Wu is with the National Key Laboratory of Science
and Technology on Communication, University of Electronic Science
and Technology of China, Chengdu 611731, China (e-mail: wugang99@uestc.edu.cn).
Part of this work will be presented in IEEE Global Communications
Conference, Rio de Janeiro, Brazil, Dec. 2022 \cite{Zhendong-CE-GC}.}}

\maketitle
\vspace{-0.8cm}

\begin{abstract}
In this paper, we adopt a three-stage based uplink channel estimation
protocol with reduced pilot overhead for an reconfigurable intelligent
surface (RIS)-aided multi-user (MU) millimeter wave (mmWave) communication
system, in which both the base station (BS) and the RIS are equipped
with a uniform planar array (UPA). Specifically, in Stage I, the channel
state information (CSI) of a typical user is estimated. To address
the power leakage issue for the common angles-of-arrival (AoAs) estimation
in this stage, we develop a low-complexity one-dimensional search
method. In Stage II, a re-parameterized common BS-RIS channel is constructed
with the estimated information from Stage I to estimate other users'
CSI. In Stage III, only the rapidly varying channel gains need to
re-estimated. Furthermore, the proposed method can be extended to
multi-antenna UPA-type users, by decomposing the estimation of a multi-antenna
channel with $J$ scatterers into estimating $J$ single-scatterer
channels for a virtual single-antenna user. An orthogonal matching
pursuit (OMP)-based method is proposed to estimate the angles-of-departure
(AoDs) at the users. Simulation results demonstrate that the proposed
algorithm significantly achieves high channel estimation accuracy,
which approaches the genie-aided upper bound in the high signal-to-noise
ratio (SNR) regime. 
\end{abstract}


\vspace{-0.2cm}

\section{Introduction}

\vspace{-0.1cm}
 Thanks to its cost-effective, power-efficient and deployment-convenient
features, reconfigurable intelligent surface (RIS) technology is envisioned
to be a promising technique for enhancing the spectrum and energy
efficiency of 6G-and-beyond communications systems\cite{Zhangrui_MAG,marco,Pan_MAG,You_6G_new_overiview,Pan2019intelleget,Pan2019multicell}.
Deploying an RIS provides additional degrees-of-freedom (DoF) that
can be used to reconfigure the wireless propagation environment, which
brings tremendous benefits for the wireless systems. To reap the benefits
promised by RIS, accurate channel state information (CSI) is required
\cite{Overview_IEEE_LEE,Beixiong2021survey,Liang_RIS_Overview}, which
is challenging to achieve for the following two reasons. First, an
RIS equipped with passive elements typically does not have a receiver,
so does not process complex baseband signals, which means that traditional
channel estimation approaches cannot be adopted in RIS-aided systems.
Due to this characteristic, it is not possible to estimate the user-RIS
channel and RIS-base station (BS) channel separately, and instead
the cascaded channel is estimated, i.e., the equivalent user-RIS-BS
channel. Second, with a large number of antennas at the BS and reflecting
elements at the RIS, the cascaded channel contains a large number
of channel coefficients, which can require a larger number of pilots.
Hence, developing an efficient channel estimation method for RIS-aided
systems with low pilot overhead is imperative.

Recently, there have been many contributions on channel estimation
for RIS-aided communication systems; see for example \cite{ON_OFF_MU,Liuliang-IRS,ris-omp-1,Jiguang_atomic,ris-omp-2,ris-omp-3,Zhou_ULA_TSP,CE_MIMO_RIS}
and the recent overview tutorial \cite{Overview_IEEE_LEE}. Early
work focused mainly on unstructured channel models, but channel estimation
for these models requires a pilot overhead that is proportional to
the number of RIS reflecting elements, which is often prohibitively
large. On the other hand, the sparse structure of high-frequency millimeter
wave (mmWave) channels, described by the angles and gains of fewer
paths, has been exploited to reduce the pilot overhead and improve
the estimation accuarcy of multiple-input multiple-output (MIMO) systems
efficiently by leveraging compressed sensing (CS) techniques, direction-of-arrival
(DOA) estimation methods and Bayesian learning frameworks \cite{UPA-MIMO,Super_Resolution_Hybrid,Fan_AoD_CE-TWC2018,Bayesian-learningTVT}.
Motivated by the works on structured channel models, the sparsity
of the user-RIS-BS cascaded channel was exploited in \cite{ris-omp-1}
using CS to reconstruct the channel. The authors in \cite{ris-omp-2}
exploited the fact that the cascaded channel matrices for multiple
users exhibit a common column-block sparsity since all users share
the same RIS-BS channel, and developed an iterative channel estimator
based on this observation. Inspired by the common column-block sparsity
property, the double-structured sparsity of the cascaded channel was
considered in \cite{ris-omp-3}, using the Discrete Fourier Transform
(DFT) to analyze the estimation of the angle parameters. The authors
of \cite{Zhou_ULA_TSP} achieved a dramatic reduction in pilot overhead
by fully utilizing the correlation among the different cascaded channels.
The above-mentioned works \cite{ris-omp-2,ris-omp-3,Zhou_ULA_TSP}
considered multiple users but assumed that they are equipped with
only a single antenna. On the other hand, the RIS-aided MIMO scenario
was considered in \cite{ris-omp-1,CE_MIMO_RIS,Jiguang_atomic}. The
authors in \cite{CE_MIMO_RIS} proposed an alternating minimization
and manifold optimization (MO) estimation protocol for this scenario.
To increase the estimation accuracy, a super-resolution CS technique
based on atomic norm minimization was applied to cascaded channel
estimation in \cite{Jiguang_atomic}. However, these three works assumed
only a single user and thus did not take advantage of the inherent
correlation among the channels of different users in an RIS-aided
system. Apart from this, \cite{ris-omp-1,Jiguang_atomic,ris-omp-3,CE_MIMO_RIS}
assumed that the number of scatterers for the user-RIS channel and
RIS-BS channels are known a priori, i.e., the sparsity level is known.
In practice, however, these parameters may not be known beforehand.
Moreover, a uniform linear array (ULA)-type BS, ULA-type users and/or
ULA-type RIS were assumed in the above mentioned works, which may
not be relevant for RIS-assisted communication systems. The extension
to the more typical uniform planar array (UPA)-type RIS-aided multi-user
(MU) system is not straightforward. First, the number of angle parameters
that must be estimated doubles that of a ULA-type system, and the
asymptotic properties exploited for large ULAs may not be applicable.
Second but important, increasing the number of parameters makes exploiting
the channel correlation among multiple users extremely complex, especially
for the cascaded channel parameters.

Against the above background, in this paper we propose an effective
three-stage channel estimation method with low pilot overhead starting
from an RIS-aided single-antenna MU mmWave communication system, in
which the BS and RIS are both equipped with a UPA. Then, we extend
the protocol to the multi-antenna user case, where the users are also
equipped with UPAs. This is the first work that investigates the UPA-type
MU MIMO case. The main contributions of this work are summarized as
follows: 
\begin{itemize}
\item We develop a three-stage uplink channel estimation protocol for an
RIS-aided mmWave communication system with a multi-antenna UPA-type
BS, a multi-element UPA-type RIS and multiple users. The protocol
is divided into two parts: full CSI estimation in the first coherence
block consisting of Stage I and Stage II, and estimation of updated
gains in the remaining coherence blocks consisting of Stage III. In
Stage I, only a typical user sends pilots to the BS for channel estimation,
from which we obtain estimated gains and angle information that is
used to reduce the pilot overhead in the next stage. In particular,
angle rotation operation is adopted to deal with the power leakage
issue when estimating common AoAs in this stage. In Stage II, we exploit
the correlation among different users' cascaded channels and construct
a re-parameterized common RIS-BS channel using the estimated CSI of
the typical user, based on which we obtain the channel estimates of
other users. Next, in Stage III during the remaining coherence blocks,
only the cascaded channel gains for different users are re-estimated
since the angle information remains constant. 
\item We propose an effective low-complexity one-dimensional (1-D) search
method to achieve the angle rotation operation in Stage I. In \cite{Fan_AoD_CE-TWC2018},
a two-dimensional (2-D) DFT together with a 2-D search method was
used to compensate for the leaked power, which has high computational
complexity. To reduce the complexity, we exploit the structure of
the steering vectors at the BS and then introduce an equivalent Fourier
matrix and rotation matrices to divide the 2-D search into two 1-D
searches. 
\item We extend the estimation protocol to the case of users with UPAs.
The angles-of-departure (AoDs) at the users and the common angles-of-arrival
(AoAs) at the BS are estimated via the proposed orthogonal matching
pursuit (OMP)-based method and DFT-based method, respectively. Then
the estimation of a multi-antenna channel with $J$ scatterers is
decomposed into the estimation of $J$ single-scatterer channels.
The cascaded AoDs at the RIS and the channel gains can be estimated
using methods similar to those developed for the single-antenna case.
This is the first approach proposed in the literature that exploits
the correlation between different users in the multi-antenna user
case. The overall number of pilots for both the single- and multi-antenna
case is also analyzed. 
\end{itemize}
The rest of this paper is organized as follows. Section \ref{sec:Model-protocol}
introduces the system model and the three-stage based channel estimation
protocol. Section \ref{sec:First_coherence} presents the full CSI
estimation algorithm in Stage I and Stage II for the single-antenna-users
case. Channel gain estimation in Stage III is discussed in Section
\ref{sec:Remaining_Coherence}. Section \ref{sec:Applying-the-Protocol}
applies the protocol to the multi-antenna-users case. Simulation results
are given in Section \ref{sec:Simulation-Results}. Finally, Section
\ref{sec:Conclusions} concludes this work.

\textit{Notations}: Vectors and matrices are denoted by boldface lowercase
letters and boldface uppercase letters, respectively. For a matrix
$\mathbf{A}$ of arbitrary size, $\mathbf{A}^{*}$, $\mathbf{A}^{\mathrm{T}}$,
$\mathbf{A}^{\mathrm{H}}$ and $\mathbf{A}^{\mathrm{\dagger}}$ stand
for the conjugate, transpose, conjugate transpose and pseudo-inverse
of $\mathbf{A}$. For a square full-rank matrix $\mathbf{A}$, $\mathbf{A}^{-1}$
denotes its inverse. The symbols $||\mathbf{A}||_{F}$, $||\mathbf{a}||$
represent the Frobenius norm of matrix $\mathbf{A}$ and the Euclidean
norm of vector $\mathbf{a}$, respectively. $\angle\left(\cdot\right)$
denotes the angle of a complex number. $\mathrm{Diag}\{\mathbf{a}\}$
is a diagonal matrix with the entries of vector $\mathbf{a}$ on its
diagonal. $\mathrm{vec}(\mathbf{A})$ denotes the vectorization of
$\mathbf{A}$ obtained by stacking the columns of matrix $\mathbf{A}$.
$\mathbb{E}\left\{ \cdot\right\} $ denotes the expectation operation.
$[\mathbf{a}]_{m}$ denotes the $m$-th element of the vector $\mathbf{a}$,
and $[\mathbf{A}]_{m,n}$ denotes the $(m,n)$-th element of the matrix
$\mathbf{A}$. The $n$-th column and the $m$-th row of matrix $\mathbf{A}$
are denoted by $\mathbf{A}_{(:,n)}$ and $\mathbf{A}_{(m,:)}$ respectively.
$\left\lceil a\right\rceil $ rounds up to the nearest integer. The
inner product between two vectors $\mathbf{a}$ and $\mathbf{b}$
is denoted by $\left\langle \mathbf{a},\mathbf{b}\right\rangle \triangleq\mathbf{a}^{\mathrm{H}}\mathbf{b}$.
Additionally, the Kronecker product, Hadamard product, Khatri-Rao
product and transposed Khatri-Rao product between two matrices $\mathbf{A}$
and $\mathbf{B}$ are denoted by $\mathbf{A}\otimes\mathbf{B}$,\textcolor{red}{{}
}$\mathbf{A}\odot\mathbf{B}$, $\mathbf{A}\diamond\mathbf{B}$ and
$\mathbf{A}\bullet\mathbf{B}$,\footnote{The transposed Khatri-Rao product is known as the ``row-wise Kronecker
product'', which utilizes the row-wise splitting of matrices with
a given quantity of rows. Specifically, for given matrices $\mathbf{A}\in\mathbb{C}^{Q\times M}$
and $\mathbf{B}\in\mathbb{C}^{Q\times N}$, $\mathbf{A}\bullet\mathbf{B}$
is a $Q\times MN$ matrix of which each row is the Kronecker product
of the corresponding rows of $\mathbf{A}$ and $\mathbf{B}$.} respectively. $\mathrm{i}\triangleq\sqrt{-1}$ is the imaginary unit.

\vspace{-0.2cm}

\section{System Model and estimation protocol \label{sec:Model-protocol}}

\vspace{-0.1cm}

\subsection{System Model\label{subsec:System-Model}}

\vspace{-0.1cm}
 We consider a narrow-band time-division duplex (TDD) mmWave system,
in which $K$ single-antenna users communicate with a BS equipped
with an $N=N_{1}\times N_{2}$ antenna UPA, where $N_{1}$ is the
number of antennas in the vertical dimension, and $N_{2}$ in the
horizontal dimension. To improve communication performance, an RIS
equipped with a passive reflecting UPA of dimension $M=M_{1}\times M_{2}$
($M_{1}$ vertical elements and $M_{2}$ hoirzontal elements) is deployed.
The channels are assumed to be block-fading, and hence constant in
each coherence block. In addition, we assume that the direct channels
between the BS and users are blocked. Otherwise first estimate the
direct channels by turning off the RIS, and then the cascaded channel
can be estimated by removing the direct channel's contribution from
the received signal.

The Saleh-Valenzuela (SV) model in \cite{mmWave_channel_overview}
is used to represent the channels due to the limited scattering characteristics
in the mmWave environment. Consider a typical $P=P_{1}\times P_{2}$
UPA whose steering vector $\mathbf{a}_{P}(z,x)\in\mathbb{C}^{P\times1}$
can be represented by\vspace{-0.1cm}
\begin{equation}
\mathbf{a}_{P}(z,x)=\mathbf{a}_{P_{1}}(z)\otimes\mathbf{a}_{P_{2}}(x),\label{eq:ax}
\end{equation}
where $\mathbf{a}_{P_{1}}(z)=[1,e^{-\mathrm{i}2\pi z},\ldots,e^{-\mathrm{i}2\pi(P_{1}-1)z}]^{\mathrm{T}}$
and $\mathbf{a}_{P_{2}}(x)=[1,e^{-\mathrm{i}2\pi x},\ldots,e^{-\mathrm{i}2\pi(P_{2}-1)x}]^{\mathrm{T}}$
are the steering vectors with respect to $z$-axis (vertical direction)
and $x$-axis (horizontal direction) of the UPA, respectively. The
variables $z$ and $x$ can be regarded as the corresponding equivalent
spatial frequency with respect to $z$-axis and $x$-axis of the UPA,
respectively. Denote $\mathfrak{\varrho}\in[-90^{\mathrm{o}},90^{\mathrm{o}})$
and $\mathfrak{\xi}\in[-180^{\mathrm{o}},180^{\mathrm{o}})$ as the
signal elevation and azimuth angles of the UPA, respectively. There
exists a relationship between the spatial frequency pair $(z,x)$
and the physical angle pair $(\mathfrak{\varrho},\mathfrak{\xi})$:\vspace{-0.1cm}
\begin{equation}
z=\frac{d}{\lambda_{c}}\cos(\mathfrak{\varrho}),~x=\frac{d}{\lambda_{c}}\sin(\varrho)\cos(\mathfrak{\xi}),\label{phy_angle}
\end{equation}
where $\lambda_{c}$ is the carrier wavelength and $d$ is the element
spacing. Assuming that $d\le\lambda_{c}/2$, there is a one-to-one
relationship between the spatial frequencies and the physical angles
on one side of the UPA. We will assume this relationship to hold in
the remainder of the paper, and we will refer to the arguments of
the steering vectors interchangeably as either angles or spatial frequencies.

Using the geometric channel model, the channel matrix between the
RIS and the BS, denoted by $\mathbf{H}\in\mathbb{C}^{N\times M}$,
and the channel matrix between user $k$ and the RIS, denoted by $\mathbf{h}_{k}\in\mathbb{C}^{M\times1}$,
can be written as \vspace{-0.1cm}

\begin{subequations}
\label{eq:H_h} 
\begin{align}
\mathbf{H}=\sum_{l=1}^{L}\alpha_{l}\mathbf{a}_{N}(\psi_{l},\nu_{l})\mathbf{a}_{M}^{\mathrm{H}}(\omega_{l},\mu_{l}),\\
\mathbf{h_{\mathit{k}}}=\sum_{j=1}^{J_{k}}\beta_{k,j}\mathbf{a}_{M}(\varphi_{k,j},\theta_{k,j}),\forall k\in\mathcal{K},
\end{align}
\end{subequations}

where $L$ denotes the number of propagation paths (scatterers) between
the BS and the RIS, and $J_{k}$ denotes the number of propagation
paths between the RIS and user $k$. In addition, $\alpha_{l}$, $(\psi_{l},\nu_{l})$
and $(\omega_{l},\mu_{l})$ are the complex path gain, AoA, and AoD
of the $l$-th path in the RIS-BS channel, respectively. Similarly,
$\beta_{k,j}$ and $(\varphi_{k,j},\theta_{k,j})$ represent the complex
path gain and AoA of the $j$-th path in the user $k$-RIS channel,
respectively. Moreover, the channel models in (\ref{eq:H_h}) can
be written in a more compact way as \vspace{-0.2cm}
\begin{align}
\mathbf{H} & =\mathbf{A}_{N}\boldsymbol{\Lambda}\mathbf{A}_{M}^{\mathrm{H}},\label{eq:H1-1}\\
\mathbf{h}_{k} & =\mathbf{A}_{M,k}\boldsymbol{\beta}_{k},\forall k\in\mathcal{K},\label{eq:hk}
\end{align}
where $\mathbf{A}_{N}=[\mathbf{a}_{N}(\psi_{1},\nu_{1}),\ldots,\mathbf{a}_{N}(\psi_{L},\nu_{L})]\in\mathbb{C}^{N\times L}$,
$\mathbf{A}_{M}=[\mathbf{a}_{M}(\omega_{1},\mu_{1}),\ldots,\mathbf{a}_{M}(\omega_{L},\mu_{L})]\in\mathbb{C}^{M\times L}$
and $\boldsymbol{\Lambda}=\mathrm{Diag}\{\alpha_{1},\ldots,\alpha_{L}\}\in\mathbb{C}^{L\times L}$
are the AoA steering (array response) matrix, AoD steering matrix
and complex gain matrix of the common RIS-BS channel, respectively,
and $\mathbf{A}_{M,k}=[\mathbf{a}_{M}(\varphi_{k,1},\theta_{k,1}),\ldots,\mathbf{a}_{M}(\varphi_{k,J_{k}},\theta_{k,J_{k}})]\in\mathbb{C}^{M\times J_{k}}$
and $\boldsymbol{\beta}_{k}=[\beta_{k,1},\ldots,\beta_{k,J_{k}}]^{\mathrm{T}}\in\mathbb{C}^{J_{k}\times1}$
are the AoA steering matrix and complex gain vector of the specific
user-RIS channel for user $k$, respectively.

Denote $\mathbf{e}_{t}\in\mathbb{C}^{M\times1}$ as the phase shift
vector of the RIS in time slot $t$ and define the user set as $\mathcal{K}=\{1,\ldots,K\}$.
Assume that users transmit pilot sequences of length $\tau_{k}$ one
by one for channel estimation. During the uplink transmission, in
time slot $t$, $1\leq t\leq\tau_{k}$, the received signal from user
$k$ at the BS can be expressed as \vspace{-0.15cm}
\begin{equation}
\mathbf{y}_{k}(t)=\mathbf{H}\mathrm{Diag}\{\mathbf{e}_{t}\}\mathbf{h}_{k}\sqrt{p}s_{k}(t)+\mathbf{n}_{k}(t),\label{transmission}
\end{equation}
where $s_{k}(t)$ is the pilot signal of the $k$-th user, $\mathbf{n}_{k}(t)\in\mathbb{C}^{N\times1}\sim\mathcal{CN}(0,\delta^{2}\mathbf{I})$
represents additive white Gaussian noise (AWGN) with power $\delta^{2}$
at the BS when user $k$ is transmitting. The scalar $p$ denotes
the transmit power of each user. Assume the pilot symbols satisfy
$s_{k}(t)=1,1\leq t\leq\tau_{k}$, so that Eq. (\ref{transmission})
can be expressed as \vspace{-0.15cm}
\begin{equation}
\begin{split}\mathbf{y}_{k}(t)=\mathbf{H}\mathrm{Diag}\{\mathbf{h}_{k}\}\mathbf{e}_{t}\sqrt{p}+\mathbf{n}_{k}(t)\triangleq\mathbf{G}_{k}\mathbf{e}_{t}\sqrt{p}+\mathbf{n}_{k}(t).\end{split}
\label{eq:2}
\end{equation}
Here, $\mathbf{G}_{k}=\mathbf{H}\mathrm{Diag}\{\mathbf{h}_{k}\}$
is regarded as the cascaded user-RIS-BS channel of user $k$, which
is the channel to be estimated in this work. Combining (\ref{eq:H1-1})
and (\ref{eq:hk}), $\mathbf{G}_{k}$ can be rewritten as \vspace{-0.1cm}
\begin{align}
\mathbf{G}_{k}=\mathbf{A}_{N}\boldsymbol{\Lambda}\mathbf{A}_{M}^{\mathrm{H}}\mathrm{Diag}\{\mathbf{A}_{M,k}\boldsymbol{\beta}_{k}\},\forall k\in\mathcal{K}.\label{eq:G1}
\end{align}
Stacking the $\tau_{k}$ time slots of (\ref{eq:2}), the received
matrix $\mathbf{Y}_{k}=\left[\mathbf{y}_{k}(1),\ldots,\mathbf{y}_{k}(\tau_{k})\right]$
is given by \vspace{-0.25cm}
\begin{equation}
\mathbf{Y}_{k}=\sqrt{p}\mathbf{G}_{k}\mathbf{E}_{k}+\mathbf{N}_{k}\in\mathbb{C}^{N\times\tau_{k}},\label{eq:m-y-2}
\end{equation}
where $\mathbf{E}_{k}=\left[\mathbf{e}_{1},\ldots,\mathbf{e}_{\tau_{k}}\right]\in\mathbb{C}^{M\times\tau_{k}}$
can be treated as the phase shift training matrix of the RIS for user
$k$ and $\mathbf{N}_{k}=\left[\mathbf{n}_{k}(1),\ldots,\mathbf{n}_{k}(\tau_{k})\right]\in\mathbb{C}^{N\times\tau_{k}}$.
\vspace{-0.25cm}

\subsection{Three-stage Channel Estimation Protocol\label{sec:protocol_overview}}

\begin{figure}
\begin{centering}
\includegraphics[width=0.9\columnwidth]{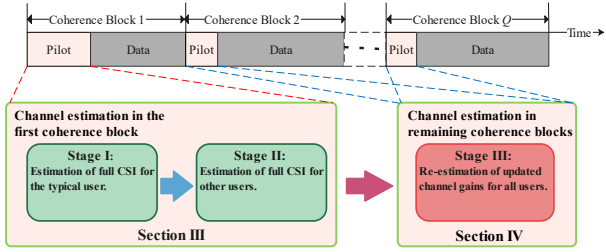} 
\par\end{centering}
\caption{The proposed three-stage channel estimation protocol.\label{flow_chart}}
\end{figure}

\vspace{-0.05cm}
 The main idea of the proposed channel estimation protocol are depicted
in Fig. \ref{flow_chart}, where ``Pilot'' and ``Data'' represent
the phases for uplink channel estimation, and downlink data transmission
at the BS side, respectively. Our work focus on the uplink channel
estimation of the cascaded channels. Specifically, in Stage I, only
one user's cascaded channel is estimated. For convenience, this user
is referred to as the typical user\textcolor{red}{.}\footnote{The user closest to the RIS is generally chosen as the typical user
since its reflected channel suffers from less severe path loss. Thus,
the received signal at the BS is stronger to ensure high estimation
performance. The location of users can be obtained using the global
position system (GPS) \cite{GPS_location_RIS}, for example.} Information regarding the common RIS-BS channel from the estimate
of the typical user's CSI is extracted in order to reduce the pilot
overhead of channel estimation for other users in the next stage.
Then, in Stage II, the cascaded channel of other users is divided
into two parts, a common part and a unique part. The common parts
can be readily obtained with the estimated angle information and cascaded
gains of the typical user obtained in the first stage. This can help
reduce the pilot overhead of estimating the other users' cascaded
channel since only a few pilots are required for estimating their
unique parts. Finally, it is observed that in the quasi-static situation,
the positions of the BS and the RIS are fixed, and the changes in
the physical positions of the users and their surrounding obstacles
are negligible over milliseconds, corresponding to several channel
coherence blocks \cite{mmWave_test_experiment_new,mmWave_test_Gao}.
This observation leads to the reasonable assumption that the angles
remain unchanged for multiple coherence blocks while the gains change
from block to block. Hence, Stage III is used for estimating the varying
channel gains for all users.

In the following sections, we can conclude that the pilots required
for different users depend on the number of paths between the user
side and the RIS, which can be estimated by the BS in this work. This
needs BS to determine the typical user, allocate the pilot slots required
for different users, and inform the users of this knowledge before
the next estimation period. The details of the adopted protocol will
be discussed later, first for the single-antenna user case and then
finally for the multi-antenna user case.

\vspace{-0cm}

\vspace{-0.1cm}

\section{Estimation in the First Coherence Block: Stage I and Stage II\label{sec:First_coherence}}

\vspace{-0.05cm}
 In this section, we start from the single-antenna user case to describe
the details of full CSI estimation of all users in the first coherence
block, formulating it as two sparse recovery problems in Stage I and
Stage II. Then, we analyze the pilot overhead and computational complexity
of the proposed method. This section lays the foundation for the extension
to the multi-antenna user case in Section \ref{sec:Applying-the-Protocol}.
\vspace{-0.2cm}

\subsection{Stage I: Estimation of Full CSI for Typical User\label{subsec:Stage-I:-Estimation}}

\vspace{-0.05cm}
 In this subsection, we provide details on full CSI estimation for
a typical single-antenna user, denoted as user $1$, where the common
AoAs are first estimated and then the cascaded gains and AoDs are
obtained.

\subsubsection{Estimation of Common AoAs}

Due to the UPA deployed at the BS and the RIS, the direct DFT approach
in \cite{Zhou_ULA_TSP,Fan_AoD_CE-TWC2018} cannot be used for AoA
estimation from $\mathbf{Y}_{1}$ in (\ref{eq:m-y-2}). Therefore,
we propose a modified DFT approach utilizing the properties of the
Kronecker product to estimate the common AoAs at the BS of the cascaded
channel, i.e., $\mathbf{A}_{N}$ in (\ref{eq:H1-1}). To this end,
we first provide two lemmas as follows. \vspace{-0.15cm}

\begin{lem}
\label{lem:1}When $N_{1}\rightarrow\infty$ and $\ensuremath{N_{2}\rightarrow\infty}$,
the following property holds 
\begin{equation}
\lim_{N\rightarrow\infty}\frac{1}{N}\mathbf{a}_{N}^{\mathrm{H}}(\psi_{j},\nu_{j})\mathbf{a}_{N}(\psi_{i},\nu_{i})=\begin{cases}
1 & \psi_{j}=\psi_{i},\nu_{i}=\nu_{j}\\
0 & \textrm{otherwise}
\end{cases},\label{eq:lemma1}
\end{equation}
where $N=N_{1}\times N_{2}$. (\ref{eq:lemma1}) implies that $\mathbf{A}_{N}^{\mathrm{H}}\mathbf{A}_{N}=N\mathbf{I}_{L}$
where $\mathbf{I}_{L}$ is the identity matrix with dimension $L\times L$. 
\end{lem}
\vspace{-0.15cm}

\begin{IEEEproof}
Please refer to Appendix A. 
\end{IEEEproof}
Define an equivalent Fourier matrix $\widetilde{\mathbf{U}}_{N}\mathbf{\triangleq U_{\mathit{N_{\mathrm{1}}}}\otimes}\mathrm{\mathbf{U}}_{\mathit{N_{\mathrm{2}}}}\in\mathbb{C}^{N\times N}$,
where $\mathrm{\mathbf{U}}_{\mathit{N_{\mathrm{1}}}}$ and $\mathrm{\mathbf{U}}_{\mathit{N_{\mathrm{2}}}}$
are the DFT matrices with $(n,m)$-th entries $[\mathrm{\mathbf{U}}_{\mathit{N_{\mathrm{1}}}}]_{n,m}=\frac{1}{\sqrt{N_{1}}}e^{-\mathrm{i}\frac{2\pi(n-1)(m-1)}{N_{1}}}$
and $[\mathrm{\mathbf{U}}_{\mathit{N_{\mathrm{2}}}}]_{n,m}=\frac{1}{\sqrt{N_{2}}}e^{-\mathrm{i}\frac{2\pi(n-1)(m-1)}{N_{2}}}$,
respectively. It can be readily verified that $\widetilde{\mathbf{U}}_{N}$
is a symmetric and unitary matrix according to its definition. Now
we show an asymptotic property of $\mathbf{A}_{N}$ via the linear
transformation $\widetilde{\mathbf{U}}_{N}^{\mathrm{H}}$. \vspace{-0.15cm}

\begin{lem}
\label{lem:3}When $N_{1}\rightarrow\infty$ and $\ensuremath{N_{2}\rightarrow\infty}$,
if the condition $\frac{d_{\mathrm{BS}}}{\lambda_{c}}\leq\frac{1}{2}$
holds,\footnote{This condition holds to avoid AoA ambiguity.} then
the linear transformation $\widetilde{\mathbf{U}}_{N}^{\mathrm{H}}\mathbf{A}_{N}$
is a tall sparse matrix with only one nonzero element in each column,
i.e., 
\begin{equation}
\lim_{N\rightarrow\infty}[\widetilde{\mathbf{U}}_{N}^{\mathrm{H}}\mathbf{A}_{N}]_{n_{l},l}\neq0,\forall l,\label{eq:lemma3}
\end{equation}
where \vspace{-0.05cm}
\begin{equation}
n_{l}=(n_{1}(l)-1)N_{\mathrm{2}}+n_{2}(l),\label{eq:lemma3_decom}
\end{equation}
and \vspace{-0.05cm}

\begin{equation}
\begin{array}{cc}
n_{1}(l) & =\begin{cases}
N_{\mathrm{1}}\psi_{l}+1 & \psi_{l}\in[0,\frac{d_{\mathrm{BS}}}{\lambda_{c}})\\
N_{\mathrm{1}}+N_{\mathrm{1}}\psi_{l}+1 & \psi_{l}\in[-\frac{d_{\mathrm{BS}}}{\lambda_{c}},0)
\end{cases},\\
n_{2}(l) & =\begin{cases}
N_{\mathrm{2}}\nu_{l}+1 & \nu_{l}\in[0,\frac{d_{\mathrm{BS}}}{\lambda_{c}})\\
N_{\mathrm{2}}+N_{\mathrm{2}}\nu_{l}+1 & \nu_{l}\in[-\frac{d_{\mathrm{BS}}}{\lambda_{c}},0)
\end{cases}.
\end{array}\label{eq:lemma3-1}
\end{equation}
\end{lem}
\begin{IEEEproof}
Please refer to Appendix B. 
\end{IEEEproof}
Since typically $L\ll N_{1},N_{2}$, Lemma \ref{lem:3} means that
matrix $\widetilde{\mathbf{U}}_{N}^{\mathrm{H}}\mathbf{A}_{N}$ is
a row sparse matrix with full column rank. By substituting (\ref{eq:G1})
into (\ref{eq:m-y-2}), we observe that $\widetilde{\mathbf{U}}_{N}^{\mathrm{H}}\mathbf{\mathbf{Y}_{\mathrm{1}}}$
is an asymptotically row-sparse matrix with $L$ nonzero rows, and
each row corresponds to one of the AoA pairs i.e., $(\psi_{l},\nu_{l})$.
Based on this fact, the estimation of the common AoAs is equivalent
to finding the indices of the nonzero rows of $\widetilde{\mathbf{U}}_{N}^{\mathrm{H}}\mathbf{\mathbf{Y}_{\mathrm{1}}}$.
Note that $n_{1}(l)$, $n_{2}(l)$ are integers, and can be derived
from (\ref{eq:lemma3_decom}) as follows \vspace{-0.1cm}
\begin{equation}
n_{1}(l)=\left\lceil \frac{n_{l}}{N_{\mathrm{2}}}\right\rceil ,~n_{2}(l)=n_{l}-N_{\mathrm{2}}(n_{1}(l)-1).\label{n_lsub}
\end{equation}
By combining (\ref{n_lsub}) with Lemma \ref{lem:3}, the AoA spatial
frequency pairs $\{(\psi_{l},\nu_{l})\}{}_{l=1}^{L}$ can be readily
estimated. Due to the fact that different scatterers have different
angles, we can draw the conclusion that any two nonzero elements are
not in the same row, i.e., $n_{l}\neq n_{i}$ for any $l\neq i$.
\vspace{-0.05cm}

\subsubsection{Low-complexity Angle Rotation for Suppressing Power Leakage}

To improve the angle estimation accuracy, the power leakage issue
\cite{Fan_AoD_CE-TWC2018} should be considered. In practice, finite
values for $N_{\mathrm{1}}$ and $N_{\mathrm{2}}$ lead to power leakage,
which means that the resolution of the estimated AoA $(\psi_{l},\nu_{l})$
is limited by half of the DFT interval, i.e., $\frac{1}{2N_{\mathrm{1}}}$
and $\frac{1}{2N_{\mathrm{2}}}$. To mitigate the power leakage, an
angle rotation operation is adopted and the rotation matrix is defined
as \vspace{-0.1cm}
\begin{equation}
\mathbf{R\mathrm{\mathbf{(}\Delta\psi,\Delta\nu\mathbf{)}}=R_{\mathrm{1}}\mathrm{(}\mathrm{\Delta\psi}\mathrm{)}\otimes R_{\mathrm{2}}\mathrm{(}\mathrm{\Delta\nu}\mathrm{)}},\label{eq:rotR-1}
\end{equation}
where the diagonal matrices $\mathbf{R_{\mathrm{1}}(\mathrm{\Delta\psi})}$
and $\mathbf{R_{\mathrm{2}}(\mathrm{\Delta}\nu)}$ are respectively
given by \vspace{-0.1cm}

\begin{subequations}
\label{rot_R-1=00003D00003D00003D00003D00003D00003D00003D00003D00003D00003D00003D0}
\begin{align}
\mathbf{R_{\mathrm{1}}\mathrm{(}\mathrm{\Delta\psi}\mathrm{)}} & =\mathrm{Diag}\{1,e^{\mathrm{-i}\Delta\psi},\ldots,e^{-\mathrm{i}(N_{\mathrm{1}}-1)\Delta\psi}\},\\
\mathbf{R_{\mathrm{2}}\mathrm{(}\mathrm{\Delta\nu}\mathrm{)}} & =\mathrm{Diag}\{1,e^{-\mathrm{i}\Delta\nu},\ldots,e^{\mathrm{-i}(N_{\mathrm{2}}-1)\Delta\nu}\},
\end{align}
\end{subequations}
where $\Delta\psi\in[-\frac{\pi}{N_{1}},\frac{\pi}{N_{1}}]$ and $\Delta\nu\in[-\frac{\pi}{N_{2}},\frac{\pi}{N_{2}}]$.
We construct $L$ rotation matrices $\mathbf{R\mathrm{(}\mathrm{\Delta\mathit{\psi}_{\mathit{l}},\Delta\mathit{\nu}_{\mathit{l}}}\mathrm{)}}$
to compensate for the $L$ estimated AoAs $(\psi_{l},\nu_{l})$. After
angle rotation, the central point, denoted as the $(n_{l},l)$-th
element of $\widetilde{\mathbf{U}}_{N}^{\mathrm{H}}\mathbf{R\mathrm{(}\mathrm{\Delta\mathit{\psi}_{\mathit{l}},\Delta\mathit{\nu}_{\mathit{l}}}\mathrm{)}}\mathbf{A}_{N}$,
is calculated as \vspace{-0.1cm}
\begin{align}
 & [\widetilde{\mathbf{U}}_{N}^{\mathrm{H}}\mathbf{R\mathrm{(}\mathrm{\Delta\mathit{\psi}_{\mathit{l}},\Delta\mathit{\nu}_{\mathit{l}}}\mathrm{)}}\mathbf{A}_{N}]_{n_{l},l}\nonumber \\
= & [\mathbf{U_{\mathit{N_{\mathrm{1}}}}^{\mathrm{H}}}\mathbf{R_{\mathrm{1}}}(\Delta\psi_{\mathit{l}})\mathbf{a}_{N_{1}}(\psi_{l})]_{n_{1}(l)}\otimes[\mathrm{\mathbf{U}}_{N_{\mathrm{2}}}^{\mathrm{H}}\mathbf{R_{\mathrm{2}}}(\Delta\nu_{\mathit{l}})\mathbf{a}_{N_{2}}(\nu_{l}))]_{n_{2}(l)}\nonumber \\
= & (\sqrt{\frac{1}{N_{1}}}\sum_{m=1}^{N_{1}}e^{-\mathrm{i}2\pi(m-1)(\psi_{l}+\frac{\Delta\psi_{l}}{2\pi}-\frac{n_{1}(l)-1}{N_{1}})})\nonumber \\
 & \times(\sqrt{\frac{1}{N_{2}}}\sum_{m=1}^{N_{2}}e^{-\mathrm{i}2\pi(m-1)(\nu_{l}+\frac{\Delta\nu_{l}}{2\pi}-\frac{n_{2}(l)-1}{N_{2}})}).\label{eq:rot_nl}
\end{align}
It can be found that the entries of $\widetilde{\mathbf{U}}_{N}^{\mathrm{H}}\mathbf{R\mathrm{(}\mathrm{\Delta\mathit{\psi}_{\mathit{l}},\Delta\mathit{\nu}_{\mathit{l}}}\mathrm{)}}\mathbf{A}_{N}$
have only $L$ nonzero elements when \vspace{-0.1cm}
\begin{equation}
\Delta\psi_{l}=2\pi\left(\frac{n_{1}(l)-1}{N_{1}}-\psi_{l}\right),~\Delta\nu_{l}=2\pi\left(\frac{n_{2}(l)-1}{N_{2}}-\nu_{l}\right).\label{parameters}
\end{equation}
The $(\Delta\psi_{l},\Delta\nu_{l})$ in (\ref{parameters}) are the
required optimal angle rotation parameters for $(\psi_{l},\nu_{l})$,
which concentrates the power of the respective frequency points and
suppress power leakage. The optimal angle rotation parameters $(\Delta\widehat{\psi}_{l},\Delta\widehat{\nu}_{l})$
can be found via a 2-D search over the very small region $\Delta\psi_{l}\in[-\frac{\pi}{N_{1}},\frac{\pi}{N_{1}}]$
and $\Delta\nu_{l}\in[-\frac{\pi}{N_{2}},\frac{\pi}{N_{2}}]$ \cite{Fan_AoD_CE-TWC2018},
as follows:\vspace{-0.15cm}
\begin{equation}
\begin{split}(\Delta\widehat{\psi}_{l},\Delta\widehat{\nu}_{l})=\mathrm{arg}\max_{\Delta\psi_{l}\in[-\frac{\pi}{N_{1}},\frac{\pi}{N_{1}}],\Delta\nu_{l}\in[-\frac{\pi}{N_{2}},\frac{\pi}{N_{2}}]}\\
||[\widetilde{\mathbf{U}}_{N}]_{:,n_{l}}^{\mathrm{H}}\mathbf{R\mathrm{(}\mathrm{\Delta\psi_{\mathit{l}},\Delta\nu_{\mathit{l}}}\mathrm{)}}\mathbf{Y}_{1}||^{2}.
\end{split}
\label{eq:2-Dsearch}
\end{equation}

The accuracy of the AoA estimation depends on the number of grid points.
The complexity of the 2-D search is approximately $\mathcal{O}(Lg_{1}g_{2})$,
where $g_{1}$ and $g_{2}$ denote the number of grid points in the
interval $[-\frac{\pi}{N_{1}},\frac{\pi}{N_{1}}]$ and $[-\frac{\pi}{N_{2}},\frac{\pi}{N_{2}}]$,
respectively. Obviously, large values for $g_{1}$ and $g_{2}$ lead
to high computational complexity. Therefore, we exploit the structure
of the steering vector and propose a 1-D search method to reduce the
complexity of angle rotation. We note that the first elements of the
steering vectors, i.e., $\mathbf{a}_{N_{1}}(\psi_{l})$ or $\mathbf{a}_{N_{2}}(\nu_{l})$,
are equal to 1. Using this fact, we can divide the 2-D search into
two 1-D searches. Specifically, we construct two rotation matrices
shown below to rotate $\psi$ and $\nu$, as\vspace{-0.15cm}
\begin{equation}
\widetilde{\mathbf{R}}_{1}(\Delta\psi)\mathbf{\triangleq\mathrm{\mathbf{R}_{1}(\Delta\psi)}\otimes}\mathbf{D}_{\mathit{N_{\mathrm{2}}}},~\widetilde{\mathbf{R}}_{2}(\Delta\nu)\mathbf{\triangleq\mathbf{D}_{\mathit{N_{\mathrm{1}}}}\otimes}\mathrm{\mathbf{R_{\mathrm{2}}\mathrm{(}\mathrm{\Delta\nu}})},\label{rot_mat}
\end{equation}
where $\mathbf{R}_{1}(\Delta\psi)$ and $\mathbf{R_{\mathrm{2}}\mathrm{(}\mathrm{\Delta\nu}})$
are defined in (\ref{rot_R-1=00003D00003D00003D00003D00003D00003D00003D00003D00003D00003D00003D0}).
The matrices $\mathbf{D}_{\mathit{N_{\mathrm{1}}}}\in\mathbb{C}^{N_{1}\times N_{1}}$
and $\mathbf{D}_{\mathit{N_{\mathrm{2}}}}\in\mathbb{C}^{N_{2}\times N_{2}}$
are diagonal whose $(1,1)$ entry is equal to $1$ and whose other
elements are $0$. Defining $\widetilde{\mathbf{U}}_{1}\mathbf{\triangleq U_{\mathit{N_{\mathrm{1}}}}\otimes}\mathbf{D}_{\mathit{N_{\mathrm{2}}}}$
and $\widetilde{\mathbf{U}}_{2}\mathbf{\triangleq D_{\mathit{N_{\mathrm{1}}}}\otimes}\mathbf{U}_{\mathit{N_{\mathrm{2}}}}$,
we have the following proposition. \vspace{-0.15cm}

\begin{proposition}\label{lem:4}The angle estimation operation for
the $l$-th AoA pair $(\psi_{l},\nu_{l})$ shown in (\ref{eq:rot_nl})
can be divided into two independent angle rotation operations with
the $(\overline{n_{1l}},l)$-th element of $\widetilde{\mathbf{U}}_{1}^{\mathrm{H}}\widetilde{\mathbf{R}}_{1}(\Delta\psi_{l})\mathbf{A}_{N}$,
and the $(\overline{n_{2l}},l)$-th element of $\widetilde{\mathbf{U}}_{2}^{\mathrm{H}}\widetilde{\mathbf{R}}_{2}(\Delta\nu_{l})\mathbf{A}_{N}$,
where $\overline{n_{1l}}$ and $\overline{n_{2l}}$ denote the nonzero
element of the $l$-th column of $\widetilde{\mathbf{U}}_{1}^{\mathrm{H}}\widetilde{\mathbf{R}}_{1}(\Delta\psi_{l})\mathbf{A}_{N}$
and $\widetilde{\mathbf{U}}_{2}^{\mathrm{H}}\widetilde{\mathbf{R}}_{2}(\Delta\nu_{l})\mathbf{A}_{N}$,
respectively, and satisfy \vspace{-0.25cm}
\begin{equation}
\begin{split}\overline{n_{1l}}=(n_{1}(l)-1)N_{\mathrm{2}}+1,~\overline{n_{2l}}=n_{2}(l).\end{split}
\label{eq:Rot_index}
\end{equation}
\end{proposition}

\vspace{-0.15cm}

\begin{IEEEproof}
Please refer to Appendix C. 
\end{IEEEproof}
Based on Proposition \ref{lem:4}, the optimal angle rotation parameters
$(\Delta\widehat{\psi}_{l},\Delta\widehat{\nu}_{l})$ for $(\psi_{l},\nu_{l})$
can be found by solving the two separate 1-D search problems shown
in (\ref{eq:opt_shift}), which significantly reduces the complexity
to $\mathcal{O}(L(g_{1}+g_{2}))$:\vspace{-0.15cm}

\begin{subequations}
\label{eq:opt_shift} 
\begin{align}
\Delta\widehat{\psi}_{l} & =\mathrm{arg}\max_{\Delta\psi_{l}\in[-\frac{\pi}{N_{1}},\frac{\pi}{N_{1}}]}||[\widetilde{\mathbf{U}}_{1}]_{:,\overline{n_{1l}}}^{\mathrm{H}}\widetilde{\mathbf{R}}_{1}(\Delta\psi_{l})\mathbf{Y}_{1}||^{2},\\
\Delta\widehat{\nu}_{l} & =\mathrm{arg}\max_{\Delta\nu_{l}\in[-\frac{\pi}{N_{2}},\frac{\pi}{N_{2}}]}||[\widetilde{\mathbf{U}}_{2}]_{:,\overline{n_{2l}}}^{\mathrm{H}}\widetilde{\mathbf{R}}_{2}(\Delta\nu_{l})\mathbf{Y}_{1}||^{2}.
\end{align}
\end{subequations}

Denote the estimated angle rotations as $\{(\Delta\hat{\psi}_{l},\Delta\hat{\nu}_{l})\}{}_{l=1}^{L}$,
then the estimated AoA spatial frequency pair of the $l$-th path
is given by \vspace{-0.15cm}

\begin{subequations}
\label{opt_psi_nu} 
\begin{align}
\hat{\psi}_{l} & =\begin{cases}
\frac{n_{1}(l)-1}{N_{1}}-\frac{\Delta\hat{\psi}_{l}}{2\pi} & n_{1}(l)\leq N_{1}\frac{d_{\mathrm{BS}}}{\lambda_{c}}\\
\frac{n_{1}(l)-1}{N_{1}}-1-\frac{\Delta\hat{\psi}_{l}}{2\pi} & n_{1}(l)>N_{1}\frac{d_{\mathrm{BS}}}{\lambda_{c}}
\end{cases},\\
\hat{\nu}_{l} & =\begin{cases}
\frac{n_{2}(l)-1}{N_{2}}-\frac{\Delta\hat{\nu}_{l}}{2\pi} & n_{2}(l)\leq N_{2}\frac{d_{\mathrm{BS}}}{\lambda_{c}}\\
\frac{n_{2}(l)-1}{N_{2}}-1-\frac{\Delta\hat{\nu}_{l}}{2\pi} & n_{2}(l)>N_{2}\frac{d_{\mathrm{BS}}}{\lambda_{c}}
\end{cases}.
\end{align}
\end{subequations}

With the estimated spatial frequency pairs for the AoAs, $\{(\widehat{\psi}_{l},\widehat{\nu}_{l})\}_{l=1}^{\widehat{L}}$,
we can obtain an estimate of the common AoA steering matrix $\mathbf{\widehat{A}}_{N}=[\mathbf{a}_{N}(\widehat{\psi}_{1},\widehat{\nu}_{1}),\ldots,\mathbf{a}_{N}(\widehat{\psi}_{\widehat{L}},\widehat{\nu}_{\widehat{L}})]\in\mathbb{C}^{N\times\widehat{L}}$.
AoA estimation of the different paths at the BS is summarized in Algorithm
\ref{algorithm-1}, where $\Gamma(\mathbf{z})$ represents the operation
of searching the peak power of vector $\mathbf{z}$ and $\widehat{L}$
is the estimated number of propagation paths in step 3.\footnote{If the power of the row is lager than that of its neighbor rows, and
far exceeds the minimum power of $\mathbf{z}(n)$ based on a adjustable
predefined ratio threshold, we put this row index into set $\Omega_{N}$.
Alternately, classical minimum description length (MDL) and novel
signal subspace matching (SSM) schemes \cite{subspace_mathcing} can
be adopted as a pre-processing operation before Algorithm \ref{algorithm-1}
to determine the $\widehat{L}$.} $\Omega_{N}$, $\Omega_{N_{1}}$, and $\Omega_{N_{2}}$ are the sets
with cardinality $\widehat{L}$ , and denote the position indices
of the nonzero rows for $\widetilde{\mathbf{U}}_{N}^{\mathrm{H}}\mathbf{A}_{N}$,
$\mathbf{U}_{N_{\mathrm{1}}}^{\mathrm{H}}\mathbf{A}_{N_{\mathrm{1}}}$,
and $\mathbf{U}_{N_{\mathrm{2}}}^{\mathrm{H}}\mathbf{A}_{N_{\mathrm{2}}}$,
respectively. 
\begin{algorithm}
\caption{Low-complexity Angle Rotation based AoA Estimation}

\label{algorithm-1}

\begin{algorithmic}[1]

\REQUIRE $\mathbf{Y}_{1}$.

\STATE Calculate linear transformation of $\mathbf{Y}_{1}$: $\widetilde{\mathbf{\mathbf{Y}}}_{1}=\widetilde{\mathbf{U}}_{N}^{\mathrm{H}}\mathbf{Y}_{1}$;\label{Complexity_1_1}

\STATE Calculate the sum power of each row: $\mathbf{z}(n)=||[\widetilde{\mathbf{\mathbf{Y}}}_{1}]_{n,:}||^{2},\forall n=1,2,\ldots,N$;

\STATE Find the rows with the peak power: $(\Omega_{N},\widehat{L})=\Gamma(\mathbf{z})$,
where $\Omega_{N}=\{n_{l},l=1,\cdots,\widehat{L}\}$;

\STATE Construct two sets: $\Omega_{N_{1}}=\{n_{1}(l),l=1,\cdots,\widehat{L}\}$,
$\Omega_{N_{2}}=\{n_{2}(l),l=1,\cdots,\widehat{L}\}$ via (\ref{n_lsub});

\FOR{$1\leq l\leq\widehat{L}$}

\STATE Calculate $\overline{n_{1l}}$ and $\overline{n_{2l}}$ respectively
via (\ref{eq:Rot_index});

\STATE Find $\Delta\widehat{\psi}_{l}$ and $\Delta\widehat{\nu}_{l}$
via (\ref{eq:opt_shift});\label{Complexity_1_2}

\STATE Estimate $\widehat{\psi}_{l}$ and $\widehat{\nu}_{l}$ according
to (\ref{opt_psi_nu});

\ENDFOR

\ENSURE $\{(\widehat{\psi}_{l},\widehat{\nu}_{l})\}_{l=1}^{\widehat{L}}$
and $\mathbf{\widehat{A}}_{N}$.

\end{algorithmic} 
\end{algorithm}

\begin{remark} \label{AoA_remark} Since the common AoA steering
matrix $\mathbf{A}_{N}$ is shared by all users in MU scenario, the
received signals from $K$ users in Stage I and Stage II during the
first coherence block can be utilized jointly to estimate $\mathbf{A}_{N}$.
Accordingly, the input of Algorithm \ref{algorithm-1} is given by
$\mathbf{Y}=[\mathbf{Y}_{1},\mathbf{Y}_{2},...,\mathbf{Y}_{K}]\in\mathbb{C}^{N\times(\sum_{k=1}^{K}\tau_{k})}$.
In this case, the number of measurements used for the estimation of
$\mathbf{A}_{N}$ increases, which enhances the estimation performance
and alleviates the error propagation effect in the following stages.\end{remark}

\subsubsection{Estimation of the Cascaded Spatial Frequencies and Gains}

By substituting $\mathbf{A}_{N}=\mathbf{\widehat{A}}_{N}+\Delta\mathbf{A}_{N}$
and applying Lemma \ref{lem:1}, we take the linear transformation
$\frac{1}{N\sqrt{p}}\mathbf{\widehat{A}}_{N}^{\mathrm{H}}$ of the
received signals to eliminate the effects of the common AoAs, i.e.,
\vspace{-0.15cm}
\begin{align}
\frac{1}{N\sqrt{p}}\mathbf{\widehat{A}}_{N}^{\mathrm{H}}\mathbf{Y}_{1}= & \boldsymbol{\Lambda}\mathbf{A}_{M}^{\mathrm{H}}\mathrm{Diag}\{\mathbf{h}_{1}\mathrm{\}}\mathbf{E}_{1}+\frac{1}{N\sqrt{p}}\mathbf{\widehat{A}}_{N}^{\mathrm{H}}(\mathbf{N}_{1}\nonumber \\
 & +\sqrt{p}\Delta\mathbf{A}_{N}\boldsymbol{\Lambda}\mathbf{A}_{M}^{\mathrm{H}}\mathrm{Diag}\{\mathbf{h}_{1}\mathrm{\}}\mathbf{E}_{1}).\label{eq:LT_Y1}
\end{align}
Here, $\Delta\mathbf{A}_{N}\triangleq\mathbf{A}_{N}-\mathbf{\widehat{A}}_{N}$
is treated as the estimation error between the common AoA and its
estimate, and the third term $(\frac{1}{N}\mathbf{\widehat{A}}_{N}^{\mathrm{H}}\Delta\mathbf{A}_{N}\boldsymbol{\Lambda}\mathbf{A}_{M}^{\mathrm{H}}\mathrm{Diag}\{\mathbf{h}_{1}\mathrm{\}}\mathbf{E}_{1})$
represents the corresponding negative error propagation effect. Clearly,
$\Delta\mathbf{A}_{N}$ can be reduced effectively via the MU joint
estimation strategy discussed in Remark \ref{AoA_remark}.

Now we define the transpose of $\frac{1}{N\sqrt{p}}\mathbf{\widehat{A}}_{N}^{\mathrm{H}}\mathbf{Y}_{1}$
as an equivalent measurement matrix $\overline{\mathbf{\mathbf{Y}}}_{1}\in\mathbb{C}^{\tau_{1}\times L}$
shown below \vspace{-0.15cm}
\begin{align}
\overline{\mathbf{\mathbf{Y}}}_{1} & \triangleq(\frac{1}{N\sqrt{p}}\mathbf{\widehat{A}}_{N}^{\mathrm{H}}\mathbf{Y}_{1})^{\mathrm{H}}\nonumber \\
 & =\mathbf{E}_{1}^{\mathrm{H}}\mathrm{Diag}\{{\bf h}_{1}^{*}\}\mathbf{A}_{M}\boldsymbol{\Lambda}^{*}+\overline{\mathbf{N}}_{1}=\mathbf{E}_{1}^{\mathrm{H}}\mathbf{H}_{\mathrm{RIS}}+\overline{\mathbf{N}}_{1},\label{eq:Y1_formula}
\end{align}
where $\mathbf{H}_{\mathrm{RIS}}\triangleq$$\mathrm{Diag}\{{\bf h}_{1}^{*}\}\mathbf{A}_{M}\boldsymbol{\Lambda}^{*}$
and $\overline{\mathbf{N}}_{1}$ is the corresponding transpose of
the second term in Eq. (\ref{eq:LT_Y1}), seen as the equivalent noise.
By exploiting the structure of $\mathbf{H}_{\mathrm{RIS}}$, we have
\vspace{-0.15cm}
\begin{align}
\mathbf{H}_{\mathrm{RIS}}={\bf h}_{1}^{*}\bullet(\mathbf{A}_{M}\boldsymbol{\Lambda}^{*}) & =(\mathbf{A}_{M,1}\boldsymbol{\beta}_{1})^{*}\bullet(\mathbf{A}_{M}\boldsymbol{\Lambda}^{*})\nonumber \\
 & =(\mathbf{A}_{M,1}^{*}\bullet\mathbf{A}_{M})(\boldsymbol{\beta}_{1}^{*}\otimes\boldsymbol{\Lambda}^{*}),\label{eq:H_ris}
\end{align}
where $\mathbf{A}_{M,1}^{*}\bullet\mathbf{A}_{M}=[\mathbf{a}_{M}(\omega_{1}-\varphi_{1,1},\mu_{1}-\theta_{1,1}),\mathbf{a}_{M}(\omega_{2}-\varphi_{1,1},\mu_{2}-\theta_{1,1})...,\mathbf{a}_{M}(\omega_{L}-\varphi_{1,J_{1}},\mu_{L}-\theta_{1,J_{1}})]\in\mathbb{C}^{M\times J_{1}L}$,
and the last equality uses the identity $(\mathbf{A}\bullet\mathbf{B})(\mathbf{C}\otimes\mathbf{D})=(\mathbf{AC})\bullet(\mathbf{BD})$
\cite{K-R_product}. To extract the cascaded directional spatial frequency
pairs $\{(\omega_{l}-\varphi_{1,j},\mu_{l}-\theta_{1,j})\}_{j=1,l=1}^{J_{1}L}$
and gains $(\boldsymbol{\beta}_{1}^{*}\otimes\boldsymbol{\Lambda}^{*})$
from $\overline{\mathbf{\mathbf{Y}}}_{1}$, (\ref{eq:Y1_formula})
could be approximated using the virtual angular domain (VAD) representation
and converted into a $J_{1}L$-sparse recovery problem via vectorization
\cite{ris-omp-1}, but this approach has high complexity and performance
loss.

Instead, another method is developed as follows. We first estimate
$J_{1}$ cascaded spatial frequency pairs and gains from a typical
column vector of $\overline{\mathbf{\mathbf{Y}}}_{1}$ using CS, and
then estimate the remaining parameters by exploiting the correlation
between the typical column and other columns. Specifically, denote
$\overline{{\bf y}}_{r}$ as the $r$-th column of $\overline{\mathbf{\mathbf{Y}}}_{1}$,
which is given by \vspace{-0.15cm}
\begin{align}
\overline{{\bf y}}_{r} & =\mathbf{E}_{1}^{\mathrm{H}}\mathrm{Diag}\{{\bf h}_{1}^{*}\}[\mathbf{A}_{M}\boldsymbol{\Lambda}^{*}]_{:\mathbf{,}r}+\overline{{\bf n}}_{r}\nonumber \\
 & =\mathbf{E}_{1}^{\mathrm{H}}{\bf h}_{1}^{*}\bullet(\alpha_{r}^{*}\mathbf{a}_{M}(\omega_{r},\mu_{r}))+\overline{{\bf n}}_{r}\nonumber \\
 & =\mathbf{E}_{1}^{\mathrm{H}}(\mathbf{A}_{M,1}^{*}\bullet\mathbf{a}_{M}(\omega_{r},\mu_{r}))\alpha_{r}^{*}\boldsymbol{\beta}_{1}^{*}+\overline{{\bf n}}_{r},\label{eq:yl_sparse}
\end{align}
where $\mathbf{A}_{M,1}^{*}\bullet\mathbf{a}_{M}(\omega_{r},\mu_{r})=[\mathbf{a}_{M}(\omega_{r}-\varphi_{1,1},\mu_{r}-\theta_{1,1}),...,\mathbf{a}_{M}(\omega_{r}-\varphi_{1,J_{1}},\mu_{r}-\theta_{1,J_{1}})]\in\mathbb{C}^{M\times J_{1}}$
and $\overline{{\bf n}}_{r}$ is the $r$-th column of $\overline{\mathbf{N}}_{1}$.
Note that $\mathrm{Diag}\{{\bf h}_{1}^{*}\}[\mathbf{A}_{M}\boldsymbol{\Lambda}^{*}]_{:\mathbf{,}r}$
is the $r$-th column of $\mathbf{H}_{\mathrm{RIS}}$, which we denote
as $\mathbf{h}_{\mathrm{RIS,\mathit{r}}}$. Since $\{(\omega_{r}-\varphi_{1,j})\}_{j=1}^{J_{1}}$
and $\{(\mu_{r}-\theta_{1,j})\}_{j=1}^{J_{1}}$ lie in the interval
$[-2\frac{d_{\mathrm{RIS}}}{\lambda_{c}},2\frac{d_{\mathrm{RIS}}}{\lambda_{c}}]$,
we can formulate (\ref{eq:yl_sparse}) as a $J_{1}$-sparse signal
recovery problem \vspace{-0.25cm}
\begin{equation}
\overline{{\bf y}}_{r}=\mathbf{E}_{1}^{\mathrm{H}}(\mathbf{A}_{\mathrm{1}}\otimes\mathbf{A}_{\mathrm{2}})\mathbf{b}_{r}+\overline{{\bf n}}_{r},\label{eq:sparse-formula}
\end{equation}
where $\mathbf{A}_{1}\in\mathbb{C}^{M_{1}\times D_{1}}$ and $\mathbf{A}_{2}\in\mathbb{C}^{M_{2}\times D_{2}}$
are overcomplete dictionary matrices $(D_{1}\geq M_{1},D_{2}\geq M_{2})$
with resolutions $\frac{1}{D_{1}}$ and $\frac{1}{D_{2}}$, respectively,
and the columns of $\mathbf{A}_{1}$ and $\mathbf{A}_{2}$ contain
values for $\mathbf{a}_{M_{1}}(\omega_{r}-\varphi_{1,j})$ and $\mathbf{a}_{M_{2}}(\mu_{r}-\theta_{1,j})$
on the angle grid, i.e., $\mathbf{\mathbf{A}_{\mathrm{1}}}=[\mathbf{a}_{M_{1}}(-2\frac{d_{\mathrm{RIS}}}{\lambda_{c}}),\mathbf{a}_{M_{1}}((-2+\frac{4}{D_{1}})\frac{d_{\mathrm{RIS}}}{\lambda_{c}}),\ldots,\mathbf{a}_{M_{1}}((2-\frac{4}{D_{1}})\frac{d_{\mathrm{RIS}}}{\lambda_{c}})]$
and $\mathbf{\mathbf{A}_{\mathrm{2}}}=[\mathbf{a}_{M_{2}}(-2\frac{d_{\mathrm{RIS}}}{\lambda_{c}}),\mathbf{a}_{M_{2}}((-2+\frac{4}{D_{2}})\frac{d_{\mathrm{RIS}}}{\lambda_{c}}),\ldots,\mathbf{a}_{M_{2}}((2-\frac{4}{D_{2}})\frac{d_{\mathrm{RIS}}}{\lambda_{c}})].$

In addition, $\mathbf{b}_{r}\in\mathbb{C}^{D_{1}D_{2}\times1}$ in
(\ref{eq:sparse-formula}) is a sparse vector with $J_{1}$ nonzero
entries corresponding to the cascaded channel path gains $\{\alpha_{r}^{*}\beta_{1,j}^{*}\}_{j=1}^{J_{1}}$.
To obtain the best possible CS performance, the RIS phase shift training
matrix $\mathbf{E}_{1}$ should be designed to ensure that the columns
of the equivalent dictionary $\mathbf{E}_{1}^{\mathrm{H}}(\mathbf{A}_{\mathrm{1}}\otimes\mathbf{A}_{\mathrm{2}})$
are orthogonal. A detailed design of $\mathbf{E}_{1}$ that achieves
this goal can be found in \cite{Zhou_ULA_TSP}. A simpler method is
to choose the random Bernoulli matrix as $\mathbf{E}_{1}$, i.e.,
randomly generate the elements of $\mathbf{E}_{1}$ from $\{-1,+1\}$
with equal probability \cite{ris-omp-3}. Later in Section \ref{sec:Simulation-Results},
we will show that this random method has near-optimal performance,
and provides a nearly orthogonal equivalent dictionary.\footnote{Please note that the number of scatterers in the user $1$-RIS channel,
i.e., the sparsity level for the sparse recovery problem associated
with (\ref{eq:sparse-formula}), denoted as $J_{1}$, is estimated
via the selected CS-based techniques. For example, in Section \ref{sec:Simulation-Results},
the proposed estimation protocol adopts OMP as the recovery algorithm.
In this case, the stopping criteria for this algorithm is based on
the power of the residual error, i.e., the algorithm is stopped when
the residual energy is smaller than a predefined threshold. Thus the
number of iterations is treated as the estimate of $J_{1}$.}

Using CS, we obtain the cascaded AoD pair, i.e., $(\omega_{r}-\varphi_{1,j},\mu_{r}-\theta_{1,j})$.
The corresponding cascaded AoD, i.e., $(\omega_{r}-\varphi_{1,j})$
and $(\mu_{r}-\theta_{1,j})$, can be obtained similarly using the
properties of the Kronecker product. Assume that the $m$-th element
of sparse vector $\mathbf{b}_{r}$ is nonzero, then the $m$-th column
of $(\mathbf{A}_{\mathrm{1}}\otimes\mathbf{A}_{\mathrm{2}})$ is the
corresponding cascaded steering vector. The corresponding indices
in $\mathbf{\mathbf{A}_{\mathrm{1}}}$ and $\mathbf{\mathbf{A}_{\mathrm{2}}}$,
denoted as $m_{1}$ and $m_{2}$, can be derived as \vspace{-0.1cm}
\begin{equation}
m_{1}=\left\lceil \frac{m}{D_{\mathrm{2}}}\right\rceil ,~m_{2}=m-D_{\mathrm{2}}(m_{1}-1).\label{m_lsub}
\end{equation}
Finally, we obtain the estimate of the cascaded AoD, i.e., $\{(\widehat{\omega_{r}-\varphi_{1,j}})\}_{j=1}^{\hat{J_{1}}}$
and $\{(\widehat{\mu_{r}-\theta_{1,j}})\}_{j=1}^{\hat{J_{1}}}$. As
a result, $\mathbf{\widehat{h}}_{\mathrm{RIS,\mathit{r}}}$ is obtained
according to (\ref{eq:yl_sparse}). Estimates of the other columns
of $\mathbf{H}_{\mathrm{RIS}}$, i.e., $\{{\bf h}_{\mathrm{RIS},l}\}_{l\neq r}^{L}$,
can be obtained by exploiting the correlation among different columns.
To illustrate the correlation relationship, a compensation matrix
$\mathbf{\mathrm{\Delta}H_{\mathit{l}}}$ with respect to the reference
index $r$ is defined as \vspace{-0.2cm}
\begin{align}
\mathbf{\mathrm{\Delta}H_{\mathit{l}}} & =\frac{\alpha_{l}^{*}}{\alpha_{r}^{*}}\mathrm{Diag}\{\mathbf{a}_{M}(\omega_{l}-\omega_{r},\mu_{l}-\mu_{r})\}\nonumber \\
 & =\gamma_{l}\mathrm{Diag}\{\mathbf{a}_{M}(\Delta\omega_{l},\Delta\mu_{l})\},\label{eq:compensate_mtx}
\end{align}
where $\Delta\omega_{l}$, $\Delta\mu_{l}$ are rotation factors and
$\gamma_{l}$ is a gain scaling factor given by \vspace{-0.15cm}
\begin{equation}
\Delta\omega_{l}=\omega_{l}-\omega_{r},~\Delta\mu_{l}=\mu_{l}-\mu_{r},~\gamma_{l}=\frac{\alpha_{l}^{*}}{\alpha_{r}^{*}}.\label{rot_fac_omega_mu_scale}
\end{equation}
Clearly, $\Delta\omega_{l},\Delta\mu_{l}\in[-2\frac{d_{\mathrm{RIS}}}{\lambda_{c}},2\frac{d_{\mathrm{RIS}}}{\lambda_{c}}]$.
Then, we have \vspace{-0.2cm}
\[
\begin{split} & \mathbf{\mathrm{\Delta}H_{\mathit{l}}}{\bf h}_{\mathrm{RIS},r}\\
= & \mathbf{\mathrm{\Delta}H_{\mathit{l}}}\mathrm{Diag}\{{\bf h}_{1}^{*}\}(\alpha_{r}^{*}\mathbf{a}_{M}(\omega_{r},\mu_{r}))\\
= & \mathbf{\mathrm{\Delta}H_{\mathit{l}}}\mathrm{Diag}\{\mathbf{a}_{M}(\omega_{r},\mu_{r})\}(\alpha_{r}^{*}{\bf h}_{1}^{*})\\
= & \gamma_{l}\mathrm{Diag}\{\mathbf{a}_{M}(\Delta\omega_{l},\Delta\mu_{l})\}\mathrm{Diag}\{\mathbf{a}_{M_{1}}(\omega_{r})\otimes\mathbf{a}_{M_{2}}(\mu_{r})\}(\alpha_{r}^{*}{\bf h}_{1}^{*})\\
= & \mathrm{(Diag}\{\mathbf{a}_{M_{1}}(\Delta\omega_{l})\}\mathrm{Diag}\{\mathbf{a}_{M_{1}}(\omega_{r})\})\\
 & \otimes\mathrm{(Diag}\{\mathbf{a}_{M_{2}}(\Delta\mu_{l})\}\mathrm{Diag}\{\mathbf{a}_{M_{2}}(\mu_{r})\})(\alpha_{l}^{*}{\bf h}_{1}^{*})\\
= & \mathrm{Diag}\{\mathbf{a}_{M_{1}}(\omega_{l})\}\otimes\mathrm{Diag}\{\mathbf{a}_{M_{2}}(\mu_{l})\}(\alpha_{l}^{*}{\bf h}_{1}^{*})={\bf h}_{\mathrm{RIS},l}.
\end{split}
\]

This equality shows that we can estimate the compensation matrix $\mathbf{\mathrm{\Delta}H_{\mathit{l}}}$
instead of directly estimating ${\bf h}_{\mathrm{RIS},l}$. Specifically,
${\bf h}_{\mathrm{RIS},r}$ is estimated by applying CS to (\ref{eq:sparse-formula}),
and ${\bf h}_{\mathrm{RIS},l}$ can be rewritten as \vspace{-0.3cm}

\begin{align}
{\bf h}_{\mathrm{RIS},l} & =\gamma_{l}\mathrm{Diag}\{\mathbf{a}_{M}(\Delta\omega_{l},\Delta\mu_{l})\}{\bf h}_{\mathrm{RIS},r}\nonumber \\
 & =\mathrm{Diag}\{{\bf h}_{\mathrm{RIS},r}\}\mathbf{a}_{M}(\Delta\omega_{l},\Delta\mu_{l})\gamma_{l}.\label{eq:hris-l}
\end{align}
We define $\mathbf{c}_{l}(\Delta\omega_{l},\Delta\mu_{l})=\mathbf{E}_{1}^{\mathrm{H}}\mathrm{Diag}\{\widehat{{\bf h}}_{\mathrm{RIS},r}\}\mathbf{a}_{M}(\Delta\omega_{l},\Delta\mu_{l})$.
Then, by replacing ${\bf h}_{\mathrm{RIS},r}$ with $\widehat{{\bf h}}_{\mathrm{RIS},r}+\Delta{\bf h}_{\mathrm{RIS},r}$,
the $l$-th column of $\overline{\mathbf{\mathbf{Y}}}_{1}$ in (\ref{eq:Y1_formula})
is given by \vspace{-0.2cm}
\begin{align}
\overline{{\bf y}}_{l} & =\mathbf{E}_{1}^{\mathrm{H}}\mathrm{Diag}\{{\bf h}_{\mathrm{RIS},r}\}\mathbf{a}_{M}(\Delta\omega_{l},\Delta\mu_{l})\gamma_{l}+\overline{{\bf n}}_{l}\nonumber \\
 & =\mathbf{E}_{1}^{\mathrm{H}}\mathrm{Diag}\{\widehat{{\bf h}}_{\mathrm{RIS},r}\}\mathbf{a}_{M}(\Delta\omega_{l},\Delta\mu_{l})+\mathbf{n}_{\mathrm{noise}},\label{eq:formula-y_l}
\end{align}
where $\mathbf{n}_{\mathrm{noise}}\triangleq\mathbf{E}_{1}^{\mathrm{H}}\mathrm{Diag}\{\Delta{\bf h}_{\mathrm{RIS},r}\}\mathbf{a}_{M}(\Delta\omega_{l},\Delta\mu_{l})+\overline{{\bf n}}_{l}$
represents the corresponding noise vector and $\Delta{\bf h}_{\mathrm{RIS},r}$
is the estimation error of ${\bf h}_{\mathrm{RIS},r}$.\footnote{To reduce the error propagation, the reference index $r$ can be chosen
based on the maximum received power criterion, i.e., $r=\mathrm{arg}\max_{i\in[1,\widehat{L}]}||\overline{{\bf y}}_{i}||^{2}$.}

To find the optimal rotation factors $(\Delta\omega_{l},\Delta\mu_{l})$,
a simple 2-D search method can be used: \vspace{-0.45cm}

\begin{equation}
(\Delta\widehat{\omega}_{l},\Delta\widehat{\mu}_{l})=\mathrm{arg}\max_{\Delta\omega,\Delta\mu\in[-2\frac{d_{\mathrm{RIS}}}{\lambda_{c}},2\frac{d_{\mathrm{RIS}}}{\lambda_{c}}]}\left|\left\langle \overline{{\bf y}}_{l},\mathbf{c}_{l}(\Delta\omega,\Delta\mu)\right\rangle \right|.\label{eq:2-D_search_Corr}
\end{equation}
The gain scaling factor $\gamma_{l}$ can be determined as the solution
to the least square (LS) problem \vspace{-0.25cm}

\begin{equation}
\widehat{\gamma}_{l}=\mathrm{arg}\min_{x}||\overline{{\bf y}}_{l}-\mathbf{c}_{l}(\Delta\widehat{\omega}_{l},\Delta\widehat{\mu}_{l})x||,\label{eq:LS-x_l}
\end{equation}
whose solution is $\widehat{\gamma}_{l}=(\mathbf{c}_{l}^{\mathrm{H}}(\Delta\widehat{\omega}_{l},\Delta\widehat{\mu}_{l})\mathbf{c}_{l}(\Delta\widehat{\omega}_{l},\Delta\widehat{\mu}_{l}))^{-1}\mathbf{c}_{l}^{\mathrm{H}}(\Delta\widehat{\omega}_{l},\Delta\widehat{\mu}_{l})\overline{{\bf y}}_{l}$.
Substituting the solutions of (\ref{eq:2-D_search_Corr}) and (\ref{eq:LS-x_l})
into (\ref{eq:hris-l}), we can obtain $\widehat{{\bf h}}_{\mathrm{RIS},l}$,
$(1\leq l\leq L,l\neq r)$. Finally, the estimated cascaded channel
of user $1$ is given by \vspace{-0.2cm}

\begin{equation}
\widehat{\mathbf{G}}_{1}=\mathbf{\widehat{A}}_{N}\widehat{{\bf H}}_{\mathrm{R\mathrm{IS}}}^{\mathrm{H}},\label{eq:G1_estimate}
\end{equation}
where $\widehat{{\bf H}}_{\mathrm{RIS}}=[\widehat{{\bf h}}_{\mathrm{RIS},1},\cdots,\widehat{{\bf h}}_{\mathrm{RIS},L}]$.
Furthermore, the cascaded AoD in ${\bf h}_{\mathrm{RIS},l}$ can be
obtained as \vspace{-0.5cm}

\begin{equation}
\omega_{l}-\varphi_{1,j}=(\omega_{r}-\varphi_{1,j})+\Delta\omega_{l},\mu_{l}-\theta_{1,j}=(\mu_{r}-\theta_{1,j})+\Delta\mu_{l},\label{cascaded1_omega_mu}
\end{equation}
where the estimate of $(\omega_{r}-\varphi_{1,j},\mu_{r}-\theta_{1,j})$
and $(\Delta\omega_{l},\Delta\mu_{l})$ can be readily obtained from
(\ref{eq:yl_sparse}) and (\ref{eq:2-D_search_Corr}), respectively.
The overall estimation of $\mathbf{G}_{1}$ is summarized in Algorithm
\ref{algorithm-2}.

\begin{algorithm}
\caption{Estimation of Full CSI for Typical User}

\label{algorithm-2}

\begin{algorithmic}[1]

\REQUIRE $\mathbf{Y}_{1}$.

\STATE Return the estimated number of paths between BS and RIS $\widehat{L}$
and AoA steering matrix $\mathbf{\widehat{A}}_{N}$ from Algorithm
\ref{algorithm-1};

\STATE Calculate equivalent measurement matrix $\overline{\mathbf{\mathbf{Y}}}_{1}=[\overline{{\bf y}}_{1},\ldots,\overline{{\bf y}}_{\widehat{L}}]$;

\STATE Choose the typical reference index $r$ and estimate ${\bf h}_{\mathrm{RIS},r}$
by solving sparse recovery problem associated with (\ref{eq:sparse-formula});\label{Complexity_2_1}

\FOR{$1\leq l\leq\widehat{L},l\neq r$}

\STATE Estimate $(\Delta\omega_{l},\Delta\mu_{l})$ according to
(\ref{eq:2-D_search_Corr});\label{Complexity_2_2}

\STATE Estimate $\gamma_{l}$ according to (\ref{eq:LS-x_l});

\STATE Estimate ${\bf h}_{\mathrm{RIS},l}$ according to (\ref{eq:hris-l});

\ENDFOR

\ENSURE $\widehat{\mathbf{G}}_{1}=\mathbf{\widehat{A}}_{N}[\widehat{{\bf h}}_{\mathrm{RIS},1},\cdots,\widehat{{\bf h}}_{\mathrm{RIS},\widehat{L}}]^{\mathrm{H}}$.

\end{algorithmic} 
\end{algorithm}

\vspace{-0.3cm}

\subsection{Stage II: Estimation of Full CSI for Other Users\label{subsec:Stage-II:-Estimation}}

\vspace{-0.1cm}
 In this subsection, the property that all users share the common
RIS-BS channel is invoked for reducing the pilot overhead of channel
estimation. First, we re-exploit the structure of the cascaded channel
$\mathbf{G}_{k}$, and then divide it into two parts, i.e., a common
part and a unique part. Then, only re-estimating the unique part is
necessary for obtaining the full CSI of the other users.

\subsubsection{Re-express Cascaded Channel\label{subsec:Re-express-Cascaded-Channel}}

In order to illustrate the necessity of re-expressing cascaded channel
$\mathbf{G}_{k}$, let us recall its structure and see why the common
RIS-BS channel $\mathbf{H}$ cannot be obtained in Stage I. According
to (\ref{eq:G1}), all users share the common $\mathbf{H}$ consisting
of three matrices, i.e., $\mathbf{A}_{N}$, $\boldsymbol{\Lambda}$
and $\mathbf{A}_{M}$. The first, $\mathbf{A}_{N}$, is estimated
in Stage I. However, $\boldsymbol{\Lambda}$ and $\mathbf{A}_{M}$
cannot be extracted separately from $\mathbf{G}_{1}$ since we can
only estimate the spatial frequencies of the cascaded AoDs, i.e.,
$(\omega_{l}-\varphi_{1,j})$, $(\mu_{l}-\theta_{1,j})$ and the cascaded
gains, i.e., $\alpha_{l}\beta_{1,j}$ for any $l$ and $j$. If other
users only utilize the obtained $\mathbf{\hat{A}}_{N}$, the estimation
for these users is the same as that of the typical user, and thus
the pilot overhead cannot be decreased further. Therefore, we aim
to fully exploit the structure of $\mathbf{H}$ so as to utilize the
common channel's information from $\boldsymbol{\Lambda}$ and $\mathbf{A}_{M}$.

Motivated by this, we decompose the cascaded channel $\mathbf{G}_{k}$
into two parts, i.e., a common part and a unique part, where the common
part can be obtained from the estimation of $\mathbf{G}_{1}$ in Stage
I. The constructed common part has the full information of $\mathbf{A}_{N}$,
and the re-parameterized information of $\boldsymbol{\Lambda}$ and
$\mathbf{A}_{M}$, so as to achieve the full exploitation of $\mathbf{H}$.
Then, we only need to re-estimate the unique part of the cascaded
channel for the other users. To this end, we denote the common part
of $\mathbf{G}_{k}$ as $\mathbf{H}_{\mathrm{s}}\in\mathbb{C}^{N\times M}$,
which can be regarded as a substitute for $\mathbf{H}$ from $\mathbf{G}_{1}$.
Similarly, the unique part of $\mathbf{G}_{k}$ is denoted by $\mathbf{h}_{\mathrm{s},k}\in\mathbb{C}^{M\times1}$,
which can be regarded as a substitute for $\mathbf{h}_{k}$. Then,
$\mathbf{G}_{k}$ can be re-expressed as \vspace{-0.35cm}

\begin{equation}
\mathbf{G}_{k}=\mathbf{H}_{\mathrm{s}}\mathrm{Diag}\{\mathbf{h}_{\mathrm{s},k}\}\in\mathbb{C}^{N\times M},\forall k\in\mathcal{K}.\label{paril-G}
\end{equation}

\vspace{-0.2cm}

In the following, we first construct the common part $\mathbf{H}_{\mathrm{s}}$
with the knowledge obtained in Stage I. Then, we estimate each user's
unique part $\mathbf{h}_{\mathrm{s},k}$.

\subsubsection{Construction of Common Part }

Define the average value of user $1$'s complex gains $\boldsymbol{\beta}_{1}$
as $\overline{\beta}=\frac{1}{J_{1}}\mathbf{1}_{J_{1}}^{\mathrm{T}}\boldsymbol{\beta}_{1}$,
then we have \vspace{-0.3cm}

\begin{align}
\boldsymbol{\Lambda} & =\mathrm{Diag}\{\alpha_{1},\alpha_{2},\ldots,\alpha_{L}\}=\alpha_{r}\mathrm{Diag}\{\gamma_{1}^{*},\gamma_{2}^{*},\ldots,\gamma_{L}^{*}\}\nonumber \\
 & =\frac{1}{\overline{\beta}}\overline{\beta}\alpha_{r}\mathrm{Diag}\{\gamma_{1}^{*},\gamma_{2}^{*},\ldots,\gamma_{L}^{*}\}\triangleq\frac{1}{\overline{\beta}}\boldsymbol{\Lambda}_{\mathrm{s}}.\label{eq:gamma_common}
\end{align}
Here, $\boldsymbol{\Lambda}_{\mathrm{s}}=(\frac{1}{J_{1}}\mathbf{1}_{J_{1}}^{\mathrm{T}}\boldsymbol{\beta}_{1}\alpha_{r})\mathrm{Diag}\{\gamma_{1}^{*},\gamma_{2}^{*},\ldots,\gamma_{L}^{*}\}$.
Obviously, $\boldsymbol{\beta}_{1}\alpha_{r}$ can be obtained by
solving the sparse recovery problem corresponding to (\ref{eq:yl_sparse})
and $\gamma_{l}$ can be obtained according to (\ref{eq:LS-x_l}).
Thus, the constructed matrix, $\boldsymbol{\Lambda}_{\mathrm{s}}$,
can be readily calculated.

Similarly, the matrix $\mathbf{A}_{M}$ can be rewritten as \vspace{-0.15cm}
\begin{align}
\mathbf{A}_{M} & =[\mathbf{a}_{M}(\omega_{1},\mu_{1}),\ldots,\mathbf{a}_{M}(\omega_{L},\mu_{L})]\nonumber \\
 & =\mathrm{Diag}\{\mathbf{a}_{M}(\omega_{r},\mu_{r})\}\mathbf{A}_{\Delta M},\label{eq:AoD_common_part}
\end{align}
where $\mathbf{A}_{\Delta M}=[\mathbf{a}_{M}(\Delta\omega_{1},\Delta\mu_{1}),\ldots,\mathbf{a}_{M}(\Delta\omega_{L},\Delta\mu_{L})]$.
Note that the rotation factors $\Delta\omega_{l}$, $\Delta\mu_{l}$
can be obtained by Algorithm \ref{algorithm-2}, but we need to find
$(\omega_{r},\mu_{r})$, which is not possible. Instead, we introduce
two parameters, $\omega_{\mathrm{s}}=\frac{1}{J_{1}}\sum_{j=1}^{J_{1}}(\omega_{r}-\varphi_{1,j})$
and $\mu_{\mathrm{s}}=\frac{1}{J_{1}}\sum_{j=1}^{J_{1}}(\mu_{r}-\theta_{1,j})$
as substitutes for $\omega_{r}$ and $\mu_{r}$, which can be readily
obtained since $(\omega_{r}-\varphi_{1,j})$ and $(\mu_{r}-\theta_{1,j})$
for $\forall j\in\{1,...,J_{1}\}$ have been estimated in Algorithm
\ref{algorithm-2}.

Then, define $\overline{\varphi_{1}}$ as $(-\frac{1}{J_{1}}\sum_{j=1}^{J_{1}}\varphi_{1,j})$
and $\overline{\theta_{1}}$ as $(-\frac{1}{J_{1}}\sum_{j=1}^{J_{1}}\theta_{1,j})$.
The following relationship exists between $(\omega_{\mathrm{s}},\mu_{\mathrm{s}})$
and $(\omega_{r},\mu_{r})$: \vspace{-0.1cm}

\begin{subequations}
\label{omega_mu_common} 
\begin{align}
\omega_{\mathrm{s}} & =\frac{1}{J_{1}}\sum_{j=1}^{J_{1}}(\omega_{r}-\varphi_{1,j})=\omega_{r}+\overline{\varphi_{1}},\\
\mu_{\mathrm{s}} & =\frac{1}{J_{1}}\sum_{j=1}^{J_{1}}(\mu_{r}-\theta_{1,j})=\mu_{r}+\overline{\theta_{1}}.
\end{align}
\end{subequations}
Based on the above definitions, $\mathrm{Diag}\{\mathbf{a}_{M}(\omega_{r},\mu_{r})\}$
in (\ref{eq:AoD_common_part}) can be represented as \vspace{-0.15cm}
\[
\begin{aligned}\mathrm{Diag}\{\mathbf{a}_{M}(\omega_{r},\mu_{r})\}= & \mathrm{Diag}\{\mathbf{a}_{M}(\omega_{\mathrm{s}}-\overline{\varphi_{1}},\mu_{\mathrm{s}}-\overline{\theta_{1}})\}\\
= & \mathrm{Diag}\{\mathbf{a}_{M_{1}}(\omega_{\mathrm{s}}-\overline{\varphi_{1}})\otimes\mathbf{a}_{M_{2}}(\mu_{\mathrm{s}}-\overline{\theta_{1}})\}\\
= & \mathrm{(Diag}\{\mathbf{a}_{M_{1}}(-\overline{\varphi_{1}})\}\otimes\mathrm{Diag}\{\mathbf{a}_{M_{2}}(-\overline{\theta_{1}})\})\\
 & \mathrm{(Diag}\{\mathbf{a}_{M_{1}}(\omega_{\mathrm{s}})\}\otimes\mathrm{Diag}\{\mathbf{a}_{M_{2}}(\mu_{\mathrm{s}})\})\\
= & \mathrm{Diag}\{\mathbf{a}_{M}(-\overline{\varphi_{1}},-\overline{\theta_{1}})\}\mathrm{Diag}\{\mathbf{a}_{M}(\omega_{\mathrm{s}},\mu_{\mathrm{s}})\}.
\end{aligned}
\]
Then, combining this equality with (\ref{eq:AoD_common_part}), $\mathbf{A}_{M}$
is rewritten as \vspace{-0.3cm}
\begin{align}
\mathbf{A}_{M} & =\mathrm{Diag}\{\mathbf{a}_{M}(-\overline{\varphi_{1}},-\overline{\theta_{1}})\}\mathrm{Diag}\{\mathbf{a}_{M}(\omega_{\mathrm{s}},\mu_{\mathrm{s}})\}\mathbf{A}_{\Delta M}\nonumber \\
 & \triangleq\mathrm{Diag}\{\mathbf{a}_{M}(-\overline{\varphi_{1}},-\overline{\theta_{1}})\}\mathbf{A}_{\mathrm{s}},\label{eq:AoD_common}
\end{align}
where $\mathbf{A}_{\mathrm{s}}=\mathrm{Diag}\{\mathbf{a}_{M}(\omega_{\mathrm{s}},\mu_{\mathrm{s}})\}\mathbf{A}_{\Delta M}$
can be readily estimated using Algorithm \ref{algorithm-2}. Based
on (\ref{eq:gamma_common}) and (\ref{eq:AoD_common}), the common
RIS-BS channel matrix $\mathbf{H}$ in (\ref{eq:H1-1})\textcolor{red}{{}
}is re-expressed as \vspace{-0.2cm}
\begin{align}
\mathbf{H}=\mathbf{A}_{N}\boldsymbol{\Lambda}\mathbf{A}_{M}^{\mathrm{H}} & =\mathbf{A}_{N}\frac{1}{\overline{\beta}}\boldsymbol{\Lambda}_{\mathrm{s}}\mathbf{A}_{\mathrm{s}}^{\mathrm{H}}\mathrm{Diag}\{\mathbf{a}_{M}(\overline{\varphi_{1}},\overline{\theta_{1}})\}\nonumber \\
 & \triangleq\frac{1}{\overline{\beta}}\mathbf{H}_{\mathrm{s}}\mathrm{Diag}\{\mathbf{a}_{M}(\overline{\varphi_{1}},\overline{\theta_{1}})\},\label{eq:H_common}
\end{align}
where $\mathbf{H}_{\mathrm{s}}=\mathbf{A}_{N}\boldsymbol{\Lambda}_{\mathrm{s}}\mathbf{A}_{\mathrm{s}}^{\mathrm{H}}$
is the common part of the cascaded channel that can be estimated using
Algorithm \ref{algorithm-1} and Algorithm \ref{algorithm-2}. Then,
combining (\ref{eq:H_common}) with (\ref{paril-G}), we have \vspace{-0.15cm}
\begin{align*}
\mathbf{G}_{k}= & \mathbf{H}\mathrm{Diag}\{\mathbf{h}_{k}\}=\frac{1}{\overline{\beta}}\mathbf{H}_{\mathrm{s}}\mathrm{Diag}\{\mathbf{a}_{M}(\overline{\varphi_{1}},\overline{\theta_{1}})\}\mathrm{Diag}\{\mathbf{h}_{k}\}\\
= & \mathbf{H}_{\mathrm{s}}\mathrm{Diag}\{\frac{1}{\overline{\beta}}\mathrm{Diag}\{\mathbf{a}_{M}(\overline{\varphi_{1}},\overline{\theta_{1}})\}\mathbf{h}_{k}\}=\mathbf{H}_{\mathrm{s}}\mathrm{Diag}\{\mathbf{h}_{\mathrm{s},k}\},
\end{align*}
where $\mathbf{h}_{\mathrm{s},k}=\frac{1}{\overline{\beta}}\mathrm{Diag}\{\mathbf{a}_{M}(\overline{\varphi_{1}},\overline{\theta_{1}})\}\mathbf{h}_{k}$
is the unique part of user $k$'s channel, that needs to be obtained.
Next we will show how to estimate the unique part and present the
channel estimation strategy for other users, leading to a significant
reduction in the pilot overhead.

\subsubsection{Estimation of Unique Part}

Denote the estimate of $\mathbf{H}_{\mathrm{s}}$ as $\mathbf{\widehat{H}}_{\mathrm{s}}=\mathbf{\widehat{A}}_{N}\widehat{\boldsymbol{\Lambda}}_{\mathrm{s}}\widehat{\mathbf{A}}_{\mathrm{s}}^{\mathrm{H}}$
where $\mathbf{\widehat{A}}_{N}$, $\widehat{\boldsymbol{\Lambda}}_{\mathrm{s}}$,
and $\widehat{\mathbf{A}}_{\mathrm{s}}$ are the estimates of $\mathbf{A}_{N}$,
$\boldsymbol{\Lambda}_{\mathrm{s}}$, and $\mathbf{A}_{\mathrm{s}}$,
respectively. By replacing $\mathbf{H}_{\mathrm{s}}$ with $\mathbf{\widehat{H}}_{\mathrm{s}}+\Delta\mathbf{H}_{\mathrm{s}}$
where $\Delta\mathbf{\mathbf{H}_{\mathrm{s}}}$ represents the error
between $\mathbf{H}_{\mathrm{s}}$ and its estimate, user $k$'s received
data $\mathbf{Y}_{k}$ after eliminating the effects of the estimated
common AoAs is expressed as \textcolor{red}{\vspace{-0.2cm}
}
\begin{align}
\frac{1}{N\sqrt{p}}\mathbf{\widehat{A}}_{N}^{\mathrm{H}}\mathbf{Y}_{k}= & \frac{1}{N}\mathbf{\widehat{A}}_{N}^{\mathrm{H}}\mathbf{H}_{\mathrm{s}}\mathrm{Diag}\{\mathbf{h}_{\mathrm{s},k}\}\mathbf{E}_{k}+\frac{1}{N\sqrt{p}}\mathbf{\widehat{A}}_{N}^{\mathrm{H}}\mathbf{N}_{k}\nonumber \\
= & \widehat{\boldsymbol{\Lambda}}_{\mathrm{s}}\widehat{\mathbf{A}}_{\mathrm{s}}^{\mathrm{H}}\mathrm{Diag}\{\mathbf{h}_{\mathrm{s},k}\}\mathbf{E}_{k}+\frac{1}{N\sqrt{p}}\mathbf{\widehat{A}}_{N}^{\mathrm{H}}\mathbf{N}_{k}\nonumber \\
 & +\frac{1}{N}\mathbf{\widehat{A}}_{N}^{\mathrm{H}}\Delta\mathbf{\mathbf{H}_{\mathrm{s}}}\mathrm{Diag}\{\mathbf{h}_{\mathrm{s},k}\}\mathbf{E}_{k}.\label{eq:LT_Yk_common}
\end{align}
For the estimation of $\mathbf{h}_{\mathrm{s},k}$, define $\mathbf{w}_{k}=\mathrm{vec}(\frac{1}{N\sqrt{p}}\mathbf{\widehat{A}}_{N}^{\mathrm{H}}\mathbf{Y}_{k})\in\mathbb{C}^{L\tau_{k}\times1}$.
Then, we have \vspace{-0.25cm}
\begin{align}
\mathbf{w}_{k}= & \mathrm{vec}(\widehat{\boldsymbol{\Lambda}}_{\mathrm{s}}\widehat{\mathbf{A}}_{\mathrm{s}}^{\mathrm{H}}\mathrm{Diag}\{\mathbf{h}_{\mathrm{s},k}\}\mathbf{E}_{k})+\widetilde{\mathbf{n}}_{k}\nonumber \\
= & (\mathbf{E}_{k}^{\mathrm{T}}\diamond\widehat{\boldsymbol{\Lambda}}_{\mathrm{s}}\widehat{\mathbf{A}}_{\mathrm{s}}^{\mathrm{H}})\mathbf{h}_{\mathrm{s},k}+\widetilde{\mathbf{n}}_{k}=\mathbf{W}_{k}\mathbf{h}_{\mathrm{s},k}+\widetilde{\mathbf{n}}_{k},\label{eq:Wk_}
\end{align}
where $\mathbf{W}_{k}\triangleq(\mathbf{E}_{k}^{\mathrm{T}}\diamond\widehat{\boldsymbol{\Lambda}}_{\mathrm{s}}\widehat{\mathbf{A}}_{\mathrm{s}}^{\mathrm{H}})$
and $\widetilde{\mathbf{n}}_{k}$ is the corresponding equivalent
noise vector given by\textcolor{red}{{} }$\mathrm{vec}(\frac{1}{N\sqrt{p}}\mathbf{\widehat{A}}_{N}^{\mathrm{H}}\mathbf{N}_{k}+\frac{1}{N}\mathbf{\widehat{A}}_{N}^{\mathrm{H}}\Delta\mathbf{\mathbf{H}_{\mathrm{s}}}\mathrm{Diag}\{\mathbf{h}_{\mathrm{s},k}\}\mathbf{E}_{k})\in\mathbb{C}^{L\tau_{k}\times1}$.
The second equality is obtained via $\mathrm{vec}(\mathbf{\mathbf{A}\mathrm{Diag}\{b\}C})=(\mathbf{\mathbf{C}^{\mathrm{T}}\diamond A})\mathbf{b}$
\cite{Xinda2017}. Then, substituting $\mathbf{h}_{k}=\mathbf{A}_{M,k}\boldsymbol{\beta}_{k}$
in (\ref{eq:hk}) into $\mathbf{h}_{\mathrm{s},k}$, we have \vspace{-0.2cm}
\begin{align}
\mathbf{h}_{\mathrm{s},k}= & \frac{1}{\overline{\beta}}\mathrm{Diag}\{\mathbf{a}_{M}(\overline{\varphi_{1}},\overline{\theta_{1}})\}\mathbf{A}_{M,k}\boldsymbol{\beta}_{k}\nonumber \\
= & (\mathbf{a}_{M}\mathrm{(}\overline{\varphi_{1}},\overline{\theta_{1}}\mathrm{)}\bullet\mathbf{A}_{M,k}\mathrm{)}\frac{1}{\overline{\beta}}\boldsymbol{\beta}_{k},\label{eq:hc_k}
\end{align}
where $\mathbf{a}_{M}(\overline{\varphi_{1}},\overline{\theta_{1}})\bullet\mathbf{A}_{M,k}=[\mathbf{a}_{M}(\varphi_{k,1}+\overline{\varphi_{1}},\theta_{k,1}+\overline{\theta_{1}}),\ldots,\mathbf{a}_{M}(\varphi_{k,J_{k}}+\overline{\varphi_{1}},\theta_{k,J_{k}}+\overline{\theta_{1}})]\in\mathbb{C}^{M\times J_{k}}$.
Since both $\varphi_{k,l}+\overline{\varphi_{1}}$ and $\theta_{k,l}+\overline{\theta_{1}}$
lie within $[-2\frac{d_{\mathrm{RIS}}}{\lambda_{c}},2\frac{d_{\mathrm{RIS}}}{\lambda_{c}}]$,
we can formulate (\ref{eq:Wk_}) as a $J_{k}$-sparse signal recovery
problem \vspace{-0.2cm}
\begin{align}
\mathbf{w}_{k}=\mathbf{W}_{k}\mathbf{h}_{\mathrm{s},k}+\widetilde{\mathbf{n}}_{k} & =\mathbf{W}_{k}\mathrm{(}\mathbf{a}_{M}(\overline{\varphi_{1}},\overline{\theta_{1}})\bullet\mathbf{A}_{M,k}\mathrm{)}\frac{1}{\overline{\beta}}\boldsymbol{\beta}_{k}+\widetilde{\mathbf{n}}_{k}\nonumber \\
 & =\mathbf{W}_{k}\mathrm{(}\mathbf{A}_{\mathrm{1}}\otimes\mathbf{A}_{\mathrm{2}}\mathrm{)}\mathbf{d}_{k}+\widetilde{\mathbf{n}}_{k}.\label{eq:wk_sparse}
\end{align}
Here $\mathbf{A}_{1}\in\mathbb{C}^{M_{1}\times D_{1}}$ and\textcolor{red}{{}
}$\mathbf{A}_{2}\in\mathbb{C}^{M_{2}\times D_{2}}$ are overcomplete
dictionary matrices similar to (\ref{eq:sparse-formula}) satisfying
$D_{1}\geq M_{1}$ and $D_{2}\geq M_{2}$, and $\mathbf{d}_{k}\in\mathbb{C}^{D_{1}D_{2}\times1}$
is a sparse vector with $J_{k}$ nonzero entries corresponding to
$\{\frac{1}{\overline{\beta}}\beta_{k,j}\}_{j=1}^{J_{k}}$. Hence,
the angle estimation problem corresponding to (\ref{eq:wk_sparse})
can be solved using CS-based techniques. To improve the estimation
performance, the alternating optimization (AO) method in \cite{Zhou_ULA_TSP}
can be adopted to optimize the RIS phase shift training matrix $\mathbf{E}_{k}$
so as to ensure the near column-orthogonality of the equivalent dictionary
$\mathbf{W}_{k}\mathrm{(}\mathbf{A}_{\mathrm{1}}\otimes\mathbf{A}_{\mathrm{2}}\mathrm{)}$.
In addition, the estimate of the number of scatterers between user-RIS
channel for user $k$, i.e., the sparsity level for the sparse recovery
problem associated with (\ref{eq:wk_sparse}) $J_{k}$, is obtained
by the selected CS-based techniques, similarly to the estimation of
$J_{1}$ discussed before. Note that we obtain the equivalent AoA
pair of user $k$'s user-RIS channel, i.e., $(\varphi_{k,j}+\overline{\varphi_{1}},\theta_{k,j}+\overline{\theta_{1}})$,
by solving angle estimation problem based on (\ref{eq:wk_sparse}).
The corresponding equivalent AoAs, i.e., $(\varphi_{k,j}+\overline{\varphi_{1}})$
and $(\theta_{k,j}+\overline{\theta_{1}})$, can be obtained similar
to (\ref{m_lsub}). Assume that the $p$-th element of sparse vector
$\mathbf{d}_{k}$ is nonzero, then the corresponding indices in $\mathbf{\mathbf{A}_{\mathrm{1}}}$
and $\mathbf{\mathbf{A}_{\mathrm{2}}}$ in (\ref{eq:wk_sparse}),
denoted by $p_{1}$ and $p_{2}$, are derived as \vspace{-0.2cm}
\begin{equation}
p_{1}=\left\lceil \frac{p}{D_{\mathrm{2}}}\right\rceil ,~p_{2}=p-D_{\mathrm{2}}(p_{1}-1).\label{pl_sub}
\end{equation}
Finally, we obtain an estimate of the equivalent AoA spatial frequencies
for user $k$'s user-RIS channel, i.e., $\{(\widehat{\varphi_{k,j}+\overline{\varphi_{1}}})\}_{j=1}^{\hat{J_{k}}}$
and $\{(\widehat{\theta_{k,j}+\overline{\theta_{1}}})\}_{j=1}^{\hat{J_{k}}}$.
Furthermore, user $k$'s cascaded AoDs, i.e., $(\omega_{l}-\varphi_{k,j})$
and $(\mu_{l}-\theta_{k,j})$, for $\forall l\in\{1,...,L\}$ and
$\forall j\in\{1,...,J_{k}\}$, can be also obtained as follows:\vspace{-0.3cm}
\begin{subequations}
\label{cascaded_omega_mu} 
\begin{align}
\omega_{l}-\varphi_{k,j} & =\omega_{r}+\overline{\varphi_{1}}-(\overline{\varphi_{1}}+\varphi_{k,j})+\omega_{l}-\omega_{r}\nonumber \\
 & =\omega_{\mathrm{s}}-(\overline{\varphi_{1}}+\varphi_{k,j})+\Delta\omega_{l},\label{eq:cascaded_omega}\\
\mu_{l}-\theta_{k,j} & =\mu_{r}+\overline{\theta_{1}}-(\overline{\theta_{1}}+\theta_{k,j})+\mu_{l}-\mu_{r}\nonumber \\
 & =\mu_{\mathrm{s}}-(\overline{\theta_{1}}+\theta_{k,j})+\Delta\mu_{l}.\label{eq:cascaded_mu}
\end{align}
\end{subequations}

\vspace{-0.25cm}

Based on (\ref{rot_fac_omega_mu_scale}), (\ref{omega_mu_common})
and (\ref{eq:wk_sparse}), the parameters $\Delta\omega_{l}$, $\Delta\mu_{l}$,
$\omega_{\mathrm{s}}$, $\mu_{\mathrm{s}}$, $(\overline{\varphi_{1}}+\varphi_{k,j})$
and $(\overline{\theta_{1}}+\theta_{k,j})$ for $\forall l\in\{1,2,...,L\}$
and $\forall j\in\{1,2,...,J_{k}\}$ can be readily estimated. Finally,
the completed CS-based estimation of $\mathbf{G}_{k}$ for $2\leq k\leq K$
is summarized in Algorithm \ref{algorithm-3}. As shown in Algorithm
\ref{algorithm-3}, the obtained common part of cascaded channel $\mathbf{H}_{\mathrm{s}}$
allows us to estimate the unique part $\mathbf{h}_{\mathrm{s},k}$
with reduced pilot overhead.

\begin{algorithm}
\caption{Estimation of Full CSI for Other Users}

\label{algorithm-3}

\begin{algorithmic}[1]

\REQUIRE $\mathbf{Y}_{k}$, $\mathbf{\widehat{A}}_{N}$.

\STATE Obtain the estimate $\widehat{\boldsymbol{\Lambda}}_{\mathrm{s}}$
based on (\ref{eq:gamma_common});

\STATE Obtain the estimate $\widehat{\mathbf{A}}_{\mathrm{s}}$ based
on (\ref{eq:AoD_common});

\STATE Obtain the estimate of the common part, i.e., $\mathbf{\widehat{H}}_{\mathrm{s}}=\mathbf{\widehat{A}}_{N}\widehat{\boldsymbol{\Lambda}}_{\mathrm{s}}\widehat{\mathbf{A}}_{\mathrm{s}}^{\mathrm{H}}$;

\FOR{$2\leq k\leq K$}

\STATE Calculate $\mathbf{w}_{k}=\mathrm{vec}(\frac{1}{N\sqrt{p}}\mathbf{\widehat{A}}_{N}^{\mathrm{H}}\mathbf{Y}_{k})$;

\STATE Calculate equivalent dictionary $\mathbf{W}_{k}(\mathbf{A}_{\mathrm{1}}\otimes\mathbf{A}_{\mathrm{2}})$
according to (\ref{eq:Wk_});

\STATE Estimate unique part $\mathbf{h}_{\mathrm{s},k}$ by solving
sparse recovery problem associated with (\ref{eq:wk_sparse});\label{complexity_3_1}

\STATE Obtain the estimate of cascaded channel, i.e., $\widehat{\mathbf{G}}_{k}=\widehat{\mathbf{H}}_{\mathrm{s}}\mathrm{Diag}\{\widehat{\mathbf{h}}_{\mathrm{s},k}\}$;

\ENDFOR

\ENSURE $\widehat{\mathbf{G}}_{k},2\leq k\leq K$.

\end{algorithmic} 
\end{algorithm}

\vspace{-0.35cm}

\subsection{Pilot Overhead and Computational Complexity Analysis\label{subsec:complexity_analysis}}

\vspace{-0.15cm}
 In this subsection, we first analyze the pilot overhead required
for the full CSI estimation. Then, the corresponding computational
complexity is evaluated. For simplicity, $J_{1}=J_{2}=\cdots=J_{K}=J$
is assumed.

\subsubsection{Pilot Overhead Analysis \label{subsec:Pilot-Overhead}}

Clearly, the number of pilot symbols directly affects the sparse recovery
performance for equations (\ref{eq:sparse-formula}) and (\ref{eq:wk_sparse}).
According to \cite{Sparse-Recovery_Measurement}, to find a $l$-sparse
complex signal (vector) with dimension $n$, the number of measurements
$m$ is required to be on the order of $\mathcal{O}(l\log(n))$, which
is proportional to the sparsity level $l$.

Based on this fact, we first analyze the number of pilots required
for the typical user, i.e., user $1$. For the sparse recovery problem
associated with (\ref{eq:sparse-formula}) in Stage I, the dimension
of the equivalent sensing matrix $\mathbf{F_{\mathrm{1}}}\triangleq\mathbf{E}_{1}^{\mathrm{H}}(\mathbf{A}_{\mathrm{1}}\otimes\mathbf{A}_{\mathrm{2}})$
is $\tau_{1}\times D_{1}D_{2}$ where $D_{1}\geq M_{1}$ and $D_{2}\geq M_{2}$,
and the corresponding sparsity level is $J_{1}$, thus the pilot overhead
required for user $1$ should satisfy{} $\tau_{1}\geq\mathcal{O}(J_{1}\log(D_{1}D_{2}))\geq\mathcal{O}(J_{1}\log(M_{1}M_{2}))=\mathcal{O}(J_{1}\log(M))$.

For the sparse recovery problem associated with (\ref{eq:wk_sparse})
in Stage II, the dimension of the equivalent sensing matrix $\mathbf{F}_{k}\triangleq\mathbf{W}_{k}(\mathbf{A}_{\mathrm{1}}\otimes\mathbf{A}_{\mathrm{2}})$
is $L\tau_{k}\times D_{1}D_{2}$ where $D_{1}\geq M_{1}$ and $D_{2}\geq M_{2}$,
and the corresponding sparsity level is $J_{k}$, thus user $k$ needs
$\tau_{k}\geq\mathcal{O}(J_{k}\log(D_{1}D_{2})/L)\geq\mathcal{O}(J_{k}\log(M_{1}M_{2})/L)=\mathcal{O}(J_{k}\log(M)/L)$
pilot symbols. Therefore, the overall required pilot overhead in the
first coherence block is $\mathcal{O}(J\log(M)+(K-1)J\log(M)/L)$.
\vspace{-0.05cm}

\subsubsection{Computational Complexity Analysis\label{subsec:Computational-Complex}}

For the estimation of the typical user in Stage I shown in Algorithm
\ref{algorithm-2}, the computational complexity mainly stems from
Algorithm \ref{algorithm-1} in Step 1, the CS-based method for the
estimation of ${\bf h}_{\mathrm{RIS},r}$ in Step \ref{Complexity_2_1}
and the correlation based scheme in Step \ref{Complexity_2_2}. Specifically,
the dominant complexity for Algorithm \ref{algorithm-1} are calculating
the matrix multiplication in its Step \ref{Complexity_1_1} with computational
complexity of $\mathcal{O}(N^{2}\tau_{1})$ and implementing the angle
rotation in its Step \ref{Complexity_1_2} with computational complexity
of $\mathcal{O}(N\tau_{1}L(g_{1}+g_{2}))$. We take OMP as the recovery
algorithm, whose corresponding dominant complexity is $\mathcal{O}(mnl)$
\cite{Zhou_ULA_TSP}, where $m$ is the length of the measurements,
and $n$ is the length of the sparse signal with sparsity level $l$.
Hence, the complexity for estimating ${\bf h}_{\mathrm{RIS},r}$ is
$\mathcal{O}(\tau_{1}D_{1}D_{2}J_{1})$. Additionally, the computational
complexity of the correlation based scheme is given by $\mathcal{O}(M\tau_{1}(L-1)d_{1}d_{2})$,
where $d_{1}$ and $d_{2}$ represent the search grids for $\Delta\omega_{l}$
and $\Delta\mu_{l}$ within $[-2\frac{d_{\mathrm{RIS}}}{\lambda_{c}},2\frac{d_{\mathrm{RIS}}}{\lambda_{c}}]$,
respectively. The overall computational complexity in Stage I is $\mathcal{O}(\tau_{1}D_{1}D_{2}J+N^{2}\tau_{1}+N\tau_{1}L(g_{1}+g_{2})+M\tau_{1}(L-1)d_{1}d_{2})$.

Then, we analyze the computational complexity for the estimation of
other users in Stage II shown in Algorithm \ref{algorithm-3}, which
mainly stems from the CS-based method for estimation of $\mathbf{h}_{\mathrm{s},k}$
in Step \ref{complexity_3_1}. Similarly, we choose OMP to solve the
sparse recovery problem associated with (\ref{eq:wk_sparse}), and
thus the corresponding computational complexity is $\mathcal{O}(\tau_{k}LD_{1}D_{2}J_{k})$.
Consider $(K-1)$ users in total, the overall computational complexity
in Stage II during the first coherence block is $\mathcal{O}((K-1)\tau_{k}LD_{1}D_{2}J)$.\vspace{-0.2cm}

\section{Channel Estimation in Remaining Coherence Blocks\label{sec:Remaining_Coherence}}

\vspace{-0.15cm}
 After the first coherence block, we adopt the LS estimator to re-estimate
the cascaded gains since the angles remain unchanged during the remaining
coherence blocks. Later we will see the required pilot overhead can
be reduced further in this stage.

Without loss of generality, we consider an arbitrary $k$ from $\mathcal{K}$
and show how to re-estimate user $k$'s channel gains. Similar to
(\ref{eq:Y1_formula}), we first take user $k$'s equivalent measurement
matrix $\overline{\mathbf{\mathbf{Y}}}_{k}$, i.e., $\overline{\mathbf{\mathbf{Y}}}_{k}=(\frac{1}{N\sqrt{p}}\mathbf{\widehat{A}}_{N}^{\mathrm{H}}\mathbf{Y}_{k})^{\mathrm{H}}\in\mathbb{C}^{\tau_{k}\times L}$,
where $\mathbf{\widehat{A}}_{N}$ has been acquired in Stage I. Then,
following the same derivations as for (\ref{eq:yl_sparse}), the $r$-th
column of $\overline{\mathbf{\mathbf{Y}}}_{k}$ , denoted as $\overline{{\bf y}}_{k,r}$,
is given by \vspace{-0.25cm}
\begin{align}
\overline{{\bf y}}_{k,r} & =\mathbf{E}_{k}^{\mathrm{H}}(\mathbf{A}_{M,k}^{*}\bullet\mathbf{a}_{M}(\omega_{r},\mu_{r}))\alpha_{r}^{*}\boldsymbol{\beta}_{k}^{*}+\overline{{\bf n}}_{k,r}\nonumber \\
 & \triangleq\mathbf{E}_{k}^{\mathrm{H}}\mathbf{V}_{k,r}\alpha_{r}^{*}\boldsymbol{\beta}_{k}^{*}+\overline{{\bf n}}_{k,r}.\label{eq:Yk_r}
\end{align}
Here, $\mathbf{V}_{k,r}\triangleq\mathbf{A}_{M,k}^{*}\bullet\mathbf{a}_{M}(\omega_{r},\mu_{r})$$=$$[\mathbf{a}_{M}(\omega_{r}-\varphi_{k,1},\mu_{r}-\theta_{k,1}),...,\mathbf{a}_{M}(\omega_{r}-\varphi_{k,J_{k}},\mu_{r}-\theta_{k,J_{k}})]\in\mathbb{C}^{M\times J_{k}}$
and $\overline{{\bf n}}_{k,r}$ is the $r$-th column of $[\frac{1}{N\sqrt{p}}\mathbf{\widehat{A}}_{N}^{\mathrm{H}}(\mathbf{N}_{k}+\sqrt{p}\Delta\mathbf{A}_{N}\boldsymbol{\Lambda}\mathbf{A}_{M}^{\mathrm{H}}\mathrm{Diag}\{\mathbf{h}_{k}\mathrm{\}}\mathbf{E}_{k})]^{\mathrm{H}}$.
We have already obtained an estimate of $\mathbf{V}_{k,r}$, denoted
by $\widehat{\mathbf{V}}_{k,r}$, in the first coherence block. Specifically,
for the typical user, i.e., user $1$, $\{(\omega_{r}-\varphi_{1,j},\mu_{r}-\theta_{1,j})\}_{j=1}^{J_{1}}$
are estimated from (\ref{eq:yl_sparse}) and (\ref{cascaded1_omega_mu})
in Stage I, while for other users, $\{(\omega_{r}-\varphi_{k,j},\mu_{r}-\theta_{k,j})\}_{j=1}^{J_{k}}$
are estimated from (\ref{cascaded_omega_mu}) in Stage II.

\vspace{-0.05cm}

The updated cascaded channel gain $\boldsymbol{\beta}_{k}^{*}\alpha_{r}^{*}$
in (\ref{eq:Yk_r}) can be found using the LS estimator \vspace{-0.15cm}
\begin{equation}
\widehat{\boldsymbol{\beta}_{k}^{*}\alpha_{r}^{*}}=(\mathbf{\widehat{V}}_{k,r}^{\mathrm{H}}\mathbf{E}_{k}\mathbf{E}_{k}^{\mathrm{H}}\widehat{\mathbf{V}}_{k,r})^{-1}\mathbf{\widehat{V}}_{k,r}^{\mathrm{H}}\mathbf{E}_{k}\overline{{\bf y}}_{k,r}.\label{eq:cascaded_gain_update}
\end{equation}
Then, following the same operation shown in (\ref{eq:G1_estimate}),
and substituting (\ref{eq:cascaded_gain_update}) into (\ref{eq:Yk_r}),
the estimate of user $k$'s cascaded channel during the remaining
coherence blocks is given by \vspace{-0.2cm}
\begin{align*}
\widehat{\mathbf{G}}_{k}=\mathbf{\widehat{A}}_{N}\mathbf{\hat{H}}_{\mathrm{RIS},k}^{\mathrm{H}} & =\mathbf{\widehat{A}}_{N}[{\bf \hat{h}}_{\mathrm{RIS}k,1},\ldots,{\bf \hat{h}}_{\mathrm{RIS}k,L}]^{\mathrm{H}}\\
 & =\mathbf{\widehat{A}}_{N}[\mathbf{\widehat{V}}_{k,1}\widehat{\boldsymbol{\beta}_{k}^{*}\alpha_{1}^{*}},\ldots,\mathbf{\widehat{V}}_{k,L}\widehat{\boldsymbol{\beta}_{k}^{*}\alpha_{L}^{*}}]^{\mathrm{H}},
\end{align*}
where ${\bf h}_{\mathrm{RIS}k,r}$ represents the $r$-th column of
$\mathbf{H}_{\mathrm{RIS},k}$.

For the pilot overhead analysis, we assume $J_{1}=J_{2}=\cdots=J_{K}=J$
as before. For the LS problem in (\ref{eq:Yk_r}), $\tau_{k}\geq J_{k}$
should hold for user $k$. Thus, the minimum number of pilot symbols
can be chosen as $\tau_{k}=J_{k}$, which is less than that required
in Stage II. Given $K$ total users, the overall minimum pilot overhead
is $JK$. On the other hand, the dominant complexity of LS problem
in (\ref{eq:Yk_r}) is $\mathcal{O}(\tau_{k}J^{2})$. Since obtaining
the entire cascaded channel, i.e., $\mathbf{G}_{k}$, needs to solve
the LS problem $L$ times, the total computational complexity for
user $k$ is $\mathcal{O}(\tau_{k}J^{2}L)$. Thus the overall computational
complexity in each remaining coherence block is $\mathcal{O}(\tau_{k}J^{2}LK)$.\vspace{-0.35cm}

\section{Extension to Multi-antenna User Case \label{sec:Applying-the-Protocol}}

\vspace{-0.15cm}
 \textcolor{black}{In this section, we extend the full CSI estimation
method in the first coherence block to the multi-antenna user case.}\footnote{The re-estimation of channel gains in the remaining coherence blocks
can be extended to the {multi-antenna-users}\textcolor{black}{{}
case} in a straightforward way, and thus will not be explicitly considered.}\textcolor{black}{{} First, the system model and corresponding two-phase
channel estimation strategy are described.} Then, we adopt an OMP-based
method to estimate the AoDs at the users in Phase I. The remaining
parameters including the common AoAs at the BS, the cascaded AoDs
at the RIS, and the cascaded gains are estimated in Phase II, similarly
to the methods developed for the single-antenna user case in Section
\ref{sec:First_coherence}. Lastly, the required pilot overhead and
computation complexity for the proposed method are \textcolor{black}{analyzed.}
\vspace{-0.45cm}

\subsection{Multi-antenna Users Model and Channel Estimation Strategy}

\vspace{-0.15cm}

\subsubsection{System Model}

We assume that $K$ users are present with an $Q_{k}=Q_{k1}\times Q_{k2}$
UPA for user $k$, while the other settings are the same as in the
single-antenna user case. Then, $\mathbf{h}_{k}$ in (\ref{eq:H_h})
and (\ref{eq:hk}) can be modified as \vspace{-0.25cm}
\begin{align}
\mathbf{H_{\mathit{k}}} & =\sum_{j=1}^{J_{k}}\beta_{k,j}\mathbf{a}_{M}(\varphi_{k,j},\theta_{k,j})\mathbf{a}_{Q_{k}}^{\mathrm{H}}(\eta_{k,j},\chi_{k,j})\nonumber \\
 & =\mathbf{A}_{M,k}\mathbf{B}_{k}\mathbf{A}_{Q,k}^{\mathrm{H}}\in\mathbb{C}^{M\times Q_{k}},\forall k\in\mathcal{K},\label{eq:hk_q}
\end{align}
where $(\eta_{k,j},\chi_{k,j})$ represents the AoD of the $j$-th
path in the user $k$-RIS channel, and $\mathbf{A}_{Q,k}=[\mathbf{a}_{Q_{k}}(\eta_{k,1},\chi_{k,1}),\ldots,\mathbf{a}_{Q_{k}}(\eta_{k,J_{k}},\chi_{k,J_{k}})]\in\mathbb{C}^{Q_{k}\times J_{k}}$
and $\mathbf{B}_{k}=\mathrm{Diag}\{\beta_{k,1},\ldots,\beta_{k,J_{k}}\}\in\mathbb{C}^{J_{k}\times J_{k}}$
are the AoD steering matrix and complex gain matrix of user $k$,
respectively. Other parameters are as defined in Section \ref{sec:Model-protocol}.
With $\mathbf{H_{\mathit{k}}}$, the transmission model in (\ref{transmission})
becomes \vspace{-0.2cm}
\begin{equation}
\mathbf{y}_{k}(t)=\mathbf{H}\mathrm{Diag}\{\mathbf{e}_{t}\}\mathbf{H_{\mathit{k}}}\sqrt{p}\mathbf{s}_{k}(t)+\mathbf{n}_{k}(t),\label{eq:Qk_transmitter}
\end{equation}
where $\mathbf{s}_{k}(t)\in\mathbb{C}^{Q_{k}\times1}$ is the pilot
vector for user $k$ in time slot $t$. Vectorizing (\ref{eq:Qk_transmitter}),
we have \vspace{-0.2cm}
\begin{align}
\mathbf{y}_{k}(t) & =\sqrt{p}(\mathbf{s}_{k}^{\mathrm{T}}(t)\otimes\mathbf{I}_{N})\mathrm{vec}(\mathbf{H}\mathrm{Diag}\{\mathbf{e}_{t}\}\mathbf{H}_{k})+\mathbf{n}_{k}(t)\nonumber \\
 & \triangleq\sqrt{p}(\mathbf{s}_{k}^{\mathrm{T}}(t)\otimes\mathbf{I}_{N})\mathbf{G}_{k}\mathbf{e}_{t}+\mathbf{n}_{k}(t),\label{eq:Q-antenna}
\end{align}
where $\mathbf{I}_{N}$ represents the $N\times N$ identity matrix,
and $\mathbf{G}_{k}=\mathbf{H}_{k}^{\mathrm{T}}\diamond\mathbf{H}$
is the cascaded user-RIS-BS channel of user $k$ that is to be estimated.
The above equality is also obtained via $\mathrm{vec}(\mathbf{\mathbf{A}\mathrm{Diag}\{b\}C})=(\mathbf{\mathbf{C}^{\mathrm{T}}\diamond A})\mathbf{b}$.
Combining (\ref{eq:hk_q}) with (\ref{eq:H1-1}), $\mathbf{G}_{k}$
can be rewritten as \vspace{-0.15cm}
\begin{align}
\mathbf{G}_{k} & =(\mathbf{A}_{M,k}\mathbf{B}_{k}\mathbf{A}_{Q,k}^{\mathrm{H}})^{\mathrm{T}}\diamond(\mathbf{A}_{N}\boldsymbol{\Lambda}\mathbf{A}_{M}^{\mathrm{H}})\nonumber \\
 & =(\mathbf{A}_{Q,k}^{\mathrm{*}}\otimes\mathbf{A}_{N})(\mathbf{B}_{k}\otimes\boldsymbol{\Lambda})(\mathbf{A}_{M,k}\bullet\mathbf{A}_{M}^{\mathrm{*}})^{\mathrm{T}},\label{eq:G_k-Q}
\end{align}
where the second equality are obtained using $(\mathbf{A}\otimes\mathbf{B})(\mathbf{C}\diamond\mathbf{D})=(\mathbf{AC})\diamond(\mathbf{BD})$
and $\mathbf{A}^{\mathrm{T}}\diamond\mathbf{B}^{\mathrm{T}}=(\mathbf{A}\bullet\mathbf{B})^{\mathrm{T}}$
{\cite{K-R_product,Xinda2017}}. The third term $(\mathbf{A}_{M,k}\bullet\mathbf{A}_{M}^{\mathrm{*}})$
accounts for the cascaded AoDs at the RIS, similar to the single-antenna
user case.

\subsubsection{Channel Estimation Strategy}

For the full-CSI estimation of any user $k$, a two-phase estimation
strategy is adopted, where the AoDs at the users, i.e., $\mathbf{A}_{Q,k}$,
is estimated in Phase I, after which the remaining parameters in (\ref{eq:G_k-Q})
are estimated in Phase II. Specifically, in this strategy, $\Upsilon_{k}$
blocks of time slots are used for the channel estimation of user $k$,
and the $i$-th block has $V_{k}^{(i)}$ time slots. The RIS phase
shift vector remains invariant for each time slot within a given block,
and is denoted by $\mathbf{e}^{(i)}$ for $\forall i\in\{1,2,...,\Upsilon_{k}\}$.
Later we will see that Phase I only occurs in the first block, and
$V_{k}^{(i)}$ can be different for different users or/and different
blocks, while Phase II consists of the whole blocks.\vspace{-0.35cm}

\subsection{Estimation in Phase I: Angle Estimation at Users}

\vspace{-0.1cm}
 In this subsection, we describe the estimation of the AoDs at the
users.

During the first block, user $k$ transmits the pilot sequence $\mathbf{S}_{k}^{(1)}=\left[\mathbf{s}_{1}^{(1)},\ldots,\mathbf{s}_{V_{k}^{(1)}}^{(1)}\right]\in\mathbb{C}^{Q_{k}\times V_{k}^{(1)}}$,
and the received signal matrix $\mathbf{Y}_{k}^{(1)}=\left[\mathbf{y}_{k}^{(1)}(1),\ldots,\mathbf{y}_{k}^{(1)}(V_{k}^{(1)})\right]\in\mathbb{C}^{N\times V_{k}^{(1)}}$
at the BS is given by \vspace{-0.2cm}
\begin{align}
\mathbf{Y}_{k}^{(1)} & =\sqrt{p}\mathbf{H}\mathrm{Diag}\{\mathbf{e}^{(1)}\}\mathbf{H}_{k}\mathbf{S}_{k}^{(1)}+\mathbf{N}_{k}^{(1)}\nonumber \\
 & =\sqrt{p}\mathbf{A}_{N}\boldsymbol{\Lambda}\mathbf{A}_{M}^{\mathrm{H}}\mathrm{Diag}\{\mathbf{e}^{(1)}\}\mathbf{A}_{M,k}\mathbf{B}_{k}\mathbf{A}_{Q,k}^{\mathrm{H}}\mathbf{S}_{k}^{(1)}+\mathbf{N}_{k}^{(1)}.\label{Y_matrix}
\end{align}
$\mathbf{A}_{Q,k}$ can be directly obtained from (\ref{Y_matrix}).

Specifically, for the estimation of $\mathbf{A}_{Q,k}$, an OMP-based
method can be adopted, which takes the transpose of (\ref{Y_matrix})
and formulates it as a simultaneously sparse approximation problem
\cite{CE_MIMO_RIS,Hybrid_beamforming}\vspace{-0.2cm}
\begin{equation}
(\mathbf{Y}_{k}^{(1)})^{\mathrm{H}}=(\mathbf{S}_{k}^{(1)})^{\mathrm{H}}\mathbf{A}_{Q,k}\mathbf{\Gamma}_{k}+(\mathbf{N}_{k}^{(1)})^{\mathrm{H}}\in\mathbb{C}^{V_{k}^{(1)}\times N},\label{AoD_UE}
\end{equation}
where $\mathbf{\Gamma}_{k}$ represents the remaining terms according
to (\ref{Y_matrix}). Similar to equations (\ref{eq:sparse-formula})
and (\ref{eq:wk_sparse}), by using the VAD representation, (\ref{AoD_UE})
can be approximated as\vspace{-0.2cm}
\begin{equation}
(\mathbf{Y}_{k}^{(1)})^{\mathrm{H}}=(\mathbf{S}_{k}^{(1)})^{\mathrm{H}}(\mathbf{A}_{Q,1}\otimes\mathbf{A}_{Q,2})\widetilde{\mathbf{\boldsymbol{\mathbf{\Gamma}}}}_{k}+(\mathbf{N}_{k}^{(1)})^{\mathrm{H}},\label{AoD_VAD}
\end{equation}
where $\mathbf{A}_{Q,1}\in\mathbb{C}^{Q_{k1}\times D_{1}}$ and $\mathbf{A}_{Q,2}\in\mathbb{C}^{Q_{k1}\times D_{2}}$
are overcomplete dictionary matrices\textcolor{red}{{} }$(D_{1}\geq Q_{k1},D_{2}\geq Q_{k1})$
similar to (\ref{eq:sparse-formula}), and contain values for $\mathbf{a}_{Q_{k1}}(\eta_{k,j})$
and $\mathbf{a}_{Q_{k2}}(\chi_{k,j})$. $\widetilde{\mathbf{\boldsymbol{\mathbf{\Gamma}}}}_{k}\in\mathbb{C}^{D_{1}D_{2}\times N}$
is a row-sparse matrix with $J_{k}$ non-zero rows. Similar to the
single-antenna user case in Section \ref{sec:First_coherence}, the
sparsity level for the sparse recovery problem associated with (\ref{AoD_VAD})
$J_{k}$, is obtained by OMP. Therefore, the AoDs at user $k$, i.e.,
$\{\eta_{k,j}\}{}_{j=1}^{J_{k}}$ and $\{\chi_{k,j}\}{}_{j=1}^{J_{k}}$
can be obtained similar to (\ref{m_lsub}). Assume the $q$-th row
of the sparse matrix $\widetilde{\mathbf{\boldsymbol{\mathbf{\Gamma}}}}_{k}$
is nonzero, then the corresponding indices in $\mathbf{A}_{Q,1}$
and $\mathbf{A}_{Q,2}$ in (\ref{AoD_VAD}), denoted by $q_{1}$ and
$q_{2}$, are derived as\vspace{-0.1cm}
\begin{equation}
q_{1}=\left\lceil \frac{q}{D_{\mathrm{2}}}\right\rceil ,~q_{2}=q-D_{\mathrm{2}}(q_{1}-1).\label{ql_sub}
\end{equation}

\vspace{-0.4cm}

\subsection{Estimation in Phase II: Estimation of Remaining Parameters}

\vspace{-0.1cm}
 In this subsection, we estimate the remaining parameters in (\ref{eq:G_k-Q})
by converting the estimation problems into several equivalent problems
as in the single-antenna user case, which can be solved using the
methods in Section \ref{sec:First_coherence}.

First, denote the typical user as user $1$ and stack the total $(\sum_{i=1}^{\Upsilon_{1}}V_{1}^{(i)})$
slots, the received signal matrix is obtained as $\mathbf{Y}_{1}=\left[\mathbf{Y}_{1}^{(1)},\ldots,\mathbf{Y}_{1}^{(\Upsilon_{1})}\right]\in\mathbb{C}^{N\times(\sum_{i=1}^{\Upsilon_{1}}V_{1}^{(i)})}$.
Then, the common AoAs in (\ref{eq:G_k-Q}), i.e., $\mathbf{A}_{N}$,
can be readily estimated via DFT-based method by calculating $\widetilde{\mathbf{U}}_{N}^{\mathrm{H}}\mathbf{Y}_{1}$
since Lemma \ref{lem:3} holds.

With $\mathbf{\widehat{A}}_{Q,k}$ obtained in Phase I and $\mathbf{\widehat{A}}_{N}$
obtained in Phase II, considering the $i$-th time block and replacing
$\mathbf{A}_{N}$ and $\mathbf{A}_{Q,k}$ with $\mathbf{\widehat{A}}_{N}+\Delta\mathbf{A}_{N}$
and $\mathbf{\widehat{A}}_{Q,k}+\Delta\mathbf{A}_{Q,k}$, respectively,
$\mathbf{Y}_{k}^{(i)}\in\mathbb{C}^{N\times V_{k}^{(i)}}$ can be
processed as \vspace{-0.2cm}
\begin{align}
\mathbf{\bar{Y}}_{k}^{(i)}\triangleq & \frac{1}{N\sqrt{p}}\mathbf{\widehat{A}}_{N}^{\mathrm{H}}\mathbf{Y}_{k}^{(i)}(\mathbf{\widehat{A}}_{Q,k}^{\mathrm{H}}\mathbf{S}_{k}^{(i)})^{\dagger}\nonumber \\
= & \frac{1}{N\sqrt{p}}\mathbf{\widehat{A}}_{N}^{\mathrm{H}}\{(\mathbf{\widehat{A}}_{N}+\Delta\mathbf{A}_{N})\boldsymbol{\Lambda}\mathbf{A}_{M}^{\mathrm{H}}\mathrm{Diag}\{\mathbf{e}^{(i)}\}\nonumber \\
 & \mathbf{A}_{M,k}\mathbf{B}_{k}(\mathbf{\widehat{A}}_{Q,k}+\Delta\mathbf{A}_{Q,k})^{\mathrm{H}}\mathbf{S}_{k}^{(i)}+\mathbf{N}_{k}^{(i)}\}(\mathbf{\widehat{A}}_{Q,k}^{\mathrm{H}}\mathbf{S}_{k}^{(i)})^{\dagger}\nonumber \\
= & \boldsymbol{\Lambda}\mathbf{A}_{M}^{\mathrm{H}}\mathrm{Diag}\{\mathbf{e}^{(i)}\}\mathbf{A}_{M,k}\mathbf{B}_{k}+\mathbf{\bar{N}}_{k}^{(i)}\in\mathbb{C}^{L\times J_{k}},\label{received_mtx_Q}
\end{align}
where $\Delta\mathbf{A}_{N}$ and $\Delta\mathbf{A}_{Q,k}$ stand
for the estimation errors of $\mathbf{A}_{N}$ and $\mathbf{A}_{Q,k}$,
respectively. $\mathbf{\check{N}}_{k}^{(i)}$ represents the remaining
terms of the second equality. As discussed in Remark \ref{AoA_remark},
all users are allowed to estimate the common $\mathbf{A}_{N}$ jointly
so as to acquire the MU diversity gains to alleviate the error propagation
effects caused by $\Delta\mathbf{A}_{N}$. Accordingly, the input
of Algorithm \ref{algorithm-1} is given by $\mathbf{Y}=[\mathbf{Y}_{1},\mathbf{Y}_{2},...,\mathbf{Y}_{K}]\in\mathbb{C}^{N\times(\sum_{k=1}^{K}\sum_{i=1}^{\Upsilon_{k}}V_{k}^{(i)})}$.

In the following, we decompose the estimation of a multi-antenna user,
i.e., user $k$, with a channel composed of $J_{k}$ scatterers, into
the estimation of $J_{k}$ channels with a single path for a virtual
single-antenna user, i.e., user $(k,j)$ for $j\in\{1,\ldots,J_{k}\}$.
The $j$-th column of $\mathbf{\check{Y}}_{k}^{(i)}$ is given by
\vspace{-0.2cm}
\begin{align}
[\mathbf{\check{Y}}_{k}^{(i)}]_{(:,j)} & =\boldsymbol{\Lambda}\mathbf{A}_{M}^{\mathrm{H}}\mathrm{Diag}\{\mathbf{e}^{(i)}\}[\mathbf{A}_{M,k}]_{(:,j)}\beta_{k,j}+[\mathbf{\check{N}}_{k}^{(i)}]_{(:,j)}\nonumber \\
 & =\boldsymbol{\Lambda}\mathbf{A}_{M}^{\mathrm{H}}\mathrm{Diag}\{[\mathbf{A}_{M,k}]_{(:,j)}\beta_{k,j}\}\mathbf{e}^{(i)}+[\mathbf{\check{N}}_{k}^{(i)}]_{(:,j)}.\label{Yk_Q_block}
\end{align}
Stacking $\Upsilon_{k}$ blocks of (\ref{Yk_Q_block}), we have \vspace{-0.15cm}
\begin{align}
 & \left[[\mathbf{\check{Y}}_{k}^{(1)}]_{(:,j)},...,[\mathbf{\check{Y}}_{k}^{(\Upsilon_{k})}]_{(:,j)}\right]\nonumber \\
= & \boldsymbol{\Lambda}\mathbf{A}_{M}^{\mathrm{H}}\mathrm{Diag}\{[\mathbf{A}_{M,k}]_{(:,j)}\beta_{k,j}\}\mathbf{\widetilde{E}}_{k}+\left[[\mathbf{\check{N}}_{k}^{(1)}]_{(:,j)},...,[\mathbf{\check{N}}_{k}^{(\Upsilon_{k})}]_{(:,j)}\right]\nonumber \\
= & \boldsymbol{\Lambda}\mathbf{A}_{M}^{\mathrm{H}}\mathrm{Diag}\{\mathbf{\widetilde{h}}_{\{k,j\}}\}\mathbf{\widetilde{E}}_{k}+\left[[\mathbf{\check{N}}_{k}^{(1)}]_{(:,j)},...,[\mathbf{\check{N}}_{k}^{(\Upsilon_{k})}]_{(:,j)}\right],\label{Q_collect}
\end{align}
where $\mathbf{\widetilde{E}}_{k}=\left[\mathbf{e}^{(1)},\ldots,\mathbf{e}^{(\Upsilon_{k})}\right]\in\mathbb{C}^{M\times\Upsilon_{k}}$.
The term $\mathbf{\widetilde{h}}_{\{k,j\}}\triangleq[\mathbf{A}_{M,k}]_{(:,j)}\beta_{k,j}\in\mathbb{C}^{M\times1}$
is treated as the channel between the RIS and the virtual single-antenna
user $(k,j)$, which only contains one scatterer.

\subsubsection{Estimation for Typical User}

This part is the extension of Section \ref{subsec:Stage-I:-Estimation}
for the typical user, i.e., user $1$. Denote the transpose of (\ref{Q_collect})
for user $1$ as $\mathbf{\widetilde{Y}}_{\{1,j\}}\in\mathbb{C}^{\Upsilon_{1}\times L}$,
which is given by\vspace{-0.15cm}
\begin{equation}
\mathbf{\widetilde{Y}}_{\{1,j\}}=\mathbf{\widetilde{E}}_{1}^{\mathrm{H}}\mathrm{Diag}\{\mathbf{\widetilde{h}}_{\{1,j\}}^{*}\}\mathbf{A}_{M}\boldsymbol{\Lambda}^{*}+\mathbf{\widetilde{N}}_{\{1,j\}}.\label{Y1_formula_Q}
\end{equation}
We note that the channel estimation problem for (\ref{Y1_formula_Q})
has a form similar to that for (\ref{eq:Y1_formula}), and can be
solved following the steps developed in Section \ref{subsec:Stage-I:-Estimation}.
Thus the virtual single-antenna cascaded AoDs for user $(1,j)$, i.e.,
$\{(\omega_{l}-\varphi_{1,j})\}_{l=1}^{L}$ and $\{(\mu_{l}-\theta_{1,j})\}_{l=1}^{L}$,
and the cascaded gains $\{\alpha_{l}\beta_{1,j}\}_{l=1}^{L}$ can
be estimated.

It is unnecessary for us to repeat the steps shown in Section \ref{subsec:Stage-I:-Estimation}
$J_{1}$ times to solve the angle estimation problem connected with
(\ref{Y1_formula_Q}). That is because we have obtained the rotation
factors $(\Delta\omega_{l},\Delta\mu_{l})$ and gain scaling factor
$\gamma_{l}$ defined in (\ref{rot_fac_omega_mu_scale}) after the
estimation procedure for the first virtual single-antenna user, user
$(1,1)$. This allows us to solve the sparse recovery problem corresponding
to (\ref{eq:yl_sparse}) without performing additional operations
for the channel estimation of the other virtual single-antenna users
$(1,j)$ for $j\neq1$.\footnote{The virtual single-antenna users $(1,j)$ for $j\neq1$ can be treated
as other users and the corresponding parameters can be estimated by
the method shown later. However, the pilot overhead for virtual users
$(1,j)$ for any $j$ should be the same, depending on the the number
of time blocks $\Upsilon_{1}$. So we still solve problem corresponding
to (\ref{eq:yl_sparse}).} In particular, for user $(1,j)$, the quantities $(\omega_{r}-\varphi_{1,j})$,
$(\mu_{r}-\theta_{1,j})$ and $\alpha_{r}^{*}\beta_{1,j}^{*}$ can
be obtained via the solution to (\ref{eq:yl_sparse}). Then, $\{(\omega_{l}-\varphi_{1,j})\}_{l\neq r}$,
$\{(\mu_{r}-\theta_{1,j})\}_{l\neq r}$ and $\{\alpha_{l}^{*}\beta_{1,j}^{*}\}_{l\neq r}$
can be directly obtained with the known $(\Delta\omega_{l},\Delta\mu_{l})$
and $\gamma_{l}$ obtained in the estimation for user $(1,1)$. Based
on this, the estimates of user $1$'s cascaded gains and cascaded
AoDs at the RIS, i.e., $\alpha_{l}\beta_{1,j}$, $(\omega_{l}-\varphi_{1,j})$
and $(\mu_{l}-\theta_{1,j})$, for $\forall l\in\{1,...,L\}$ and
$\forall j\in\{1,...,J_{1}\}$, are obtained, which allows us to determine
$\mathbf{G}_{1}$ in (\ref{eq:G_k-Q}).

\subsubsection{Estimation for Other Users}

Following the idea of the virtual single-antenna user, we convert
the channel estimation for the other multi-antenna users into the
estimation of $\sum_{k=2}^{K}J_{k}$ single scatterer channels for
the other single-antenna users. The idea of constructing the common
part as in Section \ref{subsec:Stage-II:-Estimation} still applies,
using the common RIS-BS channel to reduce the pilot overhead.

Specifically, after eliminating the effects of the common AoAs at
the BS, and the unique AoDs at the users estimated in Phase I, and
following (\ref{eq:LT_Yk_common}), $[\mathbf{\check{Y}}_{k}^{(i)}]_{(:,j)}$
in (\ref{Yk_Q_block}) can be reformulated as\vspace{-0.15cm}
\begin{align}
[\mathbf{\check{Y}}_{k}^{(i)}]_{(:,j)} & =\boldsymbol{\Lambda}\mathbf{A}_{M}^{\mathrm{H}}\mathrm{Diag}\{\mathbf{\widetilde{h}}_{\{k,j\}}\}\mathbf{e}^{(i)}+[\mathbf{\check{N}}_{k}^{(i)}]_{(:,j)}\nonumber \\
 & =\widetilde{\boldsymbol{\Lambda}}_{\mathrm{s}}\widetilde{\mathbf{A}}_{\mathrm{s}}^{\mathrm{H}}\mathrm{Diag}\{\mathbf{\widetilde{h}}_{\mathrm{s},\{k,j\}}\}\mathbf{e}^{(i)}+[\mathbf{\check{N}}_{k}^{(i)}]_{(:,j)},\label{LT_Yk_Q}
\end{align}
where $\widetilde{\boldsymbol{\Lambda}}_{\mathrm{s}}=\alpha_{r}\beta_{1,1}\mathrm{Diag}\{\gamma_{1}^{*},\gamma_{2}^{*},\ldots,\gamma_{L}^{*}\}$
and $\widetilde{\mathbf{A}}_{\mathrm{s}}=\mathrm{Diag}\{\mathbf{a}_{M}(\omega_{r}-\varphi_{1,1},\mu_{r}-\theta_{1,1})\}\mathbf{A}_{\Delta M}$
can be constructed using the estimated parameters of the virtual single-antenna
user $(1,1)$.\footnote{When user $(1,1)$ is the typical user, it can be verified that $\omega_{\mathrm{s}}$
and $\mu_{\mathrm{s}}$ defined in (\ref{omega_mu_common}) are $(\omega_{r}-\varphi_{1,1})$
and $(\mu_{r}-\theta_{1,1})$, respectively, and $\frac{1}{J_{1}}\mathbf{1}_{J_{1}}^{\mathrm{T}}\boldsymbol{\beta}_{1}\alpha_{r}$
in (\ref{eq:gamma_common}) is $\alpha_{r}\beta_{1,1}$.} The matrix $\mathbf{A}_{\Delta M}$ can be determined by (\ref{eq:AoD_common_part}).
Accordingly, $\mathbf{\widetilde{h}}_{\mathrm{s},\{k,j\}}=\frac{1}{\beta_{1,1}}\mathrm{Diag}\{\mathbf{a}_{M}(-\varphi_{1,1},-\theta_{1,1})\}\mathbf{\widetilde{h}}_{\{k,j\}}$
is the unique part of the cascaded channel for virtual single-antenna
user $(k,j)$ that is to be estimated. Stacking $\Upsilon_{k}$ time
blocks of (\ref{LT_Yk_Q}) and vectorizing, we have \vspace{-0.15cm}
\begin{align}
\mathbf{\widetilde{w}}_{\{k,j\}} & \triangleq\mathrm{vec}(\left[[\mathbf{\check{Y}}_{k}^{(1)}]_{(:,j)},...,[\mathbf{\check{Y}}_{k}^{(\Upsilon_{k})}]_{(:,j)}\right])\nonumber \\
 & =(\mathbf{\widetilde{E}}_{k}^{\mathrm{T}}\diamond\widetilde{\boldsymbol{\Lambda}}_{\mathrm{s}}\widetilde{\mathbf{A}}_{\mathrm{s}}^{\mathrm{H}})\mathbf{\widetilde{h}}_{\mathrm{s},\{k,j\}}+\widetilde{\mathbf{n}}_{\{k,j\}}\in\mathbb{C}^{L\Upsilon_{k}\times1},\label{eq:Wk_-Q}
\end{align}
where $\widetilde{\mathbf{n}}_{\{k,j\}}$ is the corresponding equivalent
noise for virtual user $(k,j)$. The last equality is obtained via
$\mathrm{vec}(\widetilde{\boldsymbol{\Lambda}}_{\mathrm{s}}\widetilde{\mathbf{A}}_{\mathrm{s}}^{\mathrm{H}}\mathrm{Diag}\{\mathbf{\widetilde{h}}_{\mathrm{s},\{k,j\}}\}\mathbf{\widetilde{E}}_{k})=(\mathbf{\widetilde{E}}_{k}^{\mathrm{T}}\diamond\widetilde{\boldsymbol{\Lambda}}_{\mathrm{s}}\widetilde{\mathbf{A}}_{\mathrm{s}}^{\mathrm{H}})\mathbf{\widetilde{h}}_{\mathrm{s},\{k,j\}}$.
Since the form of (\ref{eq:Wk_-Q}) is similar to (\ref{eq:Wk_}),
$\mathbf{\widetilde{h}}_{\mathrm{s},\{k,j\}}$ can be estimated similarly
to what was done for (\ref{eq:wk_sparse}).\vspace{-0.05cm}

With the estimates of the multi-antenna user $k$'s cascaded gains
and cascaded AoDs, i.e., $\alpha_{l}\beta_{k,j}$, $(\omega_{l}-\varphi_{k,j})$
and $(\mu_{l}-\theta_{k,j})$, for $\forall l\in\{1,...,L\}$, $\forall j\in\{1,...,J_{k}\}$,
obtained by solving the problem connected with (\ref{eq:Wk_-Q}) $J_{k}$
times, $\mathbf{G}_{k}$ in (\ref{eq:G_k-Q}) can be determined for
$\forall k\in\{2,3...,K\}$. \vspace{-0.5cm}

\subsection{Pilot Overhead Analysis}

\vspace{-0.15cm}
 In this subsection, we analyze the pilot overhead of the full CSI
estimation algorithm for the multi-antenna user case, assuming $J_{1}=J_{2}=\cdots=J_{K}=J$
and $Q_{1}=Q_{2}=\cdots=Q_{k}=Q$.

Similar to the analysis in Section \ref{subsec:Pilot-Overhead}, for
user $1$, the number of time slots in Phase I should satisfy\textcolor{red}{{}
}$V_{1}^{(1)}\geqslant\mathcal{O}(J_{1}\log(D_{1}D_{2}))\geqslant\mathcal{O}(J_{1}\log(Q_{11}Q_{12}))=\mathcal{O}(J_{1}\log(Q_{1}))$
so as to ensure the $J_{1}$-sparse recovery problem associated with
(\ref{AoD_VAD}). In Phase II, the number of time slots within each
block $V_{1}^{(i)}$, should satisfy $V_{1}^{(i)}\geqslant J_{1}$,
otherwise the right inverse $(\mathbf{A}_{Q,1}^{\mathrm{H}}\mathbf{S}_{1}^{(i)})^{\dagger}$
does not exist. On the other hand, the number of blocks, $\Upsilon_{1}$,
is determined by sparse recovery applied to (\ref{Y1_formula_Q}).
The angle estimation associated with (\ref{Y1_formula_Q}) can be
implemented by a $1$-sparse recovery problem, and thus\textcolor{red}{{}
}we have $\Upsilon_{1}\geq\mathcal{O}(\log(M))$.\textcolor{red}{{}
}As shown before, $J_{1}$ virtual single-antenna users share the
same blocks and can be processed simultaneously. In addition, the
first block is also used for Phase II. Hence the total pilot overhead
required for user $1$ should\textcolor{red}{{} }satisfy $\tau_{1}=\sum_{i=1}^{\Upsilon_{1}}V_{1}^{(i)}=V_{1}^{(1)}+\sum_{i=2}^{\Upsilon_{1}}V_{1}^{(i)}\geq\mathcal{O}(J_{1}\log(Q_{1}))+(\mathcal{O}(\log(M))-1)J_{1}$.

For the other users $2\leq k\leq K$, we have the inequalities $V_{k}^{(1)}\geqslant\mathcal{O}(J_{k}\log(Q_{k}))$\textcolor{red}{{}
}and $V_{k}^{(i)}\geqslant J_{k}$, for the same reasons as for user
$1$. As before, the angle estimation problem connected with (\ref{eq:Wk_-Q})
can be treated as a $1$-sparse recovery problem, and $J_{k}$ virtual
single-antenna users simultaneously share the same blocks, where the
number of time blocks for user $k$ satisfies\textcolor{red}{{} }$\Upsilon_{k}\geq\mathcal{O}(\log(M)/L)$.
Therefore, the total number of pilot symbols required for user $k$
should satisfy $\tau_{k}=V_{k}^{(1)}+\sum_{i=2}^{\Upsilon_{k}}V_{k}^{(i)}\geq\mathcal{O}(J_{k}\log(Q_{k}))+(\mathcal{O}(\log(M)/L)-1)J_{k}$.

Finally, the overall pilot overhead for the multi-antenna users is
given by $\mathcal{O}(JK\log(Q)+J\log(M)+(K-1)J\log(M)/L)-JK$. Table
I summarizes the\textcolor{red}{{} }total number of pilots of the proposed
method and other existing algorithms for full-CSI estimation. It is
observed that the proposed method achieves a significant reduction
in the pilot overhead for both the single-antenna and multi-antenna
user cases, owing to the exploitation of the correlation among different
users.

\begin{table*}
\centering{}\caption{Total Number of Pilots of Various Methods.}
\begin{tabular}{lll}
\toprule 
Case & Methods & Pilot Overhead\tabularnewline
\midrule 
Single-antenna User & Proposed Full-CSI Estimation & $\mathcal{O}(J\log(M)+(K-1)J\log(M)/L)$\tabularnewline
\midrule 
Single-antenna User & Direct-OMP \cite{ris-omp-1} & $\mathcal{O}(JLK\log(MN)/N)$\tabularnewline
\midrule 
Single-antenna User & DS-OMP \cite{ris-omp-3} & $\mathcal{O}(JK\log(M))$\tabularnewline
\midrule 
Single-antenna User & Row-Structure OMP\cite{ris-omp-2} & $\mathcal{O}(JK\log(M))$\tabularnewline
\midrule 
Multi-antenna User & Extension of the proposed method & $\mathcal{O}(JK\log(Q)+J\log(M)+(K-1)J\log(M)/L)-JK$\tabularnewline
\midrule 
Multi-antenna User & CS-EST OMP \cite{CE_MIMO_RIS} & $\mathcal{O}(JK\log(Q)+JLK\log(MJL)/N)$\tabularnewline
\bottomrule
\end{tabular}
\end{table*}

\vspace{-0.2cm}

\section{Simulation Results\label{sec:Simulation-Results}}

\vspace{-0.05cm}
 In this section, simulation results are provided to evaluate the
performance of the proposed three-stage channel estimation protocol
for both the single-antenna user case and multi-antenna user case.
We assume that channel gains $\alpha_{l}$ and $\beta_{k,j}$ follow
a complex Gaussian distribution with zero mean and variance of $10^{-3}d_{\mathrm{BR}}^{-2.2}$
and $10^{-3}d_{\mathrm{RU}}^{-2.8}$, respectively. Here, $d_{\mathrm{BR}}$
is defined as the distance between the BS and the RIS, while, $d_{\mathrm{RU}}$
is defined as the distance between the RIS and the users. The antenna
spacing at the BS and the element spacing at the RIS are assumed to
satisfy $d_{\mathrm{BS}}=d_{\mathrm{RIS}}=\frac{\lambda_{c}}{2}$.
The random Bernoulli matrix is chosen as the initial RIS phase shift
training matrix $\mathbf{E}$, i.e., the elements are selected from
$\{-1,+1\}$ with equal probability \cite{ris-omp-3}. The transmitted
power is set to $p=1$ W. It is assumed that the propagation angles
change every ten channel coherence blocks, while the gains change
for each coherence block. Unless otherwise specified, for the single-antenna
user case, the dimensions of the UPAs deployed on the BS and the RIS
are $N_{1}=N_{2}=10$ and $M_{1}=M_{2}=10$, respectively. $d_{\mathrm{BR}}$
and $d_{\mathrm{RU}}$ are set to $10$ m and $100$ m \cite{ris-omp-3},
respectively. The number of users is set to $K=4$. The number of
scatterers between the BS and the RIS, and that between the RIS and
users are set to $L=5$ and $J_{1}=\cdots=J_{K}=4$. For the multi-antenna
user case, the corresponding parameter settings are $N_{1}=N_{2}=8$,
$M_{1}=M_{2}=8$, $d_{\mathrm{BR}}=80$m, $d_{\mathrm{RU}}=40$m,
$L=3$ and $J_{1}=\cdots J_{k}=2$. In addition, we set the number
of users to $K=6$ and all the users adopt $36$-antenna UPAs with
$6$ rows and $6$ columns, i.e., $Q_{k1}=Q_{k2}=6$ for $\forall k\in\mathcal{K}$.
The antenna spacing at the user equipments still satisfies $d_{\mathrm{UE}}=\frac{\lambda_{c}}{2}$.
The normalized mean square error (NMSE) is chosen as the main metric
for evaluating estimation performance, which is defined by $\mathrm{NMSE}=\mathbb{E}\{(\sum_{k=1}^{K}||\widehat{\mathbf{G}}_{k}-\mathbf{G}_{k}||_{F}^{2})\mathbf{/}(\sum_{k=1}^{K}||\mathbf{G}_{k}||_{F}^{2})\}.$

We compare the proposed three-stage channel estimation protocol with
the following channel estimation methods, in which Direct-OMP \cite{ris-omp-1}
and DS-OMP \cite{ris-omp-3} were developed for the single-antenna
user case while CS-EST OMP \cite{CE_MIMO_RIS} was developed for the
multi-antenna user case.\vspace{-0.1cm}

\begin{itemize}
\item Direct-OMP \cite{ris-omp-1}: By directly formulating the VAD representation
of the cascaded channel as a sparse recovery problem using the vectorization
operation, the authors in \cite{ris-omp-1} used OMP to reconstruct
the channels. We extend this method to UPA-Type BS in our simulation. 
\item DS-OMP \cite{ris-omp-3}: By exploiting the common row-block sparsity
and common column-block sparsity of the cascaded channel to formulate
a sparse recovery problem, the authors in \cite{ris-omp-3} adopted
OMP to reconstruct the channels. 
\item CS-EST OMP \cite{CE_MIMO_RIS}: The authors of \cite{CE_MIMO_RIS}
proposed an OMP-based three-stage channel estimation in ULA-type MIMO
case, which estimated AoDs at the users in Stage I, AoAs at the BS
in Stage II, and cascaded channel gains in Stage III. We extend the
method in \cite{CE_MIMO_RIS} to UPA -type MIMO case and regard it
as the benchmark. 
\item Proposed full-CSI: During the first coherence block, full CSI for
all users is estimated using Algorithm \ref{algorithm-2} in Stage
I and Algorithm \ref{algorithm-3} in Stage II assuming a UPA-type
RIS and a UPA-type BS. OMP is adopted to solve the sparse recovery
problems in these two stages. 
\item Oracle full-CSI: This method is treated as the performance upper bound
of the Proposed full-CSI method assuming that perfect angle information
is known by the BS, providing perfect knowledge of the support of
the sparsity recovery problems. In this case, the channels are estimated
using the LS estimator in Stage I and Stage II. 
\item Proposed gains-only: During the remaining coherence blocks, only the
gains are updated using the LS method shown in Section \ref{sec:Remaining_Coherence}
for Stage III. Here, the angle information is known and estimated
using the proposed full-CSI method with an average pilot overhead
of $T=15$ (The number of pilots for typical user, i.e., $\tau_{1}$,
is set to $36$ in Stage I, while that for other users, i.e., $\tau_{k}$,
for $2\leq k\leq K$, are set to $8$ in Stage II). 
\item Oracle gains-only: This method is regarded as the performance upper
bound of the Proposed gains-only method during the remaining coherence
blocks, and assumes that the BS perfectly knows the angle information
when using the LS estimator. 
\end{itemize}
\vspace{-0.3cm}

\subsection{Single-antenna User Case}

\begin{figure}
\begin{centering}
\includegraphics[width=0.75\columnwidth]{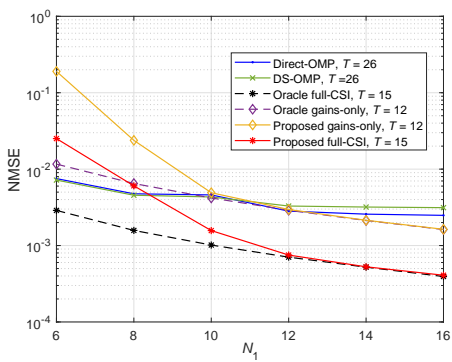}
\par\end{centering}
\centering{}\caption{NMSEs vs. Number of antennas at BS side: $N=N_{1}\times N_{2}$, $N_{1}=N_{2}$.}
\label{bs_antenna}
\end{figure}

\begin{figure}
\begin{centering}
\includegraphics[width=0.75\columnwidth]{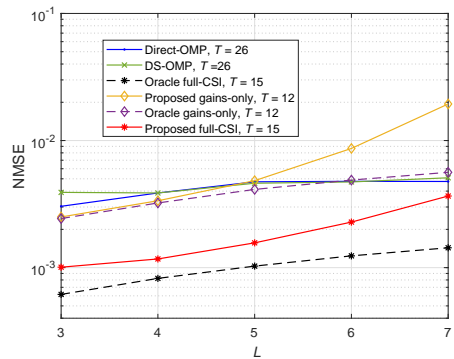}
\par\end{centering}
\centering{}\caption{NMSEs vs. Number of scatterers in RIS-BS link.}
\label{pathL}
\end{figure}

\begin{figure}
\begin{centering}
\includegraphics[width=0.75\columnwidth]{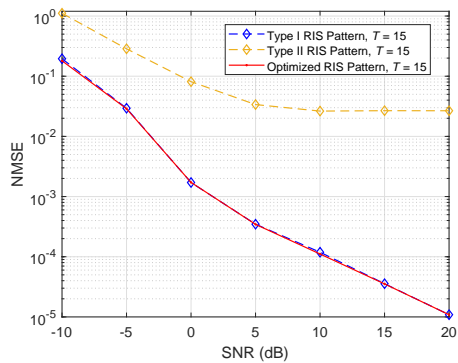}
\par\end{centering}
\centering{}\caption{Performance of Optimized vs. Non-Optimized RIS phase shift training
matrices.}
\label{opt_train}
\end{figure}

\vspace{-0.05cm}
 In this subsection, the following four figures compare the performance
of different estimation methods for the single-antenna user case.
In particular, due to the different number of pilots allocated to
the typical user and other users in the first coherence block for
the Proposed full-CSI method, we consider the users' average pilot
overhead as a measure of pilots, denoted as $T$. To reduce the error
propagation,\footnote{As shown in Section \ref{sec:First_coherence}, the estimation error
of the typical user in Stage I leads to unavoidable error propagation
for the estimation of other users in Stage II.} we allocate more pilots to the typical user and fewer pilots to the
other users. Specifically, in Fig. \ref{bs_antenna}, Fig. \ref{pathL},
and Fig. \ref{opt_train}, $36$ pilots and $8$ pilots are allocated
to the typical user and the other users, respectively, thus the average
number of pilots for the proposed method is given by $T=15$.

\begin{figure}
\begin{centering}
\includegraphics[width=0.75\columnwidth]{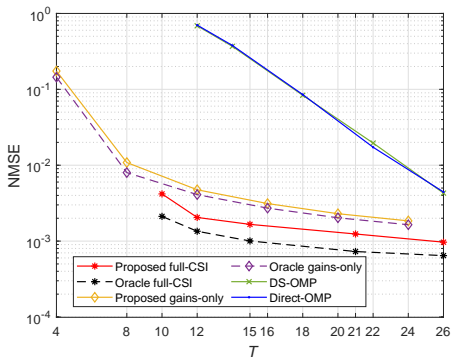}
\par\end{centering}
\centering{}\caption{NMSEs vs. Average pilot overhead $T$ of each user with SNR = $0$
dB.}
\label{pilot}
\end{figure}

Fig. \ref{pilot} illustrates the relationship between NMSE performance
and pilot overhead of the various methods, where the signal-to-noise
ratio (SNR) is set to $0$ dB. We increase the pilot overhead for
the typical user mainly for less error propagation. It can be clearly
seen that an increase in the number of pilots improves the performance
of all algorithms. In order to achieve the same estimation performance,
e.g., $\mathrm{NMSE}=10^{-2}$, the required average pilot overhead
of the Proposed full-CSI method is much lower than the methods in
\cite{ris-omp-1,ris-omp-3} during the first coherence block. On the
other hand, during the remaining coherence blocks, we note that the
Proposed gains-only method only needs $T=12$ pilots to achieve the
same performance as the Direct-OMP and DS-OMP methods with $T=26$.
Additionally, it is observed that the Proposed gains-only method performs
generally the same as its upper bound, i.e., Oracle gains-only method,
which implies that the Proposed full-CSI method with the average pilot
overhead $T=15$ during the first coherence block can provide accurate
angle estimation information for the Proposed gains-only method to
estimate the updated channel gains during the remaining coherence
blocks. \vspace{-0cm}

Fig. \ref{bs_antenna} depicts the NMSE performance as a function
of the number of antennas at the BS, where we set the SNR to $0$
dB and assume $N_{1}=N_{2}$. It can be observed that as the number
of antennas at the BS increases, the estimation accuracy of the Proposed
full-CSI method with fewer average pilots, $T=15$ ($36$ pilots allocated
to the typical user and $8$ pilots allocated to the other users),
is improved significantly, and achieves nearly the same performance
as the Oracle full-CSI method when $N$ is larger than $144$ $(12\times12)$.
This is because the Proposed full-CSI method must first estimate the
number of scatterers in the RIS-BS link from the received signal.
The estimation accuracy of this step is determined by the asymptotic
property shown in Lemma \ref{lem:3} and the resolution of the rotation
matrices defined in (\ref{rot_mat}). The asymptotic property in Lemma
\ref{lem:3} requires that both $N_{1}$ and $N_{2}$ be sufficiently
large. In addition, we observe the gap between the Proposed gains-only
method and the Oracle gains-only method is large when $N=36$ $(6\times6)$.
This behavior illustrates that with small scale antenna array, the
Proposed full-CSI method provides inaccurate angle estimation information
for the estimation of gains during the remaining coherence blocks,
which deteriorates the estimation accuracy of the Proposed gains-only
method further. Fortunately, with the increase of the number of antennas,
the gap becomes marginal, which means that the angle information has
been estimated perfectly in the first coherence block with large scale
antenna array.

\vspace{-0cm}

Fig. \ref{pathL} illustrates the NMSE performance of algorithms with
different pilot overhead versus the number of scatterers in the RIS-BS
link, where the SNR is set to $0$ dB. As shown in Fig. \ref{pathL},
the estimation accuracy decreases as the number of scatterers increases.
The reasons for this behavior can be summarized as follows. First,
the number of unknown parameters (angles and gains) to be estimated
increases, and thus the OMP-based estimation performs worse for the
same pilot overhead. Second, since the number of scatterers is unknown
in our Proposed full-CSI UPA-type based method, the estimation accuracy
of the Proposed full-CSI method is relatively more sensitive to an
increase of the number of scatterers than the other methods, which
further deteriorates the performance of the Proposed gains-only method
in the remaining coherence blocks. By contrast, the NMSEs of the DS-OMP
method and the Direct-OMP method with $T=26$ pilots increases only
moderately with the increase of the number of scatterers since the
parameters including the numbers of scatterers between the RIS-BS
link and the user-RIS link are known by BS for these two methods.

Fig. \ref{opt_train} illustrates whether the optimization of the
RIS phase shift training matrix $\mathbf{E}$ provides a significant
benefit for the estimation performance. ``Type I RIS Pattern'' refers
to choosing the random Bernoulli matrix as the training matrix, i.e.,
generating the initial training matrix with elements from $\{-1,+1\}$
with equal probability \cite{ris-omp-3}. ``Type II RIS Pattern''
refers to generating the initial training matrix with elements as
$[\mathbf{e}_{t}]_{m}=\exp\left(\mathrm{i}\angle(a+\mathrm{i}b)\right)$
where $a$ and $b$ follow independent and identically uniform distribution
$\mathcal{U}(0,1)$. It is observed that the performance of the Type
I training matrix is essentially the same as that of the optimized
training matrix, and far outperforms that of the Type II training
matrix. This behavior can be explained by exploring the mutual coherence
property of the equivalent sensing matrices for problems associated
with (\ref{eq:sparse-formula}) and (\ref{eq:wk_sparse}). For a given
matrix $\mathbf{D}$, the maximal coherence of $\mathbf{D}$, denoted
as $\mu(\mathbf{D})$, is defined as\vspace{-0.2cm}
\begin{equation}
\mu(\mathbf{D})=\max_{i\neq j}\frac{|\mathbf{D}_{(:,i)}^{\mathrm{H}}\mathbf{D}_{(:,j)}|}{||\mathbf{D}_{(:,i)}||||\mathbf{D}_{(:,j)}||},
\end{equation}
which is the largest absolute inner product between any two columns
of $\mathbf{D}$. According to the compressive sensing theory \cite{CS-Overview_Yonina},
the sensing matrix with smaller $\mu(\mathbf{D})$ could provide better
recovery performance for sparse vectors. The random Bernoulli matrix,
which is a typical sensing matrix with lower correlation of its columns
and satisfies the constant modulus constraint, is chosen as the Type
I training matrix. Furthermore, numerical results validate that the
maximal coherence of the sensing matrices generated by the Type I
training matrix is significantly lower than that generated by the
Type II training matrices, and nearly the same as that generated by
the optimized training matrix. Since optimization of the training
matrix requires extra computational complexity, this result suggests
that \textquotedbl Type I RIS Pattern\textquotedbl{} be chosen for
the RIS phase shift training matrix.\vspace{-0.3cm}

\subsection{Multi-antenna User Case}

\begin{figure}
\begin{centering}
\includegraphics[width=0.75\columnwidth]{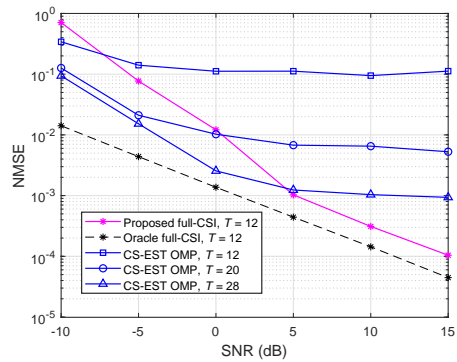}
\par\end{centering}
\centering{}\caption{NMSEs vs. SNR.}
\label{multi-user SNR}
\end{figure}

\begin{figure}
\begin{centering}
\includegraphics[width=0.75\columnwidth]{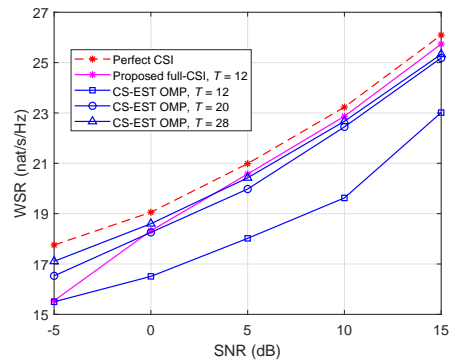}
\par\end{centering}
\centering{}\caption{WSR vs. SNR.}
\label{rate}
\end{figure}

\vspace{-0.05cm}
 In this subsection, the NMSE and weighted sum rate (WSR) of the multi-antenna
user case are respectively shown in Fig. \ref{multi-user SNR} and
Fig. \ref{rate} by using different estimation methods. The users'
average pilot overhead is considered for the proposed method in the
multi-antenna user case, similar to that in the single-antenna user
case. Specifically, for estimating the AoDs at the users, we allocate
$10$ slots to all the users including the typical user and other
users in Phase I, i.e., $V_{k}^{(1)}=10$ for $\forall k\in\mathcal{K}$.
In phase II, additional $3$ blocks of time slots are allocated to
the typical user and each block has $4$ slots, i.e., $V_{1}^{(2)}=V_{1}^{(3)}=V_{1}^{(4)}=4$.
Therefore, the pilot overhead allocated to the typical user and other
users are $22$ and $10$, respectively. The average pilot overhead
for the proposed method is given by $T=12$. In addition, for fairness,
CS-EST OMP consumes the same number of slots for the estimation of
AoDs at the users.

Fig. \ref{multi-user SNR} displays the NMSE performance of different
methods versus SNR. It is observed that the gap between the Proposed
full-CSI method and its upper bound, i.e., the Oracle full-CSI method,
becomes smaller with the increase of SNR. In particular, when the
SNR is larger than $5$ dB, the NMSE of the proposed method with $T=12$
exceeds that of the CE-EST OMP method with $T=28$, and has the same
trend as that of the Oracle full-CSI method, i.e., the NMSEs decrease
linearly with the SNR. This behavior implies the angle information
can be obtained accurately by the Proposed full-CSI method at large
SNR region. In this case, the NMSE differences between the proposed
method and its upper bound mainly results from the estimation errors
of channel gain information. By contrast, the NMSE of the CS-EST OMP
method still has a performance bottleneck in the high SNR region even
under the scenario of up to $28$ pilots per user.

\vspace{-0cm}

Fig. \ref{rate} shows the WSR performance of the MU MIMO system based
on the channels estimated using different algorithms. The weighting
factors, the maximum BS power, and the number of data streams are
set to $\varpi_{k}=1$ for $\forall k\in\mathcal{K}$, $P_{max}=1$
W, and $d=16$, respectively. The details of the calculation for WSR
can refer to {[}33, Appendix D{]}. In Fig. \ref{rate}, the case with
perfect CSI is adopted as the upper bound of the Proposed full-CSI
and CE-EST OMP methods. As can be observed, the WSR achieved by the
proposed method with $T=12$ pilots is always larger than that achieved
by the CS-EST OMP method with the same number of pilots of $T=12$.
When SNR = $5$ dB, the proposed method outperforms the other three
CS-EST OMP methods. To achieve the same WSR, the pilot overhead required
by the proposed method is less than half that of the CE-EST OMP method.
With the further increase of the SNR, the gap between the proposed
method and the upper bound becomes smaller gradually, which implies
that extension of the proposed full-CSI method to the multi-antenna
user case can achieve high estimation accuracy. \vspace{-0.3cm}

\section{Conclusions\label{sec:Conclusions}}

\vspace{-0.1cm}
 In this paper, we adopted a novel three-stage uplink channel estimation
protocol that leads to a significant reduction in the number of pilots
for a UPA-type RIS-aided mmWave system with a UPA-type BS. The proposed
estimation methods were developed starting from the single-antenna
user case, and were shown to fully exploit the correlation among the
channels of different users. To reduce the power leakage problem during
the common AoA estimation in Stage I, a low-complexity 1-D search
method was developed. Then we extended the protocol to the UPA-type
multi-antenna user case. An OMP-based method was proposed for estimation
of the AoDs at the users. Numerical results showed that choosing the
random Bernoulli matrix as the RIS training matrix has near-optimal
performance. Simulation results validated that the proposed methods
outperform other existing algorithms in terms of pilot overhead. In
addition, the proposed algorithms approach the genie-aided upper bound
in the high SNR regime.

Future studies can include the application of learning-based approaches
to our proposed channel estimation protocol. With the increase of
the number of RIS elements and BS/users antennas, the computational
complexity for the conventional model-driven estimation methods becomes
inevitably high. This fact motivates the development of data-driven
or hybrid approaches for the proposed protocol in the future, which
can obtain the estimates with reduced complexity, and the correlation
relationship among multi-user cascaded matrices is still utilized
for pilot overhead reduction.

\textcolor{red}{\vspace{-0.6cm}
}

\appendix{}

\subsection{Proof of Lemma 1}

Using $N=N_{1}\times N_{2}$, we have 
\begin{align}
 & \frac{1}{N}\mathbf{a}_{N}^{\mathrm{H}}(\psi_{j},\nu_{j})\mathbf{a}_{N}(\psi_{i},\nu_{i})\nonumber \\
= & \frac{1}{N}(\mathbf{a}_{N_{1}}(\psi_{j})\otimes\mathbf{a}_{N_{2}}(\nu_{j}))^{\mathrm{H}}(\mathbf{a}_{N_{1}}(\psi_{i})\otimes\mathbf{a}_{N_{2}}(\nu_{i}))\nonumber \\
= & \frac{1}{N}(\mathbf{a}_{N_{1}}^{\mathrm{H}}(\psi_{j})\otimes\mathbf{a}_{N_{2}}^{\mathrm{H}}(\nu_{j}))(\mathbf{a}_{N_{1}}(\psi_{i})\otimes\mathbf{a}_{N_{2}}(\nu_{i}))\nonumber \\
= & (\frac{1}{N_{1}}\mathbf{a}_{N_{1}}^{\mathrm{H}}(\psi_{j})\mathbf{a}_{N_{1}}(\psi_{i}))\otimes(\frac{1}{N_{2}}\mathbf{a}_{N_{2}}^{\mathrm{H}}(\nu_{j})\mathbf{a}_{N_{2}}(\nu_{i})).
\end{align}
In mmWave systems, $N_{1}$ and $N_{2}$ could be large, thus the
asymptotic analysis of (\ref{eq:lemma1}) is divided into two limits:
$\lim_{N_{1}\rightarrow\infty}\frac{1}{N_{1}}\mathbf{a}_{N_{1}}^{\mathrm{H}}(\psi_{j})\mathbf{a}_{N_{1}}(\psi_{i})$
and $\lim_{N_{2}\rightarrow\infty}\frac{1}{N_{2}}\mathbf{a}_{N_{2}}^{\mathrm{H}}(\nu_{j})\mathbf{a}_{N_{2}}(\nu_{i})$.
According to the conclusion in \cite{Zhou_ULA_TSP}, we have 
\begin{align}
\lim_{N_{1}\rightarrow\infty}\frac{1}{N_{1}}\mathbf{a}_{N_{1}}^{\mathrm{H}}(\psi_{j})\mathbf{a}_{N_{1}}(\psi_{i}) & =\begin{cases}
1 & \psi_{j}=\psi_{i}\\
0 & \textrm{otherwise}
\end{cases},\label{asympolic-1}\\
\lim_{N_{2}\rightarrow\infty}\frac{1}{N_{2}}\mathbf{a}_{N_{2}}^{\mathrm{H}}(\nu_{j})\mathbf{a}_{N_{2}}(\nu_{i}) & =\begin{cases}
1 & \nu_{j}=\nu_{i}\\
0 & \textrm{otherwise}
\end{cases}.\label{asympolic-2}
\end{align}
Due to the properties of the Kronecker product, $\frac{1}{N}\mathbf{a}_{N}^{\mathrm{H}}(\psi_{j},\nu_{j})\mathbf{a}_{N}(\psi_{i},\nu_{i})=1$
holds when $\psi_{j}=\psi_{i}$ and $\nu_{j}=\nu_{i}$. Furthermore,
by combining the formula of $\mathbf{A}_{N}$ in (\ref{eq:H1-1})
with (\ref{asympolic-1}) and (\ref{asympolic-2}), we have $\mathbf{A}_{N}^{\mathrm{H}}\mathbf{A}_{N}=N\mathbf{I}_{L}$.

\subsection{Proof of Lemma 2}

Based on the definition of $\widetilde{\mathbf{U}}_{N}$, the $l$-th
column of $\widetilde{\mathbf{U}}_{N}^{\mathrm{H}}\mathbf{A}_{N}$
can be calculated as 
\[
\begin{split}[\widetilde{\mathbf{U}}_{N}^{\mathrm{H}}\mathbf{A}_{N}]_{:,l} & =\widetilde{\mathbf{U}}_{N}^{\mathrm{H}}\mathbf{a}_{N}(\psi_{l},\nu_{l})\\
 & =(\mathbf{U_{\mathit{N_{\mathrm{1}}}}^{\mathrm{H}}\otimes}\mathrm{\mathbf{U}}_{\mathit{N_{\mathrm{2}}}}^{\mathrm{H}})(\mathbf{a}_{N_{1}}(\psi_{l})\otimes\mathbf{a}_{N_{2}}(\nu_{l}))\\
 & =(\mathbf{U_{\mathit{N_{\mathrm{1}}}}^{\mathrm{H}}}\mathbf{a}_{N_{1}}(\psi_{l}))\otimes(\mathrm{\mathbf{U}}_{\mathit{N_{\mathrm{2}}}}^{\mathrm{H}}\mathbf{a}_{N_{2}}(\nu_{l})).
\end{split}
\]
For large $N_{1}$ and $N_{2}$, according to \cite{Zhou_ULA_TSP},
we have 
\begin{equation}
\begin{split}\lim_{N_{\mathrm{1}}\rightarrow\infty}[\mathbf{U_{\mathit{N_{\mathrm{1}}}}^{\mathrm{H}}}\mathbf{a}_{N_{1}}(\psi_{l})]_{n_{1}(l)}\neq0,~\lim_{N_{\mathrm{2}}\rightarrow\infty}[\mathrm{\mathbf{U}}_{\mathit{N_{\mathrm{2}}}}^{\mathrm{H}}\mathbf{a}_{N_{2}}(\nu_{l})]_{n_{2}(l)}\neq0,\end{split}
\end{equation}
if $n_{1}(l)$ and $n_{2}(l)$ satisfy (\ref{eq:lemma3-1}). In other
words, only the $n_{1}(l)$-th element of $\mathbf{U_{\mathit{N_{\mathrm{1}}}}^{\mathrm{H}}}\mathbf{a}_{N_{1}}(\psi_{l})$
and the $n_{2}(l)$-th element of $\mathrm{\mathbf{U}}_{\mathit{N_{\mathrm{2}}}}^{\mathrm{H}}\mathbf{a}_{N_{2}}(\nu_{l})$
are nonzero, while the other elements of $\mathbf{U_{\mathit{N_{\mathrm{1}}}}^{\mathrm{H}}}\mathbf{a}_{N_{1}}(\psi_{l})$
and $\mathrm{\mathbf{U}}_{\mathit{N_{\mathrm{2}}}}^{\mathrm{H}}\mathbf{a}_{N_{2}}(\nu_{l})$
are nearly zero. Hence, based on the properties of the Kronecker product,
the conclusion that the $n_{l}$-th element of $\widetilde{\mathbf{U}}_{N}^{\mathrm{H}}\mathbf{a}_{N}(\psi_{l},\nu_{l})$
is nonzero holds when $n_{l}$ satisfies (\ref{eq:lemma3_decom}).

\subsection{Proof of Proposition 1}

From the definitions of $\widetilde{\mathbf{R}}_{1}$ in (\ref{rot_mat})
and $\widetilde{\mathbf{U}}_{1}$, the $l$-th column of $\widetilde{\mathbf{U}}_{1}^{\mathrm{H}}\widetilde{\mathbf{R}}_{1}(\Delta\psi_{l})\mathbf{A}_{N}$
can be calculated as 
\begin{align}
 & [\widetilde{\mathbf{U}}_{1}^{\mathrm{H}}\widetilde{\mathbf{R}}_{1}(\Delta\psi_{l})\mathbf{A}_{N}]_{:,l}\nonumber \\
= & \widetilde{\mathbf{U}}_{1}^{\mathrm{H}}\widetilde{\mathbf{R}}_{1}(\Delta\psi_{l})\mathbf{a}_{N}(\psi_{l},\nu_{l})\nonumber \\
= & (\mathbf{U_{\mathit{N_{\mathrm{1}}}}^{\mathrm{H}}\otimes}\mathbf{D}_{\mathit{N_{\mathrm{2}}}}^{\mathrm{H}})(\mathbf{R_{\mathrm{1}}}(\Delta\psi_{\mathit{l}})\otimes\mathbf{\mathbf{D}_{\mathit{N_{\mathrm{2}}}}})(\mathbf{a}_{N_{1}}(\psi_{l})\otimes\mathbf{a}_{N_{2}}(\nu_{l}))\nonumber \\
= & (\mathbf{U_{\mathit{N_{\mathrm{1}}}}^{\mathrm{H}}}\mathbf{R_{\mathrm{1}}}(\Delta\psi_{\mathit{l}})\mathbf{a}_{N_{1}}(\psi_{l}))\otimes(\mathbf{D}_{\mathit{N_{\mathrm{2}}}}^{\mathrm{H}}\mathbf{\mathbf{\mathbf{D}_{\mathit{N_{\mathrm{2}}}}}}\mathbf{a}_{N_{2}}(\nu_{l}))\nonumber \\
= & (\mathbf{U_{\mathit{N_{\mathrm{1}}}}^{\mathrm{H}}}\mathbf{R_{\mathrm{1}}}(\Delta\psi_{\mathit{l}})\mathbf{a}_{N_{1}}(\psi_{l}))\otimes(\widetilde{\mathbf{e}}_{N_{2}}),\label{eq:Rot_psi}
\end{align}
where $\widetilde{\mathbf{e}}_{N_{2}}\in\mathbb{R}^{N_{2}}$ denotes
a vector whose first element equals 1 and the other elements are 0.
As discussed in Lemma \ref{lem:3}, the index of the nonzero element
of $\mathbf{U_{\mathit{N_{\mathrm{1}}}}^{\mathrm{H}}}\mathbf{R_{\mathrm{1}}}(\Delta\psi_{\mathit{l}})\mathbf{a}_{N_{1}}(\psi_{l})$
is $n_{1}(l)$ while the index of non-zero element in $\widetilde{\mathbf{e}}_{N_{2}}$
is 1. Therefore, the index of the nonzero element of $\widetilde{\mathbf{U}}_{1}^{\mathrm{H}}\widetilde{\mathbf{R}}_{1}(\Delta\psi_{l})\mathbf{a}_{N}(\psi_{l},\nu_{l})$,
denoted as $\overline{n_{1l}}$, satisfies 
\begin{equation}
\overline{n_{1l}}=(n_{1}(l)-1)N_{\mathrm{2}}+1.\label{eq:n1l_bar}
\end{equation}
By combining (\ref{eq:n1l_bar}) with (\ref{eq:Rot_psi}), the $(\overline{n_{1l}},l)$-th
element of $\widetilde{\mathbf{U}}_{1}^{\mathrm{H}}\widetilde{\mathbf{R}}_{1}(\Delta\psi_{\mathit{l}})\mathbf{A}_{N}$
is given by 
\begin{align}
 & [\widetilde{\mathbf{U}}_{1}^{\mathrm{H}}\widetilde{\mathbf{R}}_{1}(\Delta\psi_{\mathit{l}})\mathbf{A}_{N}]_{\overline{n_{1l}},l}\nonumber \\
= & [(\mathbf{U_{\mathit{N_{\mathrm{1}}}}^{\mathrm{H}}}\mathbf{R_{\mathrm{1}}}(\Delta\psi_{\mathit{l}})\mathbf{a}_{N_{1}}(\psi_{l}))\otimes(\widetilde{\mathbf{e}}_{N_{2}})]_{\overline{n_{1l}}}\nonumber \\
= & [\mathbf{U_{\mathit{N_{\mathrm{1}}}}^{\mathrm{H}}}\mathbf{R_{\mathrm{1}}}(\Delta\psi_{\mathit{l}})\mathbf{a}_{N_{1}}(\psi_{l})]_{n_{1}(l)}\otimes[\widetilde{\mathbf{e}}_{N_{2}}]_{\mathit{\mathrm{1}}}\nonumber \\
= & \sqrt{\frac{1}{N_{1}}}\sum_{m=1}^{N_{1}}e^{-\mathrm{i}2\pi(m-1)(\psi_{l}+\frac{\Delta\psi_{l}}{2\pi}-\frac{n_{1}(l)-1}{N_{1}})}.\label{eq:psi-rot-app}
\end{align}

Similarly, the $l$-th column of $\widetilde{\mathbf{U}}_{2}^{\mathrm{H}}\widetilde{\mathbf{R}}_{2}(\Delta\nu_{\mathit{l}})\mathbf{A}_{N}$
can be calculated as 
\begin{align}
 & [\widetilde{\mathbf{U}}_{2}^{\mathrm{H}}\widetilde{\mathbf{R}}_{2}(\Delta\nu_{\mathit{l}})\mathbf{A}_{N}]_{:,l}\nonumber \\
= & \widetilde{\mathbf{U}}_{2}^{\mathrm{H}}\widetilde{\mathbf{R}}_{2}(\Delta\nu_{\mathit{l}})\mathbf{a}_{N}(\psi_{l},\nu_{l})\nonumber \\
= & (\mathbf{\mathbf{D}_{\mathit{N_{\mathrm{1}}}}^{\mathrm{H}}\otimes}\mathbf{U}_{\mathit{N_{\mathrm{2}}}}^{\mathrm{H}})(\mathbf{\mathbf{D}_{\mathit{N_{\mathrm{1}}}}\otimes}\mathbf{R_{\mathrm{2}}}(\Delta\nu_{\mathit{l}}))(\mathbf{a}_{N_{1}}(\psi_{l})\otimes\mathbf{a}_{N_{2}}(\nu_{l}))\nonumber \\
= & (\mathbf{D}_{\mathit{N_{\mathrm{1}}}}^{\mathrm{H}}\mathbf{\mathbf{\mathbf{D}_{\mathit{N_{\mathrm{1}}}}}}\mathbf{a}_{N_{1}}(\psi_{l}))\otimes(\mathbf{U_{\mathit{N_{\mathrm{2}}}}^{\mathrm{H}}}\mathbf{R_{\mathrm{2}}}(\Delta\nu_{\mathit{l}})\mathbf{a}_{N_{2}}(\nu_{l}))\nonumber \\
= & (\widetilde{\mathbf{e}}_{N_{1}})\otimes(\mathbf{U_{\mathit{N_{\mathrm{2}}}}^{\mathrm{H}}}\mathbf{R_{\mathrm{2}}}(\Delta\nu_{\mathit{l}})\mathbf{a}_{N_{2}}(\nu_{l})),\label{eq:Rot_nu}
\end{align}
where $\widetilde{\mathbf{e}}_{N_{1}}\in\mathbb{R}^{N_{1}}$ denotes
a vector with first entry 1 and 0 elsewhere. Similarly, the index
of the nonzero element of $\mathbf{U_{\mathit{N_{\mathrm{2}}}}^{\mathrm{H}}}\mathbf{R_{\mathrm{2}}}(\Delta\nu_{\mathit{l}})\mathbf{a}_{N_{2}}(\nu_{l})$
is $n_{2}(l)$ while the nonzero element index of $\widetilde{\mathbf{e}}_{N_{1}}$
is 1. Then the index of the nonzero element of $\widetilde{\mathbf{U}}_{2}^{\mathrm{H}}\widetilde{\mathbf{R}}_{2}(\Delta\nu_{\mathit{l}})\mathbf{a}_{N}(\psi_{l},\nu_{l})$,
denoted as $\overline{n_{2l}}$, satisfies 
\begin{equation}
\overline{n_{2l}}=n_{2}(l).\label{eq:n2l_bar}
\end{equation}
By combining (\ref{eq:n2l_bar}) with (\ref{eq:Rot_nu}), the $(\overline{n_{2l}},l)$-th
element of $\widetilde{\mathbf{U}}_{2}^{\mathrm{H}}\widetilde{\mathbf{R}}_{2}(\Delta\nu_{\mathit{l}})\mathbf{A}_{N}$
is given by 
\begin{align}
 & [\widetilde{\mathbf{U}}_{2}^{\mathrm{H}}\widetilde{\mathbf{R}}_{2}(\Delta\nu_{\mathit{l}})\mathbf{A}_{N}]_{\overline{n_{2l}},l}\nonumber \\
= & [(\widetilde{\mathbf{e}}_{N_{1}})\otimes(\mathbf{U_{\mathit{N_{\mathrm{2}}}}^{\mathrm{H}}}\mathbf{R_{\mathrm{2}}}(\Delta\nu_{\mathit{l}})\mathbf{a}_{N_{2}}(\nu_{l}))]_{\overline{n_{2l}}}\nonumber \\
= & [\widetilde{\mathbf{e}}_{N_{1}}]_{\mathrm{1}}\otimes[\mathbf{U_{\mathit{N_{\mathrm{2}}}}^{\mathrm{H}}}\mathbf{R_{\mathrm{2}}}(\Delta\nu_{\mathit{l}})\mathbf{a}_{N_{2}}(\nu_{l})]_{n_{2}(l)}\nonumber \\
= & \sqrt{\frac{1}{N_{2}}}\sum_{m=1}^{N_{2}}e^{-\mathrm{i}2\pi(m-1)(\nu_{l}+\frac{\Delta\nu_{l}}{2\pi}-\frac{n_{2}(l)-1}{N_{2}})}.\label{eq:nu-rot-app}
\end{align}

It can been seen that the optimal angle rotation parameters for (\ref{eq:psi-rot-app})
and (\ref{eq:nu-rot-app}) are the same as (\ref{parameters}) and
hence the position index can be obtained.

\subsection{Calculation for Weighted Sum Rate}

\subsubsection{Outline of the Calculation for the WSR}

We investigate the downlink WSR of all users based on the estimate
of the cascaded channel obtained by different algorithms. Specifically,
by jointly optimizing the precoding matrix at the BS and the phase
shift vector at the RIS, the WSR of all users can be obtained via
the following WSR maximization problem \cite{Pan2019multicell} 
\begin{align}
\max_{\mathbf{F},\mathbf{e}} & \sum_{k=1}^{K}\varpi_{k}R_{k}(\mathbf{F},\mathbf{e})\nonumber \\
\mathrm{s.t.} & |[\mathbf{e}]_{m}|=1,1\leq m\leq M,\nonumber \\
 & \sum_{k=1}^{K}||\mathbf{F}_{k}||_{F}^{2}\leq P_{max},\label{opt_rate}
\end{align}
where $\varpi_{k}$, $P_{max}$, and $\mathbf{e}\in\mathbb{C}^{M\times1}$
denote the weighting factor representing the priority of user $k$,
the maximum power provided by the BS, and the phase shift vector at
the RIS, respectively. $\mathbf{F}=\left[\mathbf{F}_{k},\forall k\right]$
where $\mathbf{F}_{k}\in\mathbb{C}^{N\times d}$ is the linear precoding
matrix used by the BS for transmitting its data vector $\mathbf{s}_{k}\in\mathbb{C}^{d}$
to user $k$. $d$ is the number of data streams and the symbol vector
$\mathbf{s}_{k}$ satisfies $\mathbb{E}\left[\mathbf{s}_{k}\mathbf{s}_{k}^{\mathrm{H}}\right]=\mathbf{I}_{d}$
and $\mathbb{E}\left[\mathbf{s}_{k}\mathbf{s}_{j}^{\mathrm{H}}\right]=\mathbf{0}$
for $k\neq j$. In (\ref{opt_rate}), $R_{k}(\mathbf{F},\mathbf{e})$
represents user $k$'s achievable rate (nat/s/Hz) given by
\begin{equation}
R_{k}(\mathbf{F},\mathbf{e})=\log|\mathbf{I}_{Q_{k}}+\bar{\mathbf{H}}_{k}\mathbf{F}_{k}\mathbf{F}_{k}^{\mathrm{H}}\mathbf{\bar{\mathbf{H}}}_{k}^{\mathrm{H}}\mathbf{J}_{k}^{\mathrm{-1}}|,\forall k\in\mathcal{K}.\label{rate_formula}
\end{equation}
In Eq. (\ref{rate_formula}), $\bar{\mathbf{H}}_{k}\triangleq\mathbf{H}_{k}^{\mathrm{H}}\mathrm{Diag}\{\mathbf{e}\}\mathbf{H}^{\mathrm{H}}$
is regarded as the effective cascaded channel matrix during downlink
transmission, which can be written in a similar form to that in (\ref{eq:Q-antenna})
as 
\begin{align}
\bar{\mathbf{H}}_{k} & =[\mathbf{H}\mathrm{Diag}\{\mathbf{e^{*}}\}\mathbf{H}_{k}]^{\mathrm{H}}\nonumber \\
 & =[\mathrm{mat}(\mathrm{vec}(\mathbf{H}\mathrm{Diag}\{\mathbf{e^{*}}\}\mathbf{H}_{k}))_{N\times Q_{k}}]^{\mathrm{H}}\nonumber \\
 & =[\mathrm{mat}((\mathbf{H}_{k}^{\mathrm{T}}\diamond\mathbf{H})\mathbf{e^{*}})_{N\times Q_{k}}]^{\mathrm{H}},\label{cascaded-channel-MIMO}
\end{align}
where $\mathrm{mat}(\cdot)_{N\times Q_{k}}$ denotes the operation
that reshapes an $NQ_{k}\times1$ vector to an $N\times Q_{k}$ matrix.
The term $\mathbf{H}_{k}^{\mathrm{T}}\diamond\mathbf{H}$ is the cascaded
user-RIS-BS channel of user $k$ that can be estimated via the proposed
method in Section \ref{sec:Applying-the-Protocol}. The other matrix
in Eq. (\ref{rate_formula}), i.e., $\mathbf{J}_{k}\in\mathbb{C}^{Q_{k}\times Q_{k}}$,
is known as the interference-plus-noise covariance matrix: 
\begin{equation}
\mathbf{J}_{k}=\sum_{j=1,j\neq k}^{K}\bar{\mathbf{H}}_{k}\mathbf{F}_{j}\mathbf{F}_{j}^{\mathrm{H}}\mathbf{\bar{\mathbf{H}}}_{k}^{\mathrm{H}}+\sigma^{2}\mathbf{I}_{Q_{k}},\label{Rate-Jk}
\end{equation}
where $\sigma^{2}$ is the power of AWGN at the users.

To tackle the non-convex optimization problem (\ref{opt_rate}) where
the precoding matrices $\mathbf{F}$ and the phase shift vector $\mathbf{e}$
are highly coupled, the efficient block coordinate descent (BCD) -
Majorization Minimization (MM) method proposed in \cite{Pan2019multicell}
is adopted. Specifically, the problem can be addressed via the following
steps. first, by exploiting the equivalence between the rate and the
weighted minimum mean-square error (MSE), the original problem is
reformulated into an equivalent and tractable form. Then, BCD algorithm
is used for alternately optimizing the precoding matrices $\mathbf{F}$
at the BS and the phase shift vector $\mathbf{e}$ at the RIS. In
particular, when $\mathbf{e}$ is fixed, the optimal $\mathbf{F}$
can be obtained in closed form by applying the Lagrangian multiplier
method. On the other hand, to address the phase shift optimization
problem, MM algorithm is introduced, which guarantees to converge
to at least a locally optimal solution. It is worth mentioning that
when optimizing the RIS phase shift vector $\mathbf{e}$ during the
optimization process, the equalities $(\mathbf{ABK})\odot(\mathbf{LMT})=(\mathbf{A}\bullet\mathbf{L})(\mathbf{B}\otimes\mathbf{M})(\mathbf{K}\diamond\mathbf{T})$,
$\mathbf{A}^{\mathrm{T}}\diamond\mathbf{B}^{\mathrm{T}}=(\mathbf{A}\bullet\mathbf{B})^{\mathrm{T}}$
and $\mathrm{vec}(\mathbf{ABC})=(\mathbf{C}^{\mathrm{T}}\otimes\mathbf{A})\mathrm{vec}(\mathbf{B})$
are utilized to obtain the explicit expression form of the cascade
channel matrix $\mathbf{H}_{k}^{\mathrm{T}}\diamond\mathbf{H}$ in
the objective function \cite{K-R_product}.

\subsubsection{Detailed Derivation of the Calculation for the WSR}

To better illustrate that how to calculate the WSR using the representation
of the cascaded channels, we present the derivation details of it.

First, by considering the linear decoding matrix for user $k$, denoted
by $\mathbf{U}_{k}\in\mathbb{C}^{Q_{k}\times d}$, the MSE matrix
of user $k$, denoted by $\mathbf{E}_{k}$, is expressed as 
\begin{align}
\mathbf{E}_{k} & =\mathbb{E}[(\mathbf{U}_{k}^{\mathrm{H}}\mathbf{y}_{k}-\mathbf{s}_{k})(\mathbf{U}_{k}^{\mathrm{H}}\mathbf{y}_{k}-\mathbf{s}_{k})^{\mathrm{H}}],\forall k\in\mathcal{K}.\label{MSE-matrix}
\end{align}
Here, $\mathbf{y}_{k}\in\mathbb{C}^{Q_{k}}$ is the received signal
vector of user $k$ given by 
\begin{equation}
\mathbf{y}_{k}=\bar{\mathbf{H}}_{k}\mathbf{F}_{k}\mathbf{s}_{k}+\sum_{j=1,j\neq k}^{K}\bar{\mathbf{H}}_{k}\mathbf{F}_{j}\mathbf{s}_{j}+\mathbf{n}_{k},\label{received-signal-vector-k}
\end{equation}
where $\mathbf{n}_{k}$ is the AWGN satisfying $\mathcal{CN}(0,\delta^{2}\mathbf{I}_{Q_{k}})$.
Then, substitute $\mathbf{y}_{k}$ in (\ref{received-signal-vector-k})
into (\ref{MSE-matrix}), $\mathbf{E}_{k}$ is further written as
\begin{align}
\mathbf{E}_{k}= & (\mathbf{U}_{k}^{\mathrm{H}}\bar{\mathbf{H}}_{k}\mathbf{F}_{k}-\mathbf{I})(\mathbf{U}_{k}^{\mathrm{H}}\bar{\mathbf{H}}_{k}\mathbf{F}_{k}-\mathbf{I})^{\mathrm{H}}+\nonumber \\
 & \sum_{j=1,j\neq k}^{K}\mathbf{U}_{k}^{\mathrm{H}}\bar{\mathbf{H}}_{k}\mathbf{F}_{j}\mathbf{F}_{j}^{\mathrm{H}}\mathbf{\bar{\mathbf{H}}}_{k}^{\mathrm{H}}\mathbf{U}_{k}+\sigma^{2}\mathbf{U}_{k}^{\mathrm{H}}\mathbf{U}_{k}\nonumber \\
= & \mathbf{U}_{k}^{\mathrm{H}}\bar{\mathbf{H}}_{k}\mathbf{F}_{k}\mathbf{F}_{k}^{\mathrm{H}}\mathbf{\bar{\mathbf{H}}}_{k}^{\mathrm{H}}\mathbf{U}_{k}-\mathbf{F}_{k}^{\mathrm{H}}\mathbf{\bar{\mathbf{H}}}_{k}^{\mathrm{H}}\mathbf{U}_{k}-\mathbf{U}_{k}^{\mathrm{H}}\bar{\mathbf{H}}_{k}\mathbf{F}_{k}+\nonumber \\
 & \mathbf{I}+\sum_{j=1,j\neq k}^{K}\mathbf{U}_{k}^{\mathrm{H}}\bar{\mathbf{H}}_{k}\mathbf{F}_{j}\mathbf{F}_{j}^{\mathrm{H}}\mathbf{\bar{\mathbf{H}}}_{k}^{\mathrm{H}}\mathbf{U}_{k}+\sigma^{2}\mathbf{U}_{k}^{\mathrm{H}}\mathbf{U}_{k}\nonumber \\
= & \mathbf{U}_{k}^{\mathrm{H}}\bar{\mathbf{H}}_{k}\sum_{j=1}^{K}\mathbf{F}_{j}\mathbf{F}_{j}^{\mathrm{H}}\mathbf{\bar{\mathbf{H}}}_{k}^{\mathrm{H}}\mathbf{U}_{k}-\mathbf{F}_{k}^{\mathrm{H}}\mathbf{\bar{\mathbf{H}}}_{k}^{\mathrm{H}}\mathbf{U}_{k}-\nonumber \\
 & \mathbf{U}_{k}^{\mathrm{H}}\bar{\mathbf{H}}_{k}\mathbf{F}_{k}+\sigma^{2}\mathbf{U}_{k}^{\mathrm{H}}\mathbf{U}_{k}+\mathbf{I},\forall k\in\mathcal{K}.\label{Ek_formula}
\end{align}
Thus, defining $\mathbf{U}=\{\mathbf{U}_{k},\forall k\in\mathcal{K}\}$
and introducing the set of auxiliary matrices $\mathbf{W}=\{\mathbf{W}_{k}\succeq\mathbf{0},\forall k\in\mathcal{K}\}$,
the problem (\ref{opt_rate}) is transformed into a new form as 
\begin{align}
\max_{\mathbf{W},\mathbf{U},\mathbf{F},\mathbf{e}} & \sum_{k=1}^{K}\varpi_{k}h_{k}(\mathbf{W},\mathbf{U},\mathbf{F},\mathbf{e})\nonumber \\
\mathrm{s.t.} & |[\mathbf{e}]_{m}|=1,1\leq m\leq M,\nonumber \\
 & \sum_{k=1}^{K}||\mathbf{F}_{k}||_{F}^{2}\leq P_{max},\label{opt_eqv-problem}
\end{align}
where $h_{k}(\mathbf{W},\mathbf{U},\mathbf{F},\mathbf{e})=\log|\mathbf{W}_{k}|-\mathrm{Tr}\{\mathbf{W}_{k}\mathbf{E}_{k}\}+d$.
Now, the BCD optimization framework is adopted to address Problem
(\ref{opt_eqv-problem}).

Specifically, when the variables $\mathbf{W}$, $\mathbf{F}$, and
$\mathbf{e}$ are fixed, the optimal $\mathbf{U}_{k}$ is obtained
as follows 
\begin{equation}
\mathbf{U}_{k}=(\mathbf{J}_{k}+\bar{\mathbf{H}}_{k}\mathbf{F}_{k}\mathbf{F}_{k}^{\mathrm{H}}\mathbf{\bar{\mathbf{H}}}_{k}^{\mathrm{H}})^{-1}\bar{\mathbf{H}}_{k}\mathbf{F}_{k}.\label{opt-Uk}
\end{equation}
While fixing the variables $\mathbf{F}$, $\mathbf{e}$, and $\mathbf{U}$,
the optimal $\mathbf{W}_{k}$ is given by 
\begin{equation}
\mathbf{W}_{k}=\mathbf{E}_{k}^{-1}.\label{auxiliary matrix}
\end{equation}
On the other hand, for the given $\mathbf{e}$, $\mathbf{U}$, and
$\mathbf{W}$, the optimal $\mathbf{F}_{k}$ can be found by minimizing
the new problem as 
\begin{align}
\min_{\mathbf{F}} & \sum_{k=1}^{K}\varpi_{k}\mathrm{Tr}\{\mathbf{W}_{k}\mathbf{E}_{k}\}\nonumber \\
\mathrm{s.t.} & \sum_{k=1}^{K}||\mathbf{F}_{k}||_{F}^{2}\leq P_{max}.\label{opt_eqv-problem-F}
\end{align}
By substituting $\mathbf{E}_{k}$ in (\ref{Ek_formula}) into $\mathrm{Tr}\{\mathbf{W}_{k}\mathbf{E}_{k}\}$
and ignoring the constant terms, the problem (\ref{opt_eqv-problem-F})
becomes 
\begin{align}
\min_{\mathbf{F}} & \sum_{j=1}^{K}\mathrm{Tr}\{\mathbf{F}_{j}^{\mathrm{H}}\mathbf{A}\mathbf{F}_{j}\}-\sum_{k=1}^{K}\varpi_{k}\mathrm{Tr}\{\mathbf{W}_{k}\mathbf{F}_{k}^{\mathrm{H}}\mathbf{\bar{\mathbf{H}}}_{k}^{\mathrm{H}}\mathbf{U}_{k}\}\nonumber \\
 & -\sum_{k=1}^{K}\varpi_{k}\mathrm{Tr}\{\mathbf{W}_{k}\mathbf{U}_{k}^{\mathrm{H}}\bar{\mathbf{H}}_{k}\mathbf{F}_{k}\}\nonumber \\
\mathrm{s.t.} & \sum_{k=1}^{K}||\mathbf{F}_{k}||_{F}^{2}\leq P_{max},\label{opt_eqv-problem-F-new}
\end{align}
where the matrix $\mathbf{A}$ is defined as 
\begin{equation}
\mathbf{A}\triangleq\sum_{k=1}^{K}\varpi_{k}\mathbf{\bar{\mathbf{H}}}_{k}^{\mathrm{H}}\mathbf{U}_{k}\mathbf{W}_{k}\mathbf{U}_{k}^{\mathrm{H}}\bar{\mathbf{H}}_{k},\label{matrix-A}
\end{equation}
and the first term $\sum_{j=1}^{K}\mathrm{Tr}\{\mathbf{F}_{j}^{\mathrm{H}}\mathbf{A}\mathbf{F}_{j}\}$
in the objective function is obtained via $\mathrm{Tr}\{\mathbf{D}_{1}\mathbf{D}_{2}\}=\mathrm{Tr}\{\mathbf{D}_{2}\mathbf{D}_{1}\}$.
It is found that Problem (\ref{opt_eqv-problem-F-new}) is a convex
problem and can be addressed by several optimization algorithms.

Now we focus on optimizing $\mathbf{e}$ when fixing the variables
$\mathbf{U}$, $\mathbf{W}$, and $\mathbf{F}$. Similar to what was
done from Problem (\ref{opt_eqv-problem-F}) to Problem (\ref{opt_eqv-problem-F-new}),
we substitute $\mathbf{E}_{k}$ in (\ref{Ek_formula}) into $\mathrm{Tr}\{\mathbf{W}_{k}\mathbf{E}_{k}\}$
and ignoring the constant terms. Then, the optimal $\mathbf{e}$ can
be found by solving the minimization problem as 
\begin{align}
\min_{\mathbf{e}} & \sum_{k=1}^{K}\mathrm{Tr}\{\varpi_{k}\mathbf{W}_{k}\mathbf{U}_{k}^{\mathrm{H}}\bar{\mathbf{H}}_{k}\mathbf{P}\mathbf{\bar{\mathbf{H}}}_{k}^{\mathrm{H}}\mathbf{U}_{k}\}-\nonumber \\
 & \sum_{k=1}^{K}\mathrm{Tr}\{\varpi_{k}\mathbf{W}_{k}\mathbf{F}_{k}^{\mathrm{H}}\mathbf{\bar{\mathbf{H}}}_{k}^{\mathrm{H}}\mathbf{U}_{k}\}-\sum_{k=1}^{K}\mathrm{Tr}\{\varpi_{k}\mathbf{W}_{k}\mathbf{U}_{k}^{\mathrm{H}}\bar{\mathbf{H}}_{k}\mathbf{F}_{k}\}\nonumber \\
\mathrm{s.t.} & |[\mathbf{e}]_{m}|=1,1\leq m\leq M,\label{opt_eqv-problem-e}
\end{align}
where $\mathbf{P}\triangleq\sum_{k=1}^{K}\mathbf{F}_{k}\mathbf{F}_{k}^{\mathrm{H}}$.
For notation simplicity, the downlink cascaded channel $\bar{\mathbf{H}}_{k}$
can be re-expressed as 
\begin{equation}
\bar{\mathbf{H}}_{k}=\mathbf{H}_{k}^{\mathrm{H}}\mathrm{Diag}\{\mathbf{e}\}\mathbf{H}^{\mathrm{H}}=\mathbf{\tilde{H}}_{k}\boldsymbol{\Phi}\mathbf{\tilde{H}},\label{cascaded-channel-new-form}
\end{equation}
where $\mathbf{\tilde{H}}_{k}\triangleq\mathbf{H}_{k}^{\mathrm{H}}$,
$\boldsymbol{\Phi}\triangleq\mathrm{Diag}\{\mathbf{e}\}$, and $\mathbf{\tilde{H}}\triangleq\mathbf{H}^{\mathrm{H}}$
are the RIS-user channel of user $k$, the phase shift matrix at the
RIS, and the BS-RIS channel during downlink transmission, respectively.

Then, substitute $\bar{\mathbf{H}}_{k}$ in (\ref{cascaded-channel-new-form})
into the objective function of Problem (\ref{opt_eqv-problem-e}),
its first term, i.e., $\sum_{k=1}^{K}\mathrm{Tr}\{\varpi_{k}\mathbf{W}_{k}\mathbf{U}_{k}^{\mathrm{H}}\bar{\mathbf{H}}_{k}\mathbf{P}\mathbf{\bar{\mathbf{H}}}_{k}^{\mathrm{H}}\mathbf{U}_{k}\}$,
can be rewritten as 
\begin{align}
 & \sum_{k=1}^{K}\mathrm{Tr}\{\varpi_{k}\mathbf{W}_{k}\mathbf{U}_{k}^{\mathrm{H}}\bar{\mathbf{H}}_{k}\mathbf{P}\mathbf{\bar{\mathbf{H}}}_{k}^{\mathrm{H}}\mathbf{U}_{k}\}\nonumber \\
= & \sum_{k=1}^{K}\mathrm{Tr}\{\varpi_{k}\mathbf{W}_{k}\mathbf{U}_{k}^{\mathrm{H}}\mathbf{\tilde{H}}_{k}\boldsymbol{\Phi}\mathbf{\tilde{H}}\mathbf{P}\mathbf{\tilde{H}^{\mathrm{H}}}\boldsymbol{\Phi}^{\mathrm{H}}\mathbf{\tilde{H}}_{k}^{\mathrm{H}}\mathbf{U}_{k}\}\nonumber \\
= & \sum_{k=1}^{K}\mathrm{Tr}\{\boldsymbol{\Phi}^{\mathrm{H}}\varpi_{k}\mathbf{\tilde{H}}_{k}^{\mathrm{H}}\mathbf{U}_{k}\mathbf{W}_{k}\mathbf{U}_{k}^{\mathrm{H}}\mathbf{\tilde{H}}_{k}\boldsymbol{\Phi}\mathbf{\tilde{H}}\mathbf{P}\mathbf{\tilde{H}^{\mathrm{H}}}\}\nonumber \\
= & \sum_{k=1}^{K}\mathrm{Tr}\{\boldsymbol{\Phi}^{\mathrm{H}}\mathbf{B}_{k}\boldsymbol{\Phi}\mathbf{C}\}=\mathrm{Tr}\{\boldsymbol{\Phi}^{\mathrm{H}}(\sum_{k=1}^{K}\mathbf{B}_{k})\boldsymbol{\Phi}\mathbf{C}\}=\mathbf{e}^{\mathrm{H}}\boldsymbol{\Xi}\mathbf{e},\label{opt-phi-first-term}
\end{align}
where the matrix $\mathbf{B}_{k}$, $\mathbf{C}$, and $\boldsymbol{\Xi}$
are all semi-definite matrices satisfying 
\begin{align}
\mathbf{B}_{k} & \triangleq\varpi_{k}\mathbf{\tilde{H}}_{k}^{\mathrm{H}}\mathbf{U}_{k}\mathbf{W}_{k}\mathbf{U}_{k}^{\mathrm{H}}\mathbf{\tilde{H}}_{k},\label{def-B_k}\\
\mathbf{C} & \triangleq\mathbf{\tilde{H}}\mathbf{P}\mathbf{\tilde{H}^{\mathrm{H}}},\label{def-C}\\
\boldsymbol{\Xi} & \triangleq(\sum_{k=1}^{K}\mathbf{B}_{k})\odot\mathbf{C}^{\mathrm{T}}.\label{def-Xi}
\end{align}

Similarly, by defining $\mathbf{T}_{k}\triangleq\varpi_{k}\mathbf{\tilde{H}}\mathbf{F}_{k}\mathbf{W}_{k}\mathbf{U}_{k}^{\mathrm{H}}\mathbf{\tilde{H}}_{k}$,
the second term of the objective function in Problem (\ref{opt_eqv-problem-e}),
i.e., $-\sum_{k=1}^{K}\mathrm{Tr}\{\varpi_{k}\mathbf{W}_{k}\mathbf{F}_{k}^{\mathrm{H}}\mathbf{\bar{\mathbf{H}}}_{k}^{\mathrm{H}}\mathbf{U}_{k}\}$,
becomes 
\begin{align}
 & -\sum_{k=1}^{K}\mathrm{Tr}\{\varpi_{k}\mathbf{W}_{k}\mathbf{F}_{k}^{\mathrm{H}}\mathbf{\bar{\mathbf{H}}}_{k}^{\mathrm{H}}\mathbf{U}_{k}\}\nonumber \\
= & -\sum_{k=1}^{K}\mathrm{Tr}\{\varpi_{k}\mathbf{W}_{k}\mathbf{F}_{k}^{\mathrm{H}}\mathbf{\tilde{H}^{\mathrm{H}}}\boldsymbol{\Phi}^{\mathrm{H}}\mathbf{\tilde{H}}_{k}^{\mathrm{H}}\mathbf{U}_{k}\}\nonumber \\
= & -\sum_{k=1}^{K}\mathrm{Tr}\{\boldsymbol{\Phi}^{\mathrm{H}}\varpi_{k}\mathbf{\tilde{H}}_{k}^{\mathrm{H}}\mathbf{U}_{k}\mathbf{W}_{k}\mathbf{F}_{k}^{\mathrm{H}}\mathbf{\tilde{H}^{\mathrm{H}}}\}\nonumber \\
= & -\sum_{k=1}^{K}\mathrm{Tr}\{\boldsymbol{\Phi}^{\mathrm{H}}\mathbf{T}_{k}^{\mathrm{H}}\}\nonumber \\
= & \mathrm{Tr}\{\boldsymbol{\Phi}^{\mathrm{H}}(-\sum_{k=1}^{K}\mathbf{T}_{k})^{\mathrm{H}}\}=\mathrm{Tr}\{\boldsymbol{\Phi}^{\mathrm{H}}\mathbf{V}^{\mathrm{H}}\}=\mathbf{v}^{\mathrm{H}}\mathbf{e}^{*},\label{opt-phi-second-term}
\end{align}
where the matrix $\mathbf{V}$ and the vector $\mathbf{v}$ are defined
as 
\begin{align}
\mathbf{V} & \triangleq(-\sum_{k=1}^{K}\mathbf{T}_{k}),\label{def-v-mtx}\\
\mathbf{v} & \triangleq\left[[\mathbf{V}]_{1,1},[\mathbf{V}]_{2,2},...,[\mathbf{V}]_{M,M}\right]^{\mathrm{T}}.\label{def-v-vector}
\end{align}
Taking the derivation similar to (\ref{opt-phi-second-term}), the
third term of the objective function in Problem (\ref{opt_eqv-problem-e}),
i.e., $-\sum_{k=1}^{K}\mathrm{Tr}\{\varpi_{k}\mathbf{W}_{k}\mathbf{U}_{k}^{\mathrm{H}}\bar{\mathbf{H}}_{k}\mathbf{F}_{k}\}$,
is naturally expressed as 
\begin{align}
 & -\sum_{k=1}^{K}\mathrm{Tr}\{\varpi_{k}\mathbf{W}_{k}\mathbf{U}_{k}^{\mathrm{H}}\bar{\mathbf{H}}_{k}\mathbf{F}_{k}\}\nonumber \\
= & -\sum_{k=1}^{K}\mathrm{Tr}\{\mathbf{T}_{k}\boldsymbol{\Phi}\}=\mathrm{Tr}\{(-\sum_{k=1}^{K}\mathbf{T}_{k})\boldsymbol{\Phi}\}=\mathbf{v}^{\mathrm{T}}\mathbf{e}.\label{opt-phi-third-term}
\end{align}
Based on the above derivation, the problem (\ref{opt_eqv-problem-e})
is reformulated as 
\begin{align}
\min_{\mathbf{e}} & \mathbf{e}^{\mathrm{H}}\boldsymbol{\Xi}\mathbf{e}+\mathbf{v}^{\mathrm{T}}\mathbf{e}+\mathbf{v}^{\mathrm{H}}\mathbf{e}^{*}\nonumber \\
\mathrm{s.t.} & |[\mathbf{e}]_{m}|=1,1\leq m\leq M.\label{opt_eqv-problem-e-new}
\end{align}
To address the non-convex problem (\ref{opt_eqv-problem-e-new}),
MM algorithm is introduced; see for example \cite{Pan2019multicell}.
However, the method in \cite{Pan2019multicell} assumes that the BS
knows the seperate channels $\mathbf{H}$ and $\mathbf{H}_{k}$, while
in Section \ref{sec:Applying-the-Protocol}, what we have only to
obtain is the cascaded channel matrix $\mathbf{H}_{k}^{\mathrm{T}}\diamond\mathbf{H}$
instead of the channel matrices $\mathbf{H}$ and $\mathbf{H}_{k}$.
Therefore, we will show that only using the cascaded channel $\mathbf{H}_{k}^{\mathrm{T}}\diamond\mathbf{H}$
is still available for the passive beamforming at the RIS during downlink
transmission, that is to say, $\boldsymbol{\Xi}$ and $\mathbf{v}$
only depend on the cascaded channel matrices $\mathbf{H}_{k}^{\mathrm{T}}\diamond\mathbf{H}$
instead of the seperate channel matrices $\mathbf{H}$ and $\mathbf{H}_{k}$
for $\forall k$.

First, we use the representation of the cascaded channel matrix $\mathbf{H}_{k}^{\mathrm{T}}\diamond\mathbf{H}$
to characterize the matrix $\boldsymbol{\Xi}$. Specifically, $\boldsymbol{\Xi}$
can be re-expressed as 
\begin{align}
\boldsymbol{\Xi} & =(\sum_{k=1}^{K}\mathbf{B}_{k})\odot\mathbf{C}^{\mathrm{T}}=\sum_{k=1}^{K}(\mathbf{B}_{k}\odot\mathbf{C}^{\mathrm{T}}),\label{def-xi-new}
\end{align}
where the term $\mathbf{B}_{k}\odot\mathbf{C}^{\mathrm{T}}$ can be
further rewritten as 
\begin{align}
\mathbf{B}_{k}\odot\mathbf{C}^{\mathrm{T}}= & (\varpi_{k}\mathbf{\tilde{H}}_{k}^{\mathrm{H}}\mathbf{U}_{k}\mathbf{W}_{k}\mathbf{U}_{k}^{\mathrm{H}}\mathbf{\tilde{H}}_{k})\odot(\mathbf{\tilde{H}}\mathbf{P}\mathbf{\tilde{H}^{\mathrm{H}}})^{\mathrm{T}}\nonumber \\
= & \varpi_{k}(\mathbf{\tilde{H}}_{k}^{\mathrm{H}}\mathbf{D}_{k}\mathbf{\tilde{H}}_{k})\odot(\mathbf{\tilde{H}^{*}}\mathbf{P}^{\mathrm{T}}\mathbf{\tilde{H}^{\mathrm{T}}})\nonumber \\
= & \varpi_{k}(\mathbf{\tilde{H}}_{k}^{\mathrm{H}}\bullet\mathbf{\tilde{H}^{*}})(\mathbf{D}_{k}\otimes\mathbf{P}^{\mathrm{T}})(\mathbf{\tilde{H}}_{k}\diamond\mathbf{\tilde{H}^{\mathrm{T}}}).\label{def-xi-new-part}
\end{align}
Here, we define $\mathbf{D}_{k}\triangleq\mathbf{U}_{k}\mathbf{W}_{k}\mathbf{U}_{k}^{\mathrm{H}}$
and the third equality is obtained via $(\mathbf{ABK})\odot(\mathbf{LMT})=(\mathbf{A}\bullet\mathbf{L})(\mathbf{B}\otimes\mathbf{M})(\mathbf{K}\diamond\mathbf{T})$.
Using the equality $\mathbf{A}^{\mathrm{T}}\diamond\mathbf{B}^{\mathrm{T}}=(\mathbf{A}\bullet\mathbf{B})^{\mathrm{T}}$,
the relationships between the cascaded channel matrices $\mathbf{H}_{k}^{\mathrm{T}}\diamond\mathbf{H}$
and the terms $\mathbf{\tilde{H}}_{k}^{\mathrm{H}}\bullet\mathbf{\tilde{H}^{*}}$
and $\mathbf{\tilde{H}}_{k}\diamond\mathbf{\tilde{H}^{\mathrm{T}}}$
in the third equality of (\ref{def-xi-new-part}) are given by 
\begin{align}
(\mathbf{\tilde{H}}_{k}^{\mathrm{H}}\bullet\mathbf{\tilde{H}^{*}})^{\mathrm{T}}= & (\mathbf{\tilde{H}}_{k}^{\mathrm{H}})^{\mathrm{T}}\diamond(\mathbf{\tilde{H}^{*}})^{\mathrm{T}}=\mathbf{\tilde{H}}_{k}^{*}\diamond\mathbf{\tilde{H}}^{\mathrm{H}}=\mathbf{H}_{k}^{\mathrm{T}}\diamond\mathbf{H}=\mathbf{G}_{k},\label{def-xi-new-part-1}\\
(\mathbf{\tilde{H}}_{k}\diamond\mathbf{\tilde{H}^{\mathrm{T}}})^{*}= & \mathbf{\tilde{H}}_{k}^{*}\diamond\mathbf{\tilde{H}}^{\mathrm{H}}=\mathbf{H}_{k}^{\mathrm{T}}\diamond\mathbf{H}=\mathbf{G}_{k}.\label{def-xi-new-part-2}
\end{align}
Thus, $\boldsymbol{\Xi}$ in (\ref{def-xi-new}) is obtained as follows
\begin{equation}
\boldsymbol{\Xi}=\sum_{k=1}^{K}(\mathbf{B}_{k}\odot\mathbf{C}^{\mathrm{T}})=\sum_{k=1}^{K}\varpi_{k}\mathbf{G}_{k}^{\mathrm{T}}(\mathbf{D}_{k}\otimes\mathbf{P}^{\mathrm{T}})\mathbf{G}_{k}^{\mathrm{*}}.\label{def-xi-final}
\end{equation}

Next, we use the representation of the cascaded channel matrix $\mathbf{H}_{k}^{\mathrm{T}}\diamond\mathbf{H}$
to characterize the vector $\mathbf{v}$. We start the derivation
from re-expressing $\mathbf{V}$ in (\ref{def-v-mtx}) as 
\begin{align}
\mathbf{V} & =-\sum_{k=1}^{K}\mathbf{T}_{k}=-\sum_{k=1}^{K}\mathbf{\tilde{H}}\varpi_{k}\mathbf{F}_{k}\mathbf{W}_{k}\mathbf{U}_{k}^{\mathrm{H}}\mathbf{\tilde{H}}_{k}=-\sum_{k=1}^{K}\mathbf{\tilde{H}}\mathbf{C}_{k}\mathbf{\tilde{H}}_{k},\label{def-V}
\end{align}
where we define $\mathbf{C}_{k}\triangleq\varpi_{k}\mathbf{F}_{k}\mathbf{W}_{k}\mathbf{U}_{k}^{\mathrm{H}}$.
Due to the fact that $\mathbf{T}_{k}=\mathbf{\tilde{H}}\mathbf{C}_{k}\mathbf{\tilde{H}}_{k}$,
considering the partitions of $\mathbf{\tilde{H}}$ by rows and the
partitions of $\mathbf{\tilde{H}}_{k}$ by columns, it can be readily
verified that we have 
\begin{equation}
[\mathbf{T}_{k}]_{i,i}=[\mathbf{\tilde{H}}]_{i,:}\mathbf{C}_{k}[\mathbf{\tilde{H}}_{k}]_{:,i}=([\mathbf{\tilde{H}}_{k}]_{:,i}^{\mathrm{T}}\otimes[\mathbf{\tilde{H}}]_{i,:})\mathrm{vec}(\mathbf{C}_{k}),\label{def-Tk-ii}
\end{equation}
where the second equality is obtained by vectorizing the element $[\mathbf{T}_{k}]_{i,i}$
and using the equality $\mathrm{vec}(\mathbf{ABC})=(\mathbf{C}^{\mathrm{T}}\otimes\mathbf{A})\mathrm{vec}(\mathbf{B})$.
Let $\mathbf{t}_{k}$ be the stack of the diagonal elements of the
matrix $\mathbf{T}_{k}$, denoted by $\mathbf{t}_{k}\triangleq\left[[\mathbf{T}_{k}]_{1,1},[\mathbf{T}_{k}]_{2,2},...,[\mathbf{T}_{k}]_{M,M}\right]^{\mathrm{T}}$,
we have 
\begin{align}
\mathbf{t}_{k} & =\left[\begin{array}{c}
([\mathbf{\tilde{H}}_{k}]_{:,1}^{\mathrm{T}}\otimes[\mathbf{\tilde{H}}]_{1,:})\mathrm{vec}(\mathbf{C}_{k})\\
([\mathbf{\tilde{H}}_{k}]_{:,2}^{\mathrm{T}}\otimes[\mathbf{\tilde{H}}]_{2,:})\mathrm{vec}(\mathbf{C}_{k})\\
\vdots\\
([\mathbf{\tilde{H}}_{k}]_{:,M}^{\mathrm{T}}\otimes[\mathbf{\tilde{H}}]_{M,:})\mathrm{vec}(\mathbf{C}_{k})
\end{array}\right]\nonumber \\
 & =\left[\begin{array}{c}
([\mathbf{\tilde{H}}_{k}]_{:,1}^{\mathrm{T}}\otimes[\mathbf{\tilde{H}}]_{1,:})\\
([\mathbf{\tilde{H}}_{k}]_{:,2}^{\mathrm{T}}\otimes[\mathbf{\tilde{H}}]_{2,:})\\
\vdots\\
([\mathbf{\tilde{H}}_{k}]_{:,M}^{\mathrm{T}}\otimes[\mathbf{\tilde{H}}]_{M,:})
\end{array}\right]\mathrm{vec}(\mathbf{C}_{k})\nonumber \\
 & =(\mathbf{\tilde{H}}_{k}^{\mathrm{T}}\bullet\mathbf{\tilde{H}})\mathrm{vec}(\mathbf{C}_{k})=(\mathbf{H}_{k}^{\mathrm{T}}\diamond\mathbf{H}){}^{\mathrm{H}}\mathrm{vec}(\mathbf{C}_{k}),\label{def-tk}
\end{align}
where the forth equality is obtained via the simplification of the
term $(\mathbf{\tilde{H}}_{k}^{\mathrm{T}}\bullet\mathbf{\tilde{H}})$
as follows 
\begin{align}
\mathbf{\tilde{H}}_{k}^{\mathrm{T}}\bullet\mathbf{\tilde{H}} & =[(\mathbf{\tilde{H}}_{k}^{\mathrm{T}}\bullet\mathbf{\tilde{H}})^{\mathrm{T}}]^{\mathrm{T}}\nonumber \\
 & =[\mathbf{\tilde{H}}_{k}\diamond\mathbf{\tilde{H}^{\mathrm{T}}}]^{\mathrm{T}}=[\mathbf{H}_{k}^{\mathrm{H}}\diamond\mathbf{H}^{*}]^{\mathrm{T}}=[\mathbf{H}_{k}^{\mathrm{T}}\diamond\mathbf{H}]{}^{\mathrm{H}}.\label{derivation-cascaded-v}
\end{align}
Note that there is a relationship between $\mathbf{v}$ in (\ref{opt_eqv-problem-e-new})
and $\mathbf{t}_{k}$ in (\ref{def-tk}) as 
\begin{equation}
\mathbf{v}=-\sum_{k=1}^{K}\mathbf{t}_{k}=-\sum_{k=1}^{K}[\mathbf{H}_{k}^{\mathrm{T}}\diamond\mathbf{H}]{}^{\mathrm{H}}\mathrm{vec}(\mathbf{C}_{k})=-\sum_{k=1}^{K}\mathbf{G}_{k}^{\mathrm{H}}\mathrm{vec}(\mathbf{C}_{k}),\label{def_v_final}
\end{equation}
thus the vector $\mathbf{v}$ can be obtained based on the estimate
of the cascaded channel matrices. Finally, the problem (\ref{opt_eqv-problem-e-new})
is formulated as 
\begin{align}
\min_{\mathbf{e}} & \mathbf{e}^{\mathrm{H}}\boldsymbol{\Xi}\mathbf{e}+2\mathrm{Re}\{\mathbf{v}^{\mathrm{H}}\mathbf{e}^{*}\}\nonumber \\
\mathrm{s.t.} & |[\mathbf{e}]_{m}|=1,1\leq m\leq M.\label{opt_eqv-problem-e-final}
\end{align}
Where $\boldsymbol{\Xi}$ and $\mathbf{v}$ are determined by the
obtained cascaded channels. They are given by 
\begin{align}
\boldsymbol{\Xi} & =\sum_{k=1}^{K}\varpi_{k}\mathbf{G}_{k}^{\mathrm{T}}(\mathbf{D}_{k}\otimes\mathbf{P}^{\mathrm{T}})\mathbf{G}_{k}^{\mathrm{*}},\label{Xi-opt_final}\\
\mathbf{v} & =-\sum_{k=1}^{K}\mathbf{G}_{k}^{\mathrm{H}}\mathrm{vec}(\mathbf{C}_{k}).\label{v-opt_final}
\end{align}
Finally, Problem (\ref{opt_eqv-problem-e-final}) can be solved effectively
by introducing the MM algorithm.

\subsubsection{Summary of the BCD Optimization Framework}

Now we conclude the BCD optimization framework for solving the equivalent
maximization problem (\ref{opt_eqv-problem}) using the the representation
of the cascaded channels. Given the variables $\mathbf{F}$ and $\mathbf{e}$,
the optimal $\mathbf{U}$ is obtained according to (\ref{opt-Uk});
Given the variables $\mathbf{F}$, $\mathbf{e}$, and $\mathbf{U}$,
the optimal $\mathbf{W}$ is calculated via (\ref{auxiliary matrix});
Given the variables $\mathbf{e}$, $\mathbf{U}$, and $\mathbf{W}$,
the optimal $\mathbf{F}$ is given by the solution to Problem (\ref{opt_eqv-problem-F-new});
Given the variables $\mathbf{U}$, $\mathbf{W}$, and $\mathbf{F}$,
the optimal $\mathbf{e}$ is found by solving the problem (\ref{opt_eqv-problem-e-final}).

Finally, by adopting the BCD-MM method and using the obtained cascaded
channel estimated by different estimation algorithms, we can calculate
the WSR achieved by different estimation methods shown in Fig. \ref{rate}.

\bibliographystyle{IEEEtran}
\bibliography{bibfile}

\end{document}